\documentclass[twocolumn,aps,prx,floatfix,superscriptaddress]{revtex4-2}

\usepackage{amsmath,amssymb,amsfonts,graphicx,braket}

\usepackage[colorlinks=true,citecolor=blue,linkcolor=magenta]{hyperref}
\hypersetup{colorlinks = true}
\usepackage[usenames]{color}
\usepackage{relsize}
\usepackage[english]{babel}
\usepackage{enumerate}

\usepackage{xspace}

\graphicspath{{figs/},{figs_main/}{figs_SI/}{./}}

\usepackage{algorithm}
\usepackage[noend]{algpseudocode}

\usepackage{physics}

\newcommand{\nocontentsline}[3]{}
\newcommand{\tocless}[2]{\bgroup\let\addcontentsline=\nocontentsline #1{#2}\egroup}

\usepackage{bbold}

\usepackage[normalem]{ulem}

\newcommand{\SI}{Supplemental Material:\xspace}

\begin{document}
	
	\newcommand {\titletext} {Self-Correcting Quantum Many-Body Control  \\
		using Reinforcement Learning with Tensor Networks }
	
	\def \abstracttext {
		Quantum many-body control is a central milestone en route to harnessing quantum technologies. However, the exponential growth of the Hilbert space dimension with the number of qubits makes it challenging to classically simulate quantum many-body systems and consequently, to devise reliable and robust optimal control protocols. Here, we present a novel framework for efficiently controlling quantum many-body systems based on reinforcement learning (RL). We tackle the quantum control problem by leveraging matrix product states (i) for representing the many-body state and, (ii) as part of the trainable machine learning architecture for our RL agent. 
		The framework is applied to prepare ground states of the quantum Ising chain, including states in the critical region. It allows us to control systems  far larger than neural-network-only architectures permit, while retaining the advantages of deep learning algorithms, such as generalizability and trainable robustness to noise. In particular, we demonstrate that RL agents are capable of finding universal controls, of learning how to optimally steer previously unseen many-body states, and of adapting control protocols on-the-fly when the quantum dynamics is subject to stochastic perturbations. Furthermore, we map the QMPS framework to a hybrid quantum-classical algorithm that can be performed on noisy intermediate-scale quantum devices and test it under the presence of experimentally relevant sources of noise.
	}
	
	\author{Friederike Metz}
	\email[Current affiliations:~Institute of Physics, École Polytechnique Fédérale de Lausanne (EPFL), CH-1015 Lausanne, Switzerland\\Center for Quantum Science and Engineering, École Polytechnique Fédérale de Lausanne (EPFL), CH-1015 Lausanne, Switzerland\\]{friederike.metz@epfl.ch}
	\affiliation{Quantum Systems Unit, Okinawa Institute of Science and Technology Graduate University, 1919-1 Tancha, Onna, Okinawa 904-0495, Japan}
	\affiliation{Max Planck Institute for the Physics of Complex Systems, N\"othnitzer Str.~38, 01187 Dresden, Germany}
	
	\author{Marin Bukov}
	\email{mgbukov@pks.mpg.de}
	\affiliation{Max Planck Institute for the Physics of Complex Systems, N\"othnitzer Str.~38, 01187 Dresden, Germany}
	\affiliation{Department of Physics, St.~Kliment Ohridski University of Sofia, 5 James Bourchier Blvd, 1164 Sofia, Bulgaria}

	\title{\titletext}

	\begin{abstract} 
		\abstracttext
	\end{abstract}
	
	\maketitle
	
	\tocless{
		
\section{Introduction}

Quantum many-body control is an essential prerequisite for the reliable operation of modern quantum technologies which are based on harnessing quantum correlations. 
For example, quantum computing often involves high-fidelity state manipulation as a necessary component of most quantum algorithms~\cite{farhi2014,kandala2017}; In quantum simulation, the underlying AMO platforms require to prepare the system in a desired state before its properties can be measured and studied~\cite{lewenstein2007,blatt2012,casola2018}; And quantum metrology relies on the controlled engineering of (critical) states to maximize the sensitivity to physical parameters~\cite{rams2018,pang2017}.
Controlling many-body systems can also be considered in its own right, as a numerical tool which offers insights into concepts such as quantum phases and phase transitions~\cite{matos2021}. Moreover, it can reveal novel theoretical phenomena such as phase transitions in the control landscape~\cite{day2019}, and bears a direct relation to our understanding of quantum complexity~\cite{farhi2016quantum}.

\begin{figure}[t!]
	\centering
	\includegraphics[width=0.99\columnwidth]{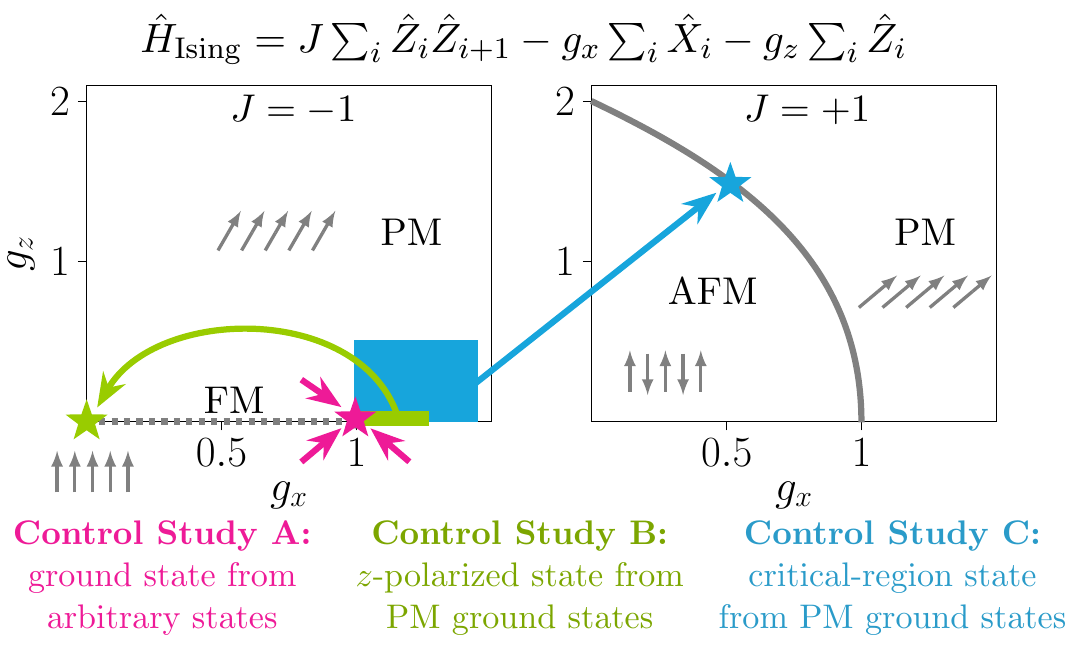}\vspace{3mm}
	\includegraphics[width=0.95\columnwidth]{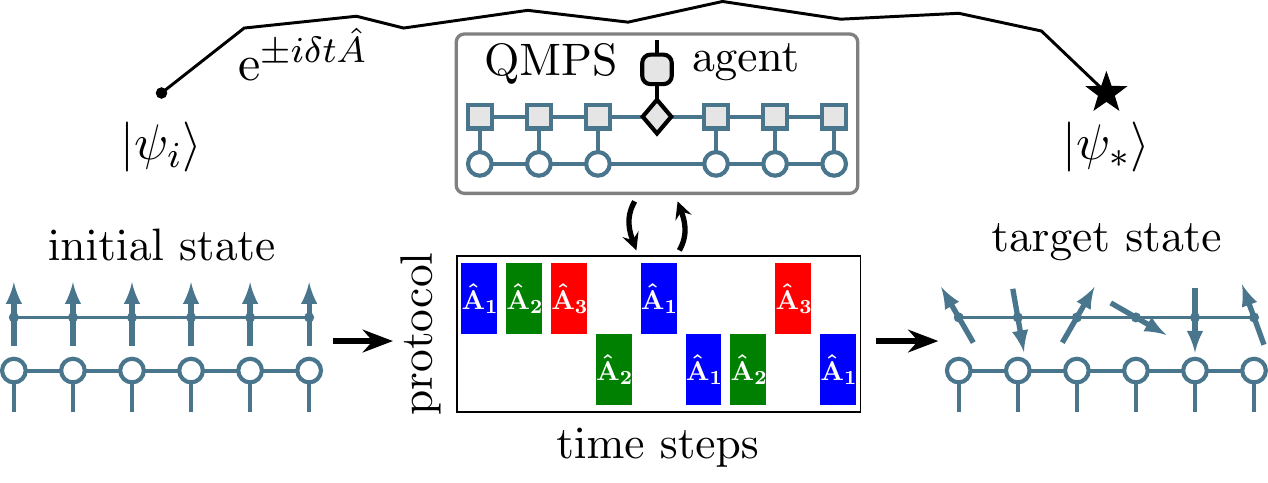}
	\caption{\label{fig:intro}
		{\bf Many-body control studies}
		in the ground state phase diagram of the quantum Ising model, analyzed in this work:~an RL agent is trained to prepare a ground state of the transverse field Ising model from random initial states (Ctrl Study A, magenta), the $z$-polarized product state from a class of paramagnetic ground states (Ctrl Study B, green), and a ground state in the critical region of the mixed field Ising model from paramagnetic ground states of opposite interaction strength (Ctrl Study C, cyan). The optimized agent outputs a control protocol as a sequence of operators $\hat A_j$, which time evolve the initial spin state into the desired target state (marked by a star).
	}
\end{figure}

Compared to single- and few-particle physics, working in the quantum \textit{many-body} domain introduces the formidable difficulty of dealing with an exponentially large Hilbert space. A specific manifestation is the accurate description and manipulation of quantum entanglement shared between many degrees of freedom. 
This poses a limitation for classical simulation methods, since memory and compute time resources scale exponentially with the system size. 

Fortunately, there exists a powerful framework to simulate the physics of one-dimensional (1d) quantum many-body systems, based on matrix product states (MPS)~\cite{white1992,ostlund1995,schollwock2011,orus2014}. 
MPS provide a compressed representation of many-body wave functions
and allow for efficient computation with resources scaling only linearly in the system size for area-law entangled states~\cite{hastings2007,schuch2008}.

While MPS-based algorithms have been used in the context of optimal many-body control to find high-fidelity protocols
~\cite{doria2011,vanFrank2016,jensen2021,Luchnikov22}, 
the advantages of deep reinforcement learning (RL) for quantum control \cite{Krenn2023}, have so far been investigated using exact simulations of only a small number of interacting quantum degrees of freedom.
Nevertheless, policy-gradient and value-function RL algorithms have recently been established as useful tools in the study of quantum state preparation~\cite{bukov2018,bukov2018b,Haug2020,mackeprang2020,niu2019,yao2020,yao2020b,haug2021,guo2021,yao2021,bolens2021,he2021,guo2021,cao2021,Porotti2022,Porotti2022b,Sivak2022,Reuer2022,Yao22},
quantum error correction and mitigation \cite{fosel2018,nautrup2019,andreasson2019,sweke2021}, quantum circuit design~\cite{zhang2020,moro2021,he2021b,fosel2021}, quantum metrology~\cite{xu2019,schuff2020}, and quantum heat engines \cite{Erdman22,Erdman22b}; quantum reinforcement learning algorithms have been proposed as well~\cite{chen2021,lockwood2020,dunjko2017,jerbi2021a,saggio2021}. 
Thus, in times of rapidly developing quantum simulators which exceed the computational capabilities of classical computers~\cite{ebadi2021quantum}, the natural question arises regarding scaling up the size of quantum systems in RL control studies beyond exact diagonalization methods.
We discuss a proof-of-principle implementation of the algorithm on noisy intermediate-scale quantum (NISQ) devices for small system sizes.

In this work, we develop a new deep RL framework for quantum many-body control, based on MPS in two complementary ways.
First, we adopt the MPS description of quantum states: this allows us to control large interacting 1d systems, whose quantum dynamics we simulate within the RL environment. Second, representing the RL state in the form of an MPS, naturally suggests the use of tensors network as (part of) the deep learning architecture for the RL agent, e.g., instead of a conventional neural network (NN) ansatz. Therefore, inspired by earlier examples of tensor-network-based machine learning~\cite{miles2016,han2018,glasser2019b}, we approximate the RL agent as a hybrid MPS-NN network, called QMPS. With these innovations at hand, the required computational resources scale linearly with the system size, in contrast to learning from the full many-body wave function. Ultimately, this allows us to train an RL agent to control a larger number of interacting quantum particles, as required by present-day quantum simulators.

As a concrete example, we consider the problem of state preparation and present three case studies in which we prepare different ground states of the paradigmatic mixed field Ising chain [Fig.~\ref{fig:intro}]. We train QMPS agents to prepare target states from a class of initial (ground) states, and devise universal controls with respect to experimentally relevant sets of initial states. 
In contrast to conventional quantum control algorithms (such as CRAB or GRAPE \cite{vanFrank2016,doria2011,grape}), once the optimization is complete, RL agents retain information during the training process in form of a policy or a value function. When enhanced with a deep learning architecture, the learned control policy generalizes to states not seen during training.
We demonstrate how this singular feature of deep RL allows our agents to efficiently control quantum Ising chains
(i) starting from various initial states that the RL agent has never encountered, and
(ii) in the presence of faulty or noisy controls and stochastic dynamics.
Thus, even in analytically intractable many-body regimes, an online RL agent produces particularly robust control protocols.

\section{\label{sec:stateprep}Quantum Many-body Control}

Consider a quantum many-body system in the initial state $\ket{\psi_i}$. Our objective is to find optimal protocols that evolve the system into a desired target state $\ket{\psi_\ast}$. We construct these protocols as a sequence of $q$ consecutive unitary operators $U(\tau)=\prod_{j=1}^{q} U_{\tau_{j}}$, where $U_{\tau_{j}}\!\!\in\!\mathcal{A}$ are chosen from a set $\mathcal{A}$. 
To assess the quality of a given protocol, we compute the fidelity of the evolved state w.r.t.~the target state:  
\begin{equation}
	F(\tau) = |\bra{\psi_\ast} U(\tau) \ket{\psi_i}|^2.
\end{equation}

Throughout the study, we focus on spin-$1/2$ chains of size $N$ with open boundary conditions. The system on lattice site $j$ is described using the Pauli matrices $X_j, Y_j, Z_j$. As initial and target states we select area-law states, e.g., ground states of the quantum Ising model [Sec.~\ref{sec:studies}]. In order to control chains composed of many interacting spins, we obtain the target ground state using DMRG \cite{white1992,schollwock2011}, and represent the quantum state as an MPS throughout the entire time evolution [\SI, Sec.~\ref{app:mps}].

We choose a set of experimentally relevant control unitaries $\mathcal{A}$ which contains uniform nearest-neighbor spin-spin interactions, and global rotations: $\mathcal{A}=\{\mathrm{e}^{\pm i \delta t_\pm \hat{A}_{j}}\}$, with 
\begin{eqnarray}
	\label{eq:actions}
	\hat{A}_{j}\in \mathcal{A} &=& \bigg\{ \sum_{i} \hat{X}_{i}, \sum_{i} \hat{Y}_{i}, \sum_{i} \hat{Z}_{i},  \\
	&&\quad \sum_{i} \hat{X}_{i} \hat{X}_{i+1}, \sum_{i} \hat{Y}_{i} \hat{Y}_{i+1}, \sum_{i} \hat{Z}_{i} \hat{Z}_{i+1} \bigg\}.\nonumber
\end{eqnarray}
Two-qubit unitaries are capable of controlling entanglement in the state. 
Note that MPS-based time evolution is particularly efficient for such locally applied operators and the resulting protocols can be considered as a series of quantum gates.

The time duration (or angle) $\delta t_\pm$ of all unitary operators is fixed and slightly different in magnitude for positive and negative generators $\hat A_j$, and kept constant throughout the time evolution.
Hence, the problem of finding an optimal sequence reduces to a discrete combinatorial optimization in the exponentially large dimensional space of all possible sequences: 
For a fixed sequence length $q$, the number of all distinct sequences is $|\mathcal{A}|^{q}$; therefore, a brute-force search quickly becomes infeasible and more sophisticated algorithms, such as RL, are needed. By fixing both $q$ and $\delta t_\pm$ prior to the optimization, in general, we may not be able to come arbitrarily close to the target state, but these constraints can be easily relaxed.

\begin{figure*}[t!]
	\centering
	\includegraphics[width=1.0\textwidth]{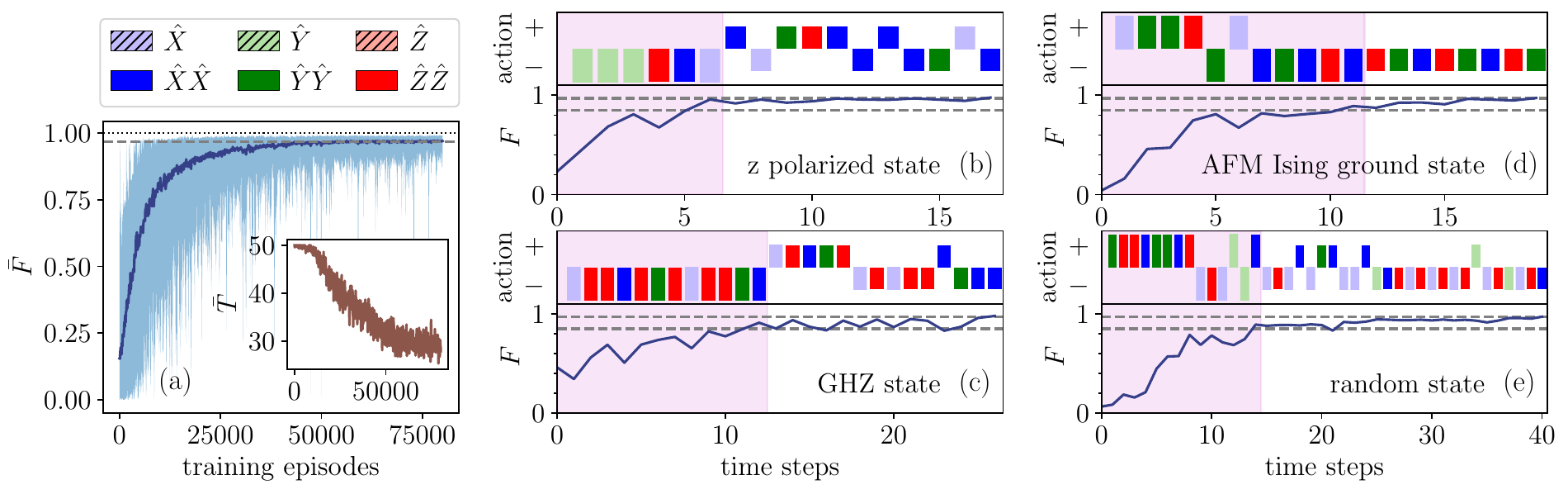}
	\caption{\label{fig:case1}
		{\bf Universal four-qubit control --- (a)}~Achieved many-body fidelity $\bar{F}$ between final and target state during training averaged over 100 training episodes (dark-blue curve). The best and worst fidelity within each episode window is indicated by the light-blue-shaded area. The fidelity threshold, $F^\ast\sim 0.97$, is marked by a gray dashed line. The inset shows the mean number of episode steps $\bar{T}$ during training (averaged over 100 episodes). The maximum number of allowed steps is set to 50.
		{\bf(b)-(e)}~Two QMPS agents [see text] are trained with fidelity thresholds $F^\ast\!\sim\!0.85,0.97$ (gray dashed lines), to prepare the Ising ground state ($J\!=\!-1, g_x\!=\!1, g_z\!=\!0$), starting from (b) the $z$-polarized product state, (c) the GHZ state, (d) an Ising antiferromagnetic ground state at ($J\!=\!+1, g_x\!=\!g_z\!=\!0.1$), and (e) a -random state. 
		The QMPS-2 agent starts from the final state reached by the QMPS-1 agent (purple shaded area). The many-body fidelity $F$ between the instantaneous and the target state, is shown in the lower part of each panel. 
		The upper part of the panels shows the control protocol.
		The colors and shading of each rectangle indicate the applied action $A$ (see legend); $\pm$ stands for the sign of the action generator, i.e., $\mathrm{exp}(\pm i \delta t_\pm \hat{A})$. 
		The QMPS-1,2 agents use fixed time steps of $\delta t_\pm=(\pi/8,\pi/16)_+,(\pi/13,\pi/21)_-$, indicated by action rectangles of different sizes in the protocol.
		$N\!=\!4$ spins.
	}
\end{figure*}

\section{\label{sec:studies}State Informed Many Body Control}
Our MPS-based RL framework is specifically designed for preparing low-entangled states in 1d, such as ground states of local gapped Hamiltonians. Hence, in the subsequent case studies we consider ground states of the 1d mixed field Ising model as an exemplary system:
\begin{equation}
	\label{eq:ising}
	\hat H_\mathrm{Ising}=J\sum_{j=1}^{N-1}\hat{Z}_{j} \hat{Z}_{j+1}-g_{x} \sum_{j=1}^{N} \hat{X}_{j}-g_{z} \sum_{j=1}^{N} \hat{Z}_{j},
\end{equation}
where $g_x$ ($g_z$) denotes a transverse (longitudinal) field. In the case of negative interaction strength and in the absence of a longitudinal field $g_z=0$, the system is integrable, and has a critical point at $g_x=1$ in the thermodynamic limit, separating a paramagnetic (PM) from a ferromagnetic phase (FM) [Fig.~\ref{fig:intro}]. 
For $g_z\!>\!0$, the model has no known closed-form expressions for its eigenstates and eigenenergies. In addition, for positive interactions, the phase diagram features a critical line from $(g_x,g_z)\!=\!(1,0)$ to $(g_x,g_z)\!=\!(0,2)$ exhibiting a transition from a paramagnetic to an antiferromagnetic phase (AFM) [see \SI Sec.~\ref{app:many-body} for a brief introduction to quantum many-body physics and phase transitions].

In the rest of this section we will analyze three different control scenarios involving ground states of the mixed field Ising model. In Sec.~\ref{sec:case1} we consider the problem of universal state preparation for $N\!=\!4$ spins and train a QMPS agent to prepare a specific target ground state starting from arbitrary initial quantum states. This example will serve as a first benchmark of the QMPS framework. In Sec.~\ref{sec:case2} we use the QMPS algorithm to prepare a spin-polarized state starting from a class of paramagnetic ground states which shows that our approach produces reliable protocols in the many-body regime. Finally, in Sec.~\ref{sec:case3} we consider $N\!=\!16$ spins and target a state in the critical region of the mixed field Ising model demonstrating that the QMPS framework can also be employed for highly non-trivial control tasks such as critical state preparation. Furthermore, we show that the obtained QMPS agent has the ability to self-correct protocols in the presence of noisy time evolution.

\subsection{\label{sec:case1}Universal ground state preparation from arbitrary initial quantum states for $N=4$ spins}

In the noninteracting limit, $J\!=\!0$, the QMPS agent readily learns how to control a large number of spins [\SI, Sec.~\ref{app:case0}].
Instead, as a nontrivial benchmark of the QMPS framework, here we teach an agent to prepare the ground state of the $4$-spin transverse field Ising model at ($J\!=\!-1, g_x\!=\!1, g_z\!=\!0$), starting from \textit{randomly drawn} initial states. 
While this control setup can be solved using the full wave function and a conventional neural network ansatz [see \SI Sec.~\ref{app:case1b}], the uniform initial state distribution over the entire continuous Hilbert space creates a highly non-trivial learning problem and presents a first benchmark for our QMPS framework. Moreover, system sizes of $N\!\sim\!4$ spins already fall within the relevant regime of most present-day studies using quantum computers, where gate errors and decoherence currently prevent exact simulations at larger scales~\cite{kandala2017,alba2018,lamm2018}.

We first train an agent (QMPS-1) to prepare the target ground state within 50 protocol steps or less, setting a many-body fidelity threshold of $F^\ast\!\approx\!0.85$. The initial states during training are chosen to be (with probability $p\!=\!0.25$) random polarized product states, or (with probability $p\!=\!0.75$) random reflection-symmetric states drawn from the full $2^4\!=\!16$ dimensional Hilbert space \footnote{Due to the enforced reflection symmetry the state space we effectively sample from is 10 dimensional. We sample (Haar) random states by drawing the wave function amplitudes from a normal distribution followed by normalization.}. In this way the QMPS-1 agent has to learn to both disentangle highly entangled states to prepare the Ising ground state, but also to appropriately entangle product states to reach the entangled target [the learning curves of the QMPS-1 agent are shown in \SI, Sec.~\ref{app:case1}]. After this training stage, we test the QMPS-1 agent on a set of $10^3$ random initial states and find that in $\sim 99.8\%$ of the cases the fidelity threshold is successfully reached within the 50 allowed steps. A (much) better fidelity cannot be achieved by the QMPS-1 agent alone, due to the discreteness of the action space and the constant step size used, rather than limitations intrinsic to the algorithm. Note that when following a conventional approach of training a neural network directly on the quantum wave function, we were not able to match the performance of the QMPS agent given the same number of parameters and training episodes [see \SI Sec.~\ref{app:case1b} for details]. This suggests that the QMPS architecture has a more natural structure for extracting relevant features from quantum state data and can already be advantageous for small system sizes.

To improve the fidelity between the final and the target state, we now train a second, independent agent (QMPS-2) with a tighter many-body fidelity threshold of $F^\ast\!\approx\! 0.97$. The initial states are again sampled randomly as mentioned above; however, we first use the already optimized QMPS-1 agent to reach the vicinity of the target state within $F\!>\! 0.85$. Then, we take those as initial states for the training of the second QMPS-2 agent. 

This two-stage learning schedule can in principle be continued to increase the fidelity threshold even further. The learning curves of the QMPS-2 optimization are shown in Fig.~\ref{fig:case1}(a).
In Fig.~\ref{fig:case1}(b)-(c) we present the obtained protocols for four exemplary initial states. 
Overall, the combined two-agent QMPS is able to reach the fidelity threshold of $F^\ast\sim 0.97$ for $\sim 93\%$ of the randomly drawn initial states within the 50 episode steps that were imposed during training. We emphasize that this result is already nontrivial, given the restricted discrete action space, and the arbitrariness of the initial state. 

Let us now exhibit two major advantages of RL against conventional quantum control algorithms. (i) After training we can double the allowed episode length for each agent to 100 steps. Since this allows for longer protocols, we find that the target state can be successfully prepared for $99.5\%$ of the initial states (compared to the previously observed $93\%$). Note that this feature is a unique advantage of (deep) RL methods, where the policy depends explicitly on the quantum state: during training, the agent learns how to take optimal actions starting from \textit{any} quantum state and hence, it is able to prepare the target state if it is given sufficient time.
Moreover, (ii) in this example we achieve universal quantum state preparation, i.e., the trained RL agent succeeds in preparing the target state irrespective of the initial state. This is not possible with conventional control techniques where the optimal protocol is usually tailored to a specific initial state, and the optimization has to be rerun when starting from a different state. 
In Sec.~\ref{NISQ}, we show how the trained QMPS architecture can be implemented to apply the RL agent on NISQ devices.

\subsection{\label{sec:case2}Preparation of a polarized product state from paramagnetic ground states for $N=32$ spins}

\begin{figure}[t!]
	\centering
	\includegraphics[width=1.0\columnwidth]{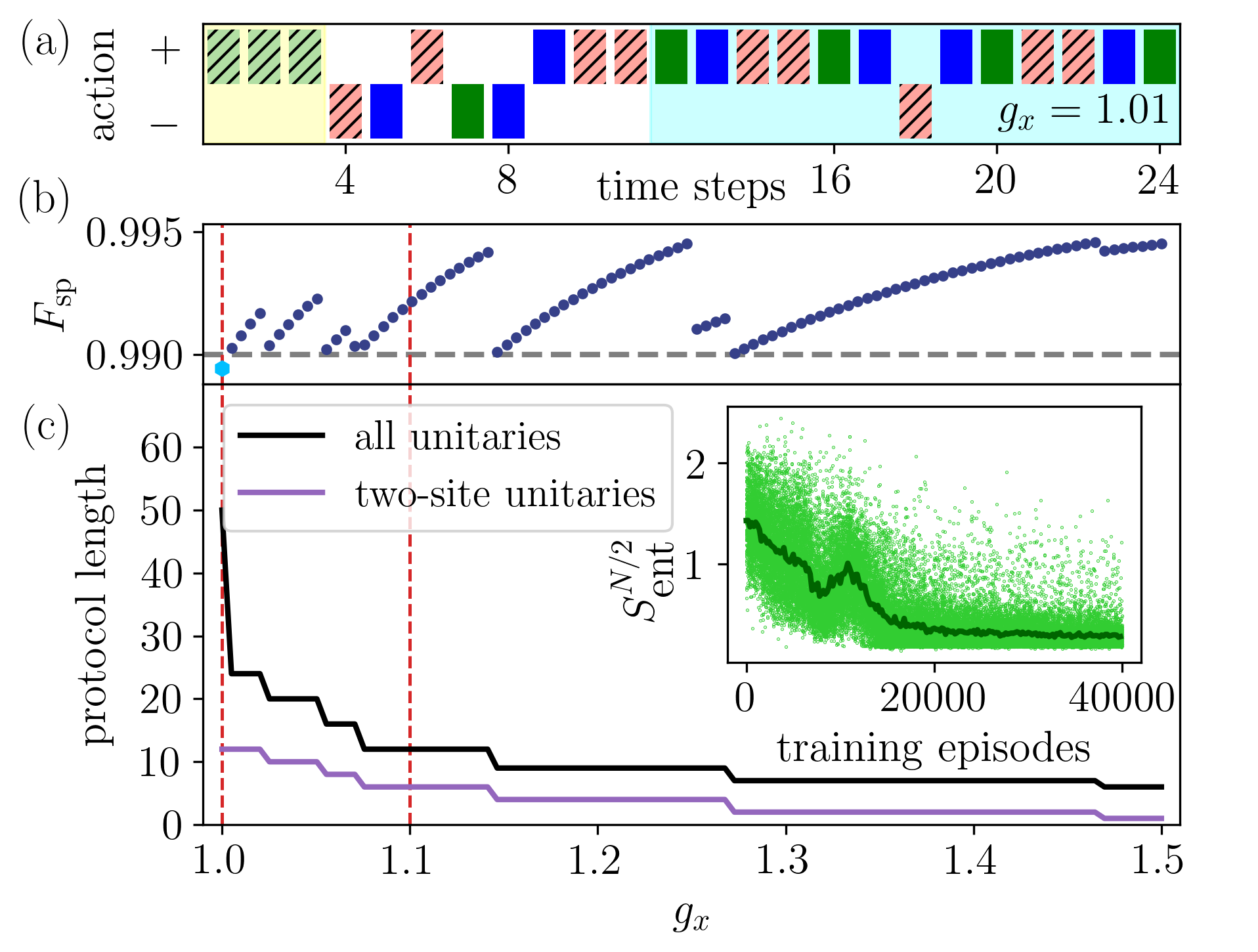}
	\caption{\label{fig:case2a}
		{\bf Transverse-field Ising control --- (a)}~Optimal protocol obtained, starting from an initial ground state at $g_x\!=\!1.01$ [cf.~Fig.~\ref{fig:case1} for action legend]. The cyan shaded segment indicates a generalized Euler-angle-like many-body rotation.
		{\bf (b)}~Final single-particle fidelities $F_{\mathrm{sp}}=\sqrt[N]{F}$ starting from initial ground states with transverse field value $g_x$. The target state is the $z$-polarized product state. The gray dashed line denotes the fidelity threshold ($F^\ast_\mathrm{sp}\!=\!0.99$, $F^\ast\!\sim\!0.72$): it is surpassed for most initial states except at the critical point $g_x\!\sim\!1$ (cyan dot). The red vertical dashed lines contain the training region.
		{\bf (c)}~The number of actions (unitaries) in the QMPS protocols versus the initial state parameter $g_x$. The protocol starting from the critical state ($g_x\!\sim\!1$) does not reach the fidelity threshold and is truncated after 50 episode steps. Inset:~the half-chain von Neumann entanglement entropy of final states during training decreases as learning improves. The dark green curve denotes the average over 200 episodes. $N\!=\!32$ spins. 
		See also Video~\protect\hyperlink{video:1}{1}.
	}
\end{figure}

In general, a (Haar) random quantum many-body state is volume-law entangled and, hence, it cannot be approximated by a MPS of a fixed bond dimension. Moreover, it becomes increasingly difficult to disentangle an arbitrarily high entangled state for larger system sizes \cite{poulin2011}. Therefore, when working in the truly quantum many-body regime, we have to restrict to initial and target states that are not volume-law entangled. 
	
As an example, here we consider a many-body system of $N\!=\!32$ spins and learn to prepare
the $z$-polarized state from a class of transverse field Ising ground states ($J\!=\!-1,g_z\!=\!0$).
Once high-fidelity protocols are found, they can be inverted to prepare any such Ising ground state from the $z$-polarized state, which presents a relevant experimental situation [\SI, Sec.~\ref{app:case2}]. 
Many-body ground state preparation is a prerequisite for both analog and digital quantum simulation, and enables the study of a variety of many-body phenomena such as the properties of equilibrium and nonequilibrium quantum phases and phase transitions.

To train the agent, we randomly sample initial ground states on the paramagnetic side of the critical point: $1.0\!<\!g_x\!<\!1.1$.
The difficulty in this state preparation task is determined by the parameter $g_x$ defining the initial state: states deeper into the paramagnetic phase are more easy to `rotate' into the product target state, while states close to the critical regime require the agent to learn how to fully disentangle the initial state in order to reach the target.
We train a QMPS agent on a system of $N\!=\!32$ spins which is infeasible to simulate using the full wavefunction, and is far out-of-reach for neural-network based approaches. We set the single-particle fidelity threshold [cf.~Sec.~\ref{sec:methods}] to $F^\ast_\mathrm{sp}\!=\!\sqrt[N]{F^\ast}\!=\!0.99$ (corresponding many-body fidelity $F^\ast\!\sim\!0.72$) and allow at most $50$ steps per protocol.

Figure~\ref{fig:case2a}(b) shows the successfully reached final fidelity when the trained QMPS agent is tested on unseen initial states, for various values of $g_x$.
First, notice that the agent is able to prepare the target state also for initial states with $g_x\!>\!1.1$ that lie outside of the training region (dashed vertical lines). Hence, we are able to extrapolate optimal control protocols well beyond the training data distribution, without additional training. Similar generalization capabilities have already been demonstrated for supervised learning tasks such as Hamiltonian parameter estimation \cite{Ma2021}. However, this is not true for states inside the critical region, $g_x\!\lesssim\!1$, and in the ferromagnetic phase ($g_x\!\ll\!1$); such a behavior is not surprising, since these many-body states have very different properties compared to those used for training. Note that in contrast to the previous control study of Sec.~\ref{sec:case1}, the initial training data states are not i.i.d.~over the full $2^{32}$ dimensional Hilbert space as we only train on PM ground states of the Ising model. Therefore the agent cannot be expected to generalize to arbitrary initial states in this case. Interestingly, it follows that the onset of criticality can be detected in the structure of control protocols, as the number of required gates (actions) and, in particular, of entangling unitaries, increases rapidly as one approaches the critical point [Fig.~\ref{fig:case2a}(c)].

Discontinuities in the achieved fidelity [Fig.~\ref{fig:case2a}(b)] arise due to the fixed, constant step size $\delta t_\pm$: we observe distinct jumps in the final fidelity, whenever the length of the protocol sequence increases. This is a primary consequence of the discrete control action space. Its physical origin can be traced back to the need for a more frequent use of disentangling two-site unitaries, for initial states approaching the critical region.

Figure~\ref{fig:case2a}(a) shows the optimal protocol at $g_x\!=\! 1.01$: first, the agent concatenates three $\hat Y$-rotations ($\delta t_+\!=\!\pi/12$) in a global gate, which shows that it learns the orientations of the initial $x$-paramagnet and the $z$-polarized target [yellow shaded region]. 
This is succeeded by a non-trivial sequence containing two-body operators. A closer inspection [Fig.~\ref{fig:app:correlations} in \SI, Sec.~\ref{app:case2} and Video~\protect\hyperlink{video:1}{1}] reveals that the agent discovered a generalization of Euler-angle rotations in the multi-qubit Hilbert space [blue shaded region]. This is remarkable, since it points to the ability of the agent to construct compound rotations, which is a highly non-trivial combinatorial problem for experimentally relevant constrained action spaces. This can be interpreted as a generalization of dynamical decoupling sequences introduced in state-of-the-art NMR experiments and used nowadays in quantum simulation, optimal quantum sensing, and to protect quantum coherence~\cite{Choi2020,Viola02,Haeberlen76}. 
We verified that this protocol is a local minimum of the control landscape.

\begin{figure*}[t!]
	\centering
	\includegraphics[width=1.0\textwidth]{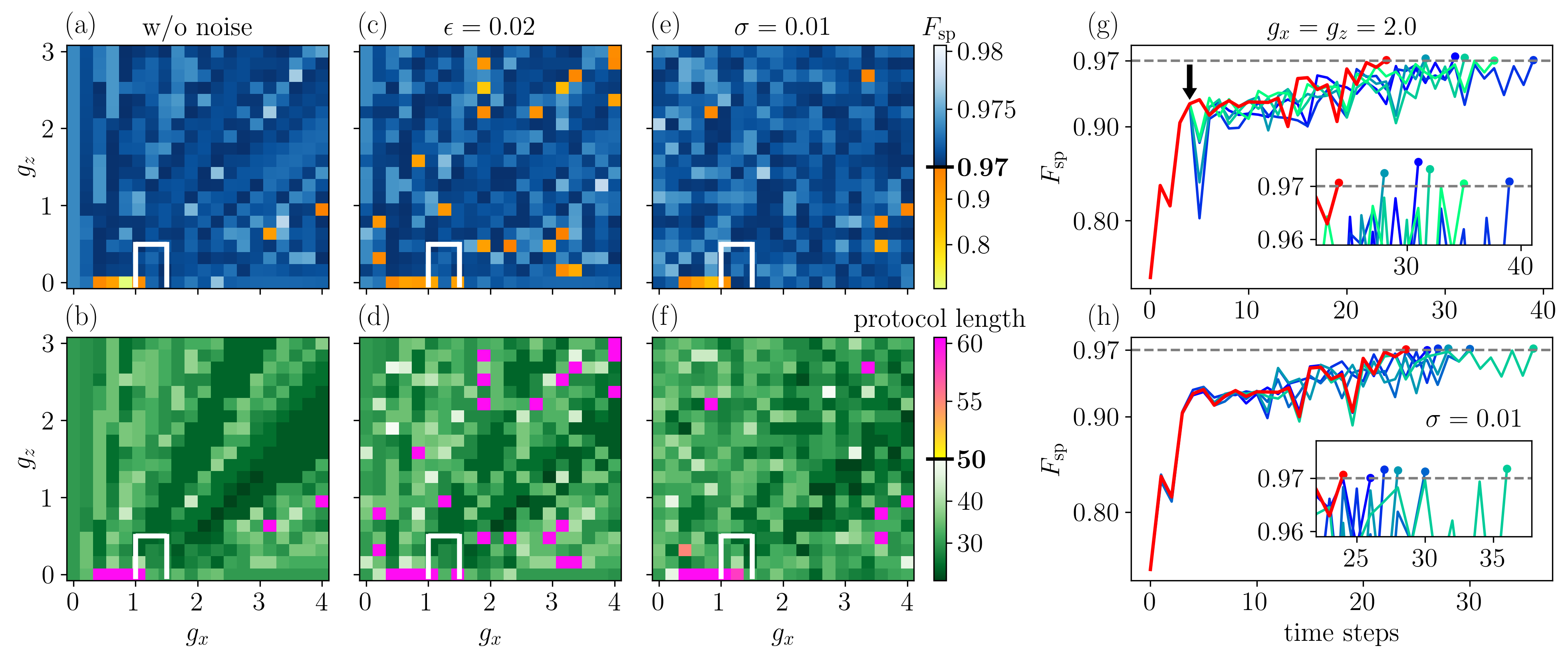}
	\caption{\label{fig:case3a}
		{\bf Self-correcting mixed-field Ising control --- (a),(b)}~Final single-particle fidelity $F_\mathrm{sp}=\sqrt[N]{F}$ [top] and corresponding protocol length [bottom] -- see colorbars -- versus the initial Ising ground state parameter values $g_x,g_z$. The target is a state in the critical region of the Ising model at ($J\!=\!+1, g_x\!=\!0.5, g_z\!=\!1.5$). Training started only from initial states sampled randomly from the enclosed white rectangle. Each of the two parts of a colorbar is shown on a linear scale with the fidelity threshold ($F^\ast_{\mathrm{sp}}\!=\!0.97$, $F^\ast\!\sim\!0.61$) and the maximum episode length during training (50), indicated by short black lines. 
		{\bf (c)-(f)}~Same as (a)-(b) but for noisy evolution -- (c),(d): At each time step, actions other than the one selected by the agent, are taken with probability $\epsilon\!=\!0.02$; (e),(f):~White Gaussian noise with standard deviation $\sigma\!=\!0.01$ is added to the time step $\delta t_\pm$ of all applied unitaries. 
		{\bf (g)}~Time-dependence of the single-particle fidelity starting from an arbitrary initial ground state, and following the trained agent. The red curve denotes the unperturbed (noise-free) QMPS protocol. At time step 5 [indicated by the black arrow], the QMPS protocol is perturbed by enforcing 5 suboptimal actions. All subsequent actions are selected again according to the trained QMPS policy without perturbation (blue curves). The inset displays a zoom in the vicinity of the fidelity threshold (dashed gray line), showing that each protocol terminates successfully. 
		{\bf (h)}~Same as in (g) but for dynamics subject to Gaussian noise at every time step $\delta t_\pm$, for 5 different random seeds giving rise to 5 distinct protocols.
		$N\!=\!16$ spins.
		See Videos~\protect\hyperlink{video:2}{2} and~\protect\hyperlink{video:3}{3}.
	}
\end{figure*}

We also investigated the system size dependence of optimal protocols in this control study. 
To our surprise, we find that agents trained on the $N\!=\!32$ spin system produce optimal protocols that perform reasonably well on smaller ($N\!=\!8$) as well as larger ($N\!=\!64$) systems. Hence, this control problem admits a certain degree of transferability, which worsens for initial states closer to the finite-size dominated critical region [\SI, Sec.~\ref{app:case2}]. 

The MPS-based control framework enables us to readily analyze the physical entanglement growth during training, via the bond dimension of the quantum state $\chi_\psi$. 
The protocol exploration mechanism in QMPS causes the agent to act mostly randomly during the initial stages of learning. This translates to random sequences of unitary gates that can lead to an increase of quantum entanglement [Fig.~\ref{fig:case2a}(c), inset]. In our simulations, we set the maximum allowed bond dimension to $\chi_{\psi}\!=\!16$, which is sufficient for the considered initial and target states to be approximated reliably. However, not all states encountered can be represented with such a small bond dimension, as reflected by large truncation errors during training [\SI, Sec.~\ref{app:case2}]. 
Nonetheless, as training progresses, the agent learns to take actions that do not create excessive entanglement [Fig.~\ref{fig:case2a}(c)]. Therefore, the truncation error naturally decreases, as training nears completion. As a consequence, the final converged protocols visit states that lie within a manifold of low-entangled states. 
Moreover, increasing $\chi_\psi$ does not change these observations.
We believe that this mechanism relies on the area-law nature of the initial and target states, and we expect it to open up the door towards future control studies deeper in the genuine many-body regime.

\subsection{\label{sec:case3}Learning robust critical-region state preparation for $N=16$ spins}

States in the critical region possess non-trivial correlations and show strong system-size dependence, which make manipulating them highly non-trivial. In particular, the required time duration to adiabatically prepare critical states diverges with the number of particles, whereas sweeping through critical points reveals properties of their universality classes~\cite{zurek2005}. Therefore, finding optimal control strategies away from the adiabatic limit is an important challenge. 
Critical state preparation is also of practical relevance for modern quantum metrology, where the enhanced sensitivity of critical states to external fields is leveraged to perform more precise measurements~\cite{rams2018}.

Our final objective is to prepare a ground state in the critical region of the mixed field Ising chain ($J\!=\!+1, g_x\!=\!0.5, g_z\!=\!1.5$) starting from non-critical paramagnetic ground states of the same model with flipped interaction strength: $J\!=\!-1$, $1.0\!<\!g_x\!<\!1.5$, $0\!<\!g_z\!<\!0.5$ [Fig.~\ref{fig:intro}].
Hence, the agent has to learn to connect ground states of two distinct Hamiltonians. This scenario is often relevant in typical experimental setups where only a single-sign interaction strength can be realized: e.g., the initial state comes from the $J\!<\!0$ Ising model, while the ground state of interest belongs to the antiferromagnatic Ising Hamiltonian. In general, one can use two completely distinct parent Hamiltonians for the initial and target states, one of which being inaccessible in the quantum simulator platform at hand, while the other being the object of interest.

We train our QMPS agent on $N\!=\!16$ spins with a single-particle fidelity threshold of $F^\ast_\mathrm{sp}\!=\!0.97$ ($F^\ast\!\sim\!0.61$), and a maximum episode length of 50. Figure~\ref{fig:case3a}(a) shows the achieved fidelity between the target state and the final state, for different initial ground states corresponding to a rectangular region in the $(g_x,g_z)$-plane. Notice that the agent is able to generalize to unseen initial states lying far outside the training region (white rectangle), and fails only close to the critical point of the transverse field Ising model ($g_x\!=\!1,g_z\!=\!0$) and for a few isolated initial states well outside of the training region.

We now demonstrate that our QMPS agent shows remarkable generalization capabilities in noisy environments. In particular, we analyze how robust the trained QMPS agent is to stochastic perturbations in the time evolution of the state -- a common problem in noisy intermediate-scale quantum (NISQ) computing devices~\cite{preskill2018}. In what follows, we consider two different sources of noise independently: 
(i) At each time step, with probability $\epsilon$, a random action rather than the selected one is enforced. This type of noise mimics bit- or phase-flip errors, which occur in quantum computing. 
(ii) Gaussian random noise with zero mean and standard deviation $\sigma$, is added to the time duration $\delta t_\pm$ of each unitary operator; this can, for instance, result from imperfect controls in the experimental platform.

\begin{figure*}[t!]
	\centering
	\includegraphics[width=1.0\textwidth]{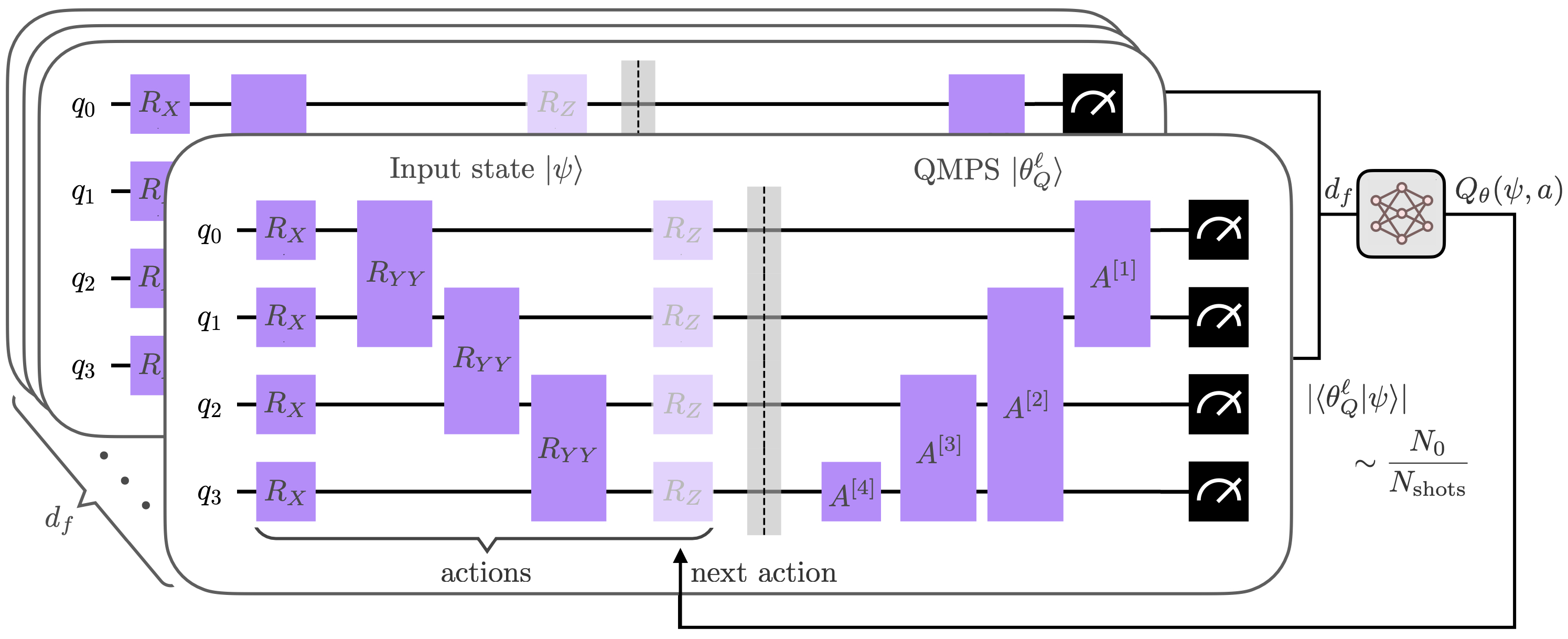}
	\caption{\label{fig:qmps_nisq}	{\bf QMPS circuit framework} as a hybrid quantum-classical algorithm. On the quantum device, we first prepare the initial state and apply the already inferred protocol actions as gates. In the example above, the initial state is the fully z-polarized state, i.e., $\ket{0}^{\otimes 4}$, and two actions are performed: A global rotation around $\hat{X}$ followed by a global two-qubit $\hat{Y}\hat{Y}$ rotation. The resulting state $\ket{\psi}$ represents the input to the QMPS network. The QMPS tensors $\theta_Q = A^{[1]}\cdots A^{[N]}$ can be mapped to unitary gates on a quantum circuit. To compute the Q-values, we first apply the inverse of the QMPS circuit unitary $U^\dagger_\theta$ and measure the output in the computational basis. The fraction of all-zero measurement outcomes is an approximation to the fidelity $|\!\braket{\theta_Q}{\psi}\!|^2$. Note that this denotes the fidelity with respect to the Q-value network state $\ket{\theta_Q}$ and not the target quantum state which is not required during protocol inference. The fidelity estimates are then fed into the neural network on a classical computer. From the resulting Q-values we can infer the next action and repeat these steps until the target state is reached.
	}
\end{figure*}

Noise type (i) is equivalent to using an $\epsilon$-greedy policy. Hence, the states encountered when acting with such a policy, could have (in principle) been visited during training. Due to the generalization capabilities of RL, it is reasonable to expect that an agent will act optimally after non-optimal actions have occurred, attempting to correct the `mistake'. In Fig.~\ref{fig:case3a}(c)-(d) we show the achieved final fidelity [top] and the required number of steps [bottom] for $\epsilon\!=\!0.02$. Overall, the fidelity threshold can still be reached in the majority of test cases. The randomness typically results in longer protocols indicating that the agent indeed adapts to the new states encountered. Interestingly, in the noise-free case the agent fails to prepare the target state for a few points outside the training region [orange points in Fig.~\ref{fig:case3a}(a)]; this can be attributed to incorrectly estimated Q-values which have not fully converged to the optimal ones outside of the training interval. However, when adding the perturbation, the agent is able to correct its mistake in one of the shown instances and prepares the target state successfully [Fig.~\ref{fig:case3a}(c)].

Recall that we use a different time step $\delta t_\pm$ for positive and negative actions. This way the agent is not just able to undo a non-optimal action by performing its inverse; it rather has to adjust the entire sequence of incoming unitaries in a non-trivial way. The ability of the QMPS agent to adapt is demonstrated in Fig.~\ref{fig:case3a}(g) where we plot the fidelity during time evolution starting from an arbitrary initial ground state. At time step $t\!=\!5$, we perturb the protocol by taking 6 different actions, and let the agent act according to the trained policy afterwards; this results in 6 distinct protocol branches. In each of them, the agent tries to maximize the fidelity and successfully reaches the fidelity threshold after a few extra protocol steps [see also Video~\protect\hyperlink{video:2}{2}]. In \SI, Sec.~\ref{app:case3} we provide further examples showing that this behavior is generic, and can also be observed for different initial states.

In contrast to the $\epsilon$-noise, adding Gaussian random noise $\sigma$, (ii), to the time step duration $\delta t_\pm$, results in states that the agent has not seen during training. This source of noise, therefore, explicitly tests the ability of the agent to generalize beyond the accessible state space, and in particular to interpolate between quantum many-body states. Fig.~\ref{fig:case3a}(e)-(f) displays the achieved fidelity and the corresponding protocol length for $\sigma = 0.01$. We find that the QMPS agent is also robust to this type of noise. In Fig.~\ref{fig:case3a}(h) we plot the fidelity trajectories starting from the same initial state using 5 different random seeds; this illustrates that our agent adapts successfully to previously unencountered many-body states, and steers the protocol on-line to reach beyond the fidelity threshold [see Video~\protect\hyperlink{video:3}{3}].

The robustness of QMPS agents to noise and, in general, to stochasticity in the quantum gates, demonstrates yet another advantage of deep RL methods over conventional quantum control techniques. The latter typically perform suboptimal in noisy systems since the optimization does not take into account the quantum state information during the time evolution, and the optimal protocols are specifically optimized for a fixed trajectory of quantum states~\cite{yao2020}. By contrast, QMPS value functions are optimized on a large class of states and, as shown above, can interpolate and extrapolate to new, seen and unseen states as long as the deep learning approximation stays sufficiently accurate. Therefore, unlike conventional quantum control algorithms, QMPS agents have the ability to automatically self-correct their protocols on-the-fly, i.e., while the system is being time evolved.

\begin{figure}[t!]
	\centering
	\includegraphics[width=1.0\columnwidth]{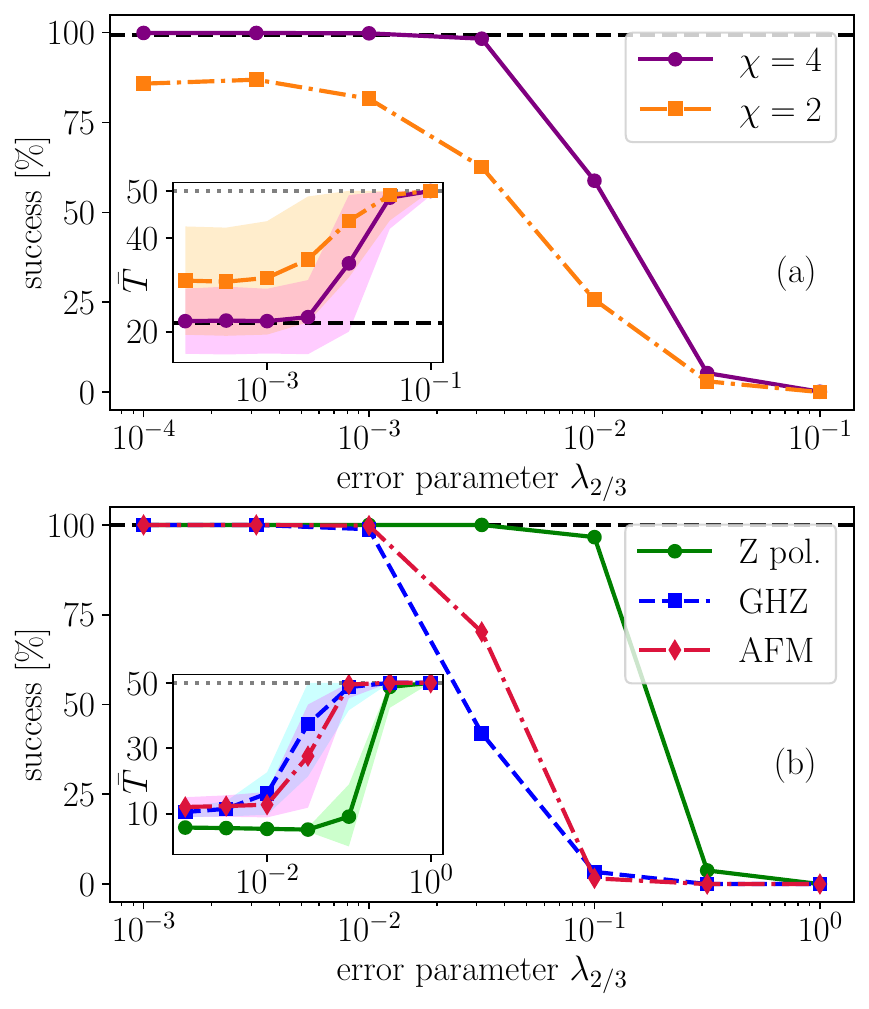}
	\caption{\label{fig:case4b}
		{\bf Universal four-qubit control --- (a)}~We sample 1000 random initial states and apply the QMPS circuit framework in the presence of depolarizing noise with error parameter $\lambda_{2/3}$ for all two-and three-qubit gates. Note that the noise parameter for all single-qubit gates is always fixed to $\lambda_1 = 10^{-4}$. We set the number of measurement shots to 4096 and plot the percentage of runs in which the target state is successfully reached against the noise strength. The success rate under exact, noise-free computation (without sampling) is shown as a black dashed line. The success probability when acting completely random is always zero. We provide both, the results for the full $\chi=4$ QMPS circuit (purple solid line) and the truncated $\chi=2$ QMPS (orange dash-dotted line). Insets: The corresponding average number of protocols steps $\bar{T}$ as a function of the noise strength $\lambda$. The standard deviation is indicated by the shaded areas.
		{\bf (b)} Same as in (a) except that we start each protocol from a fixed initial state: the fully z-polarized state (green solid line), the GHZ state (blue dashed line), and a ground state of the mixed field Ising model at $J=+1, g_x=g_z=0.1$ (red dashed-dotted line). To compute the success rates and the average protocol length we average over 500 different runs. Note that the x-axis scale is shifted by one order of magnitude compared to (a).
	}
\end{figure}

\section{\label{NISQ}Implementation on NISQ devices}

The present QMPS framework requires the quantum state to be accessible at each time step for both training and inference purposes; yet, quantum states are not observable in experiments without performing expensive quantum state tomography. On the other hand, MPS-based quantum state tomography presents a possible and efficient way of retrieving the quantum state in a form that can be straightforwardly integrated in the QMPS framework \cite{Baumgratz13,Lanyon17,Cramer10}. Alternatively, there already exist efficient encoding strategies that map MPS into quantum circuits~\cite{barratt2021,lin2021,ran2020,Rudolph2022,Dov2022,Foss-Feig2022,Wall2022}. Moreover, several proposals were recently developed in which MPS are harnessed for quantum machine learning tasks, for example as part of hybrid classical-quantum algorithms~\cite{huggins2019,chen2021a,Chen2020z} or as classical pre-training methods~\cite{dborin2021,wall2021a}. 
	
Similar ideas can be applied to the QMPS architecture by mapping the trainable MPS to a parametrized quantum circuit, thus directly integrating the QMPS framework in quantum computations with NISQ devices and hence, eliminating the need for quantum state tomography. This allows us to perform the expensive training routine on readily available classical computers while the inexpensive inference step can be performed on quantum hardware.

The mapping of the QMPS to a quantum circuit is described in detail in  \SI, Sec.~\ref{subsec:circuit_mapping}. Figure~\ref{fig:qmps_nisq} shows the resulting QMPS circuit framework for the case of $N\!=\!4$ spins/qubits in which the original QMPS state $\ket{\theta_Q^\ell}$ is represented as unitary gates (purple boxes). To calculate the Q-values $Q_{\theta}(\psi,a)$ given an input quantum state $\ket{\psi}$, we first compute the fidelity between the input and the QMPS state
\begin{equation}\label{eq:fidelity}
    \left|\braket{\theta_Q^\ell}{\psi}\right|^2 = \big|\!\bra{0}U_{\theta}^\dagger U_\psi\ket{0}\!\big|^2,
\end{equation}
which can be obtained via sampling on a quantum computer. Alternatively, the overlap can also be accessed by performing a SWAP test, albeit requiring additional ancilla qubits and non-local gates \cite{Buhrman2001,Gottesman2001}. The computed fidelities for each QMPS circuit are then fed into the classical neural network giving rise to a hybrid quantum-classical machine learning architecture as shown in Fig.~\ref{fig:qmps_nisq}. If necessary, the parameters of the QMPS circuit $U_{\theta}$ can be fine-tuned by performing some additional optimization.

We test the QMPS circuit framework on the universal ground state preparation task of Section~\ref{sec:case1}. In what follows we only report results for the QMPS-1 agent trained on a fidelity threshold of $F^\ast\!\sim\!0.85$; the generalization to include the QMPS-2 agent is straightforward. We translate the optimized QMPS to the corresponding quantum circuit and investigate the effects of noise in the quantum computation on the QMPS framework. 

To simulate incoherent errors, we consider a depolarizing noise channel $E$
\begin{equation}\label{eq:noise}
    E(\rho)=(1-\lambda) \rho+\lambda \frac{I}{2^N},
\end{equation}
and apply it after each action and QMPS gate. Here, $\rho$ denotes the quantum state density matrix and $\lambda$ is the depolarizing noise parameter which is set to $\lambda_1 = 10^{-4}$ for all single-qubit gates. We plot the success rate as a function of the two- and three-qubit gate errors $\lambda_{2/3}$ for 1000 randomly sampled initial states in Fig.~\ref{fig:case4b}(a) (purple line). For error rates $\lambda_{2/3}<10^{-3}$ the QMPS agent is able to successfully prepare the target state in almost all runs. However, the performance deteriorates with increasing error parameter $\lambda_{2/3}$. Let us note that we have used the same error rate for both, two-and three-qubit gates. On a physical quantum device, the three-qubit gate will be decomposed into a sequence of two-qubit gates and hence the introduced noise will be amplified. However, the decomposition, the resulting circuit depth, and the sources of noise vary with the hardware type \footnote{For example, transpiling (i.e., decomposing) the QMPS circuit on an IBM Quantum device results in $\sim 100$ two-qubit gates, while on an IONQ device we require only $\sim 30$ two-qubit gates.}. Thus, we chose the simplified noise model of Eq.~\eqref{eq:noise} in order to study the QMPS circuits in a hardware agnostic way.

We also report the results obtained when truncating the bond dimension $\chi=4$ QMPS to a $\chi=2$ QMPS which gives rise to at most two-qubit gates in the final circuit. In this case the success probabilities [orange dashed line in Fig.~\ref{fig:case4b}(a)] do not reach $100\%$ even for small error rates. This indicates that a bond dimension of $\chi=4$ is indeed required to faithfully represent the QMPS state.

Finally, in Fig.~\ref{fig:case4b}(b) we show the success rates when starting from three different initial states:~the fully z-polarized state (green), the GHZ state (blue), and a ground state of the mixed field Ising model at $J=+1, g_x=g_z=0.1$ (red). The success probability of unity can be maintained for error rates $\lambda$ one order of magnitude larger than for the random initial state case. Physical states such as the GHZ or ground states possess only a small amount of entanglement and hence allow us to prepare the target state using a relatively small number of two-qubit gates [c.f.~Fig.~\ref{fig:case1}(b)-(c)]. Thus, the resulting protocols are automatically more robust to two-qubit gate errors.

The effect of other decoherence channels (amplitude and phase damping) on the QMPS circuit framework leads to qualitative similar results and is further discussed in \SI Sec.~\ref{app:case4}. Furthermore, we also analyze the self-correcting property of the agent in the presence of coherent gate errors.

Let us briefly discuss the bottlenecks of the current QMPS circuit framework. The QMPS circuit depicted in Fig.~\ref{fig:qmps_nisq} is composed of a three qubit-gate and more generally an MPS with bond dimension $\chi=2^n$ will naturally give rise to $(n+1)$-qubit gates. While three-qubit gates will likely be implemented in near-term quantum computers, the gates realized in current NISQ devices are commonly composed of at most two-qubit gates. Hence, any gates acting on more than two qubits would need to be decomposed into two-qubit gates which usually gives rise to deep circuits. Instead, we can use alternative MPS-to-circuit mappings that would lead to at most two-qubit gates in the final circuit [see \SI Sec.~\ref{subsec:circuit_mapping} for a detailed discussion]~\cite{barratt2021,lin2021,ran2020,Rudolph2022,Dov2022,Foss-Feig2022,Wall2022}. Finally, the sampling of the fidelity in Eq.~\eqref{eq:fidelity} requires a number of measurement shots that could in principle grow exponentially in the system size. One possible solution is to choose a different Q-value network ansatz such as a matrix product operator [see \SI Sec.~\ref{qmpo}]. The resulting computation can then be interpreted as measuring an observable which can be performed efficiently on larger systems. We find that for the specific $N=4$ QMPS example, the number of measurements required for successfully preparing the target state can be chosen relatively small, i.e.,~$\sim 500$ to reach success rates close to unity [see Fig.~\ref{fig:app:case4}(a) in \SI Sec.~\ref{app:case4} and corresponding discussion].

\section{Discussion/Outlook}

In this work we introduced a tensor network-based Q-learning framework to control quantum many-body systems. Incorporating an MPS into the deep learning architecture allows part of the Q-value computation to be efficiently expressed as an overlap between two MPS wave functions. As a result, larger system sizes can be reached compared to learning with the full wave function. We emphasize that standard RL algorithms with conventional neural network architectures cannot handle quantum many-body states, whose number of components scale exponentially with the number of spins: e.g., for $N=32$ spins, there are $2^{32}\approx 10^{10}$ wavefunction components to store which is prohibitively expensive. By contrast, our MPS learning architecture only requires linear scaling with the system size $N$. 
Furthermore, we found that the hyperparameters of the optimization and, in particular, the number of training episodes do not require finetuning with the system size, and stayed roughly constant [\SI, Sec.~\ref{app:qmps}].
Summarizing, QMPS proposes the use of a tensor-network variational ansatz inspired by quantum many-body physics to offer a novel RL learning architecture.

QMPS-based RL is designed for solving the quantum many-body control problem by learning a value function that explicitly depends on the quantum state. Therefore, a successfully trained QMPS agent is capable of devising optimal protocols for a continuous set of initial states, and selects actions on-the-fly according to the current state visited. As a result, a QMPS agent has the ability to self-correct mistakes in the protocols when the dynamics is stochastic, even before the protocols have come to an end. Moreover, we illustrated that the agent can interpolate and extrapolate to new quantum states not seen during training.
Remarkably, we observed this behavior over test regions several times the size of the training region. 
To the best of our knowledge, there does not exist a quantum control algorithm that exhibits such desirable features, as these are based on deep learning capabilities: conventional quantum control algorithms require to re-run the optimization when the initial state has been changed, and thus lack any learning capabilities.

The generalization capabilities, the robustness to noise, and the feasibility of universal state preparation (for small system sizes) are advantages of the QMPS framework over competitive optimal control algorithms. 
These features are especially relevant for experiments and modern quantum technologies that heavily rely on quantum many-body control, and in particular for NISQ devices. 
Moreover, we demonstrated that the present QMPS framework can be integrated in quantum device simulations by mapping the optimized MPS ansatz to gates in a quantum circuit. The resulting hybrid quantum-classical algorithm allows us to control quantum states directly on the device without the need of performing expensive quantum state tomography. Thus, unlike neural networks, using an MPS learning architecture also facilitates the use of RL-agents on NISQ devices.

Our work opens up the door to further research on tensor network-based RL algorithms for quantum (many-body) control. Due to the modular structure of the architecture, the QMPS can be replaced by various tensor networks, such as tree tensor networks (TTN) \cite{shi2006} or the multi-scale entanglement renormalization ansatz (MERA)~\cite{vidal2007}; these would allow different classes of states to be represented efficiently, and affect the expressivity of the ansatz. Moreover, the infinite-system size description of iMPS can be used to devise control strategies in the thermodynamic limit for which an efficient mapping to quantum circuits exist as well \cite{Foss-Feig2022}. Similarly, systems with periodic boundary conditions can be studied \cite{Wall2022}. Furthermore, tensor networks come with a comprehensive toolbox for analyzing their properties, such as the entanglement structure and correlations. Hence, tensor-network-based reinforcement learning will enable us to study the data, the training, and the expressivity of the ansatz using well-understood concepts from quantum many-body physics~\cite{martyn2020,lu2021}.

Finally, we mention that it is straightforward to use RL algorithms other than Q-learning in conjunction with our MPS-based ansatz. While we chose the DQN framework since it is off-policy and, therefore, more data-efficient compared to policy gradient methods [Sec.~\ref{app:dqn}], the latter would more naturally allow for continuous action spaces. In turn, with continuous controls, 
target states can be reached with higher fidelity~\cite{yao2021}. One can also adapt the reward function and, for instance, consider the energy density or various distance measures beyond the fidelity.

\noindent\textit{\textbf{Acknowledgments}} --- We wish to thank Thomas Busch and Pankaj Mehta for stimulating discussions. 
M.B.~was supported by the Marie Sklodowska-Curie grant agreement No 890711, and the Bulgarian National Science Fund within National Science Program VIHREN, contract number KP-06-DV-5 (until 25.06.2021).
The authors are pleased to acknowledge that the computational work reported on in this paper was performed on the MPI PKS and OIST HPC clusters. We are grateful for the help and support provided by the Scientific Computing and Data Analysis section of Research Support Division at OIST. For the tensor network simulations we used the \textsc{tensornetwork} library~\cite{tensornetwork_google}. For performance enhancements via just-in-time compilation we employed \textsc{jax}~\cite{jax2018github}; \textsc{tenpy} \cite{tenpy} and \textsc{quspin} \cite{quspin} were used for benchmarking and testing.

\noindent\textit{\textbf{Data availability}} --- The data used in the figures are available upon reasonable request. The trained model parameters and corresponding learning curve data are available on GitHub \cite{github}.

\noindent\textit{\textbf{Code availability}} --- The software code can be accessed on GitHub \cite{github}.

\noindent\textit{\textbf{Author contributions}} --- 
F.M.~and M.B.~conceived the research and wrote the manuscript.
F.M.~performed the numerical simulations and the theoretical analysis. 
M.B.~supervised the work.

\noindent\textit{\textbf{Competing interests}} --- The authors declare no competing interests.

\section*{Videos}

\hypertarget{video:1}{ \textbf{Video~1}} -- {\bf Transverse-field Ising control}\\
Bloch sphere trajectory of the reduced density matrix of a single spin in the bulk ($i\!=\!15$) when starting from the initial ground state at $g_x\!=\!1.01$ and acting according to the optimal QMPS protocol. Also shown are the single-particle and many-body fidelities, the half-chain von Neumann entanglement entropy, the local magnetizations $\langle \sigma_i\rangle$, and the local spin-spin correlations $\langle \sigma_i \sigma_{i+1}\rangle$ during each step of the protocol. The protocol can be divided into three segments involving an initial single-particle rotation and a final many-body Euler-angle-like rotation which reduces unwanted correlations \cite{movies}.

\hypertarget{video:2}{ \textbf{Video~2}} -- {\bf Self-correcting Ising control}\\
Bloch sphere trajectories of the reduced density matrix of a single spin at the edge ($i\!=\!0$) and in the bulk ($i\!=\!15$) when starting from the initial ground state at $J\!=\!-1, g_x\!=\!1.2, g_z\!=\!0.2$ and preparing the critical-region target state. The red arrow/curve denotes the trajectory obtained via acting with the optimal QMPS protocol. The blue arrow/curve shows a suboptimal trajectory where at time step $t\!=\!5$ a different than the optimal action is chosen. Afterwards, the systems is again evolved according to the optimally acting QMPS agent. We show two examples of perturbed trajectories consecutively. In both cases the agent is able to successfully prepare the target state despite the perturbation, thus illustrating the ability of the agent to generalize and adapt its protocols on-the-fly \cite{movies}.

\hypertarget{video:3}{ \textbf{Video~3}} --  {\bf Self-correcting Ising control}\\
Same as Video~\protect\hyperlink{video:2}{2}. However, in this case the blue arrow/curve displays the Bloch sphere trajectory subject to noisy dynamics. Specifically, at each time step we add white Gaussian noise with std $\sigma\!=\!0.05$ to the time step duration $\delta t_{\pm}$. We show two examples of noisy trajectories with different random seeds consecutively. In both cases the agent is again able to successfully prepare the target state which illustrates the robustness of the QMPS agent to noisy dynamics \cite{movies}.

\section*{\label{sec:methods}Methods}

\subsection*{\label{sec:rl}Reinforcement learning (RL) framework}

In RL, a control problem is defined within the framework of an environment that encompasses the physical system to be controlled, and a trainable agent which chooses control actions to be applied to the system [Fig.~\ref{fig:qmps}] \cite{sutton2018}. The environment is described by a state space $\mathcal{S}$ and a set of physical laws that govern the dynamics of the system. 
We consider episodic training, and reset the environment after a maximum number $T$ of time steps. 
At each time step $t$ during the episode, the agent observes the current state $s_t\in\mathcal{S}$ and receives a scalar reward signal $r_t$.
Depending on the current state $s_t$, the agent chooses the next action $a_{t+1}$ from a set of allowed actions $\mathcal{A}$; this, in turn, alters the state to $s_{t+1}$. 
The feedback loop between agent and environment is known as a Markov decision process. 
The goal of the agent is to find an optimal policy (a function mapping states to actions) that maximizes the expected cumulative reward in any state [Sec.~\ref{app:dqn}].

\textit{\bf States ---} In our quantum many-body control setting, the RL state space $\mathcal{S}$ comprises all quantum states $\ket{\psi}$ of the $2^N$-dimensional many-body Hilbert space. 
Here, we consider states in the form of an MPS with a fixed bond dimension $\chi_{\psi}$: if $\chi_{\psi}\!<\!\chi_{\max}$ is smaller than the maximum bond dimension $\chi_{\max}\!=\! 2^{N/2}$, long-range entanglement cannot be fully captured, and the resulting MPS becomes a controlled approximation to the true quantum state [Sec.~\ref{app:mps}].
Hence, state preparation of volume-law entangled states is restricted to intermediate system sizes when using MPS. 
On the other hand, for large system sizes, the control problems of interest typically involve initial and target states that are only weakly entangled such as ground states of local many-body Hamiltonians. 
In these cases, the optimal protocol may not create excessive entanglement suggesting that the system follows the ground state of a family of local effective Hamiltonians \cite{Ljubotina2022,Lami2022}, similar to shortcuts-to-adiabaticity control \cite{odelin2019}, and thus, justifying a MPS-based description.

\textit{\bf Actions ---} If not specified otherwise, the set of available actions $\mathcal{A}$ contains local spin-spin interactions and single-particle rotations,  as defined in Eq.~\eqref{eq:actions}.

\textit{\bf Rewards ---} Since our goal is to prepare a specific target state, a natural figure of merit to maximize is the fidelity $F_t=|\langle\psi_t|\psi_\ast\rangle|^2$ between the current state $\ket{\psi_t}$ and the target state $\ket{\psi_\ast}$. 
To avoid a sparse-reward problem caused by exponentially small overlaps in many-body systems, we choose the log-fidelity per spin at each time step as a reward:  $r_t={N}^{-1}\log(F_t)$. Moreover, we set a fidelity threshold $F^\ast$, which the agent has to reach for an episode to be terminated successfully. Note that the agent receives a negative reward at each step; this provides an incentive to reach the fidelity threshold in as few steps as possible, in order to avoid accruing a large negative return $R = \sum_{t=1}^T r_t$, thus leading to short optimal protocols. For assessing the performance of the QMPS agent to prepare the target state, we show the final single-particle fidelity $F_{\mathrm{sp}}\!=\! \sqrt[N]{F}$
as it represents a more intuitive quantity than the related log fidelity used in quantum simulation experiments. A detailed comparison of the control study results in terms of the achieved single-and many-body fidelities can be found in the \SI, Sec.~\ref{details}.

In the case where the target state is the ground state of a Hamiltonian $H$, we can also define the reward in terms of the energy expectation value $E_t=\braket{\psi_t}{H|\psi_t}$. Specifically, we can choose $r_t = N^{-1}(E_{0} - E_t)$, where $E_{0} = \braket{\psi_*}{H|\psi_*}$ is the ground state energy. Similarly to the log-fidelity, this reward is always negative and becomes zero when evaluated on the target ground state. If the target state and therefore also its energy is a priori not known, one can alternatively replace $E_0$ with a large negative baseline which ensures that the rewards are always negative during training. Another advantage of the energy reward is the fact that expectation values can be efficiently calculated on a quantum device (in contrast to the fidelity). 
We report results obtained with this reward definition in the \SI Sec.~\ref{app:case1b}.

\textit{\bf Training ---} Each training episode starts by sampling an initial state followed by taking environment (state evolution) steps. An episode terminates once the fidelity threshold is reached. After every environment step an optimization step is performed [see Sec.~\ref{app:dqn} for a detailed explanation of the algorithm].

We note in passing that we do not fix the length of an episode (the number of protocol steps) beforehand and the agent is always trying to find the shortest possible protocol to prepare the target state. However, we terminate each episode after a maximum number of allowed steps even if the target state has not been successfully prepared yet: otherwise episodes, especially at the beginning of training, can become exceedingly long leading to unfeasible training times.

\begin{figure}[t!]
	\includegraphics[width=1.0\columnwidth]{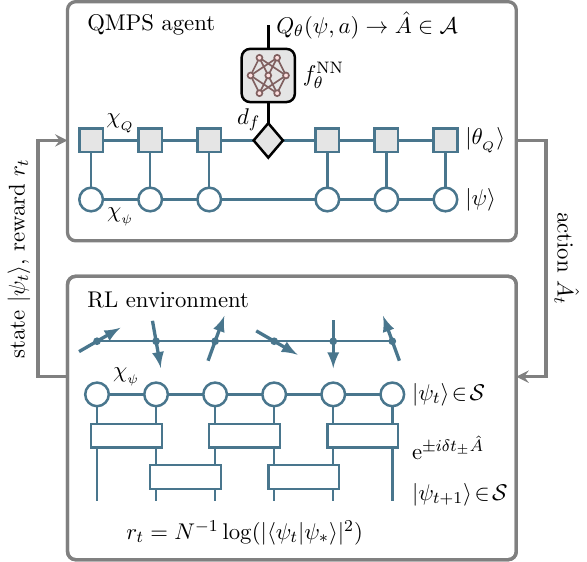}
	\caption{\label{fig:qmps} {\bf Q-learning framework (QMPS)} based on matrix product states. The RL environment encompasses a quantum many-body spin chain represented in compressed MPS form which is time evolved according to globally applied unitary operators chosen from a predefined set $\mathcal{A}$. The reward $r_t$ is given by the normalized log-fidelity between the current state $\ket{\psi_t}$ and the target $\ket{\psi_\ast}$. The QMPS agent is represented by a parameterized $Q$-value function $Q_\theta(\psi, a)$ composed of a MPS $\ket{\theta_Q}$ which is contracted with the quantum state MPS $\ket{\psi_t}$, and a subsequent neural network (NN) which outputs a $Q$-value for each different action $\hat{A}$. The trainable parameters of the QMPS are determined by the feature vector dimension $d_f$ and the bond dimension $\chi_Q$.
	}
\end{figure}

\subsection*{\label{sec:qmps}Matrix product state ansatz for Q-learning (QMPS)}

We choose Q-learning to train our RL agent [Sec.~\ref{app:dqn}], since it is off-policy and, thus, more data-efficient compared to policy gradient methods.
The optimal Q-function $Q^\ast(\psi, a)$ defines the total expected return starting from the state $|\psi\rangle$, selecting the action $a$ and then following the optimal protocol afterwards. Intuitively, the optimal action in a given state maximizes $Q^\ast(\psi, a)$. Hence, if we know $Q^\ast(\psi, a)$ for every state-action pair, we can solve the control task. In Q-learning this is achieved indirectly, by first finding $Q^\ast$. This approach offers the advantage to re-use the information stored in $Q^\ast$ even after training is complete.

Since the state space is continuous, it becomes infeasible to learn the exact $Q^\ast$-values for each state. 
Therefore, we approximate  $Q^\ast \approx Q^\ast_\theta$ using a function parametrized by variational parameters $\theta$, and employ the DQN algorithm to train the RL agent~\cite{mnih2015}.
In this work, we introduce a novel architecture for the $Q^\ast$-function, based on a combination of a MPS and a NN, called QMPS, which is specifically tailored for quantum many-body states that can be expressed as a MPS [Fig.~\ref{fig:qmps}]. We emphasize that the QMPS is independent of the MPS representation of the quantum state, and has its own bond dimension $\chi_Q$. 

To calculate $Q_{\theta}(\psi,a)$ for each possible action $a$ in a quantum state $|\psi\rangle$, we first compute the overlap between the quantum state MPS and the QMPS. The contraction of two MPS can be performed efficiently and scales only linearly in the system size for fixed bond dimensions. The output vector of the contraction corresponding to the dangling leg of the central QMPS tensor, is then interpreted as a feature vector of dimension $d_f$, which is used as an input to a small fully-connected neural network [Fig.~\ref{fig:qmps}]. Adding a NN additionally enhances the expressivity of the $Q^\ast_\theta$ ansatz by making it nonlinear.
The final NN output contains the $Q^\ast$-values for each different action. 

The QMPS feature vector can be naturally written as an overlap between the quantum state MPS $\ket{\psi}$ and the QMPS $\ket{\theta_Q}$. Thus, the $Q^\ast$-value can be expressed as
\begin{equation}
	\label{eq:qmps}
	Q_{\theta}(\psi, a)=f_{\theta}\left(N^{-1}\log \left(|\langle\theta_Q|\psi\rangle|^{2}\right) \right),
\end{equation}
where $f_{\theta}(\cdot)$ denotes the neural network. We additionally apply the logarithm and divide by the number of spins $N$ in order to scale the QMPS framework to a larger number of particles. 
Note also that the QMPS does not represent a physical wave function (it is not normalized); however, for ease of notation, we still express it using the bra-ket formalism.

Thus, the trainable parameters $\theta$ of the $Q^\ast$-function contain the $N\!+\!1$ complex-valued QMPS tensors $\ket{\theta_Q}$, plus the real-valued weights and biases of the subsequent NN. The QMPS feature dimension $d_f$ and the QMPS bond dimension $\chi_Q$ are hyperparameters of the optimization, which determine the number of variational parameters of the MPS in analogy to the hidden dimension of neural networks. An advantage of the MPS architecture is that we can open up the black-box of the ansatz and training using well-understood concepts from the MPS toolbox. For example, it allows us to analyze the correlations in the quantum state and the QMPS by studying its entanglement properties.

Alternatively to the MPS ansatz, we can also represent the parameters of the Q-value network in terms of a matrix product operator (MPO) $\hat{\theta}_Q$. The Q-value computation then amounts to computing the expectation value of the MPO ansatz with respect to the input quantum state, i.e., $Q_{\theta}(\psi, a)=f_{\theta}(\langle \psi |\hat{\theta}_Q|\psi\rangle) $. For a detailed explanation of this architecture, we refer to the \SI, Sec.~\ref{qmpo}. Furthermore, in \SI, Sec.~\ref{app:case1b} we provide a performance comparison of different QMPS/NN architecture choices for the $N\!=\!4$ qubit problem discussed in Sec.~\ref{sec:case1}.

Note that the resources (time and memory) for training the QMPS framework scale at worst polynomially in any of the parameters of the system and the ansatz, such as the QMPS bond dimension $\chi_Q$, the feature dimension $d_f$, and the local Hilbert space dimension $d\!=\!2$. Furthermore, QMPS reduces an exponential scaling of the resources with the system size $N$ to a linear scaling in $N$, therefore, allowing efficient training on large spin systems.

	}

        \let\oldaddcontentsline\addcontentsline
        \renewcommand{\addcontentsline}[3]{}
        \bibliography{bibliography}
        \let\addcontentsline\oldaddcontentsline
	
	
	
	
	\cleardoublepage
	\onecolumngrid
	
	\begin{center}
		\textbf{\large{\textit{\SI} \\ \smallskip
				\titletext}}\\
		\hfill \break
		\smallskip
	\end{center}
	
	\renewcommand{\thefigure}{S\arabic{figure}}
	\setcounter{figure}{0}
	\renewcommand{\theequation}{S.\arabic{equation}}
	\setcounter{equation}{0}
	\renewcommand{\thesection}{S.\arabic{section}}
	\setcounter{section}{0}
	
	\twocolumngrid
	\tableofcontents

	\section{Quantum many-body physics\label{app:many-body}}

Quantum many-body physics deals with the behavior of large collections of interacting quantum particles, such as electrons, atoms, or spins \cite{bruus2004,mahan1990}. Hence, quantum many-body physics is used to study materials at the atomic scale, which can provide insights into their macroscopic features, such as conductivity, polarizability, or magnetism. Quantum many-body systems can exhibit a wide range of phenomena, from the emergence of novel phases of matter to the occurrence of quantum phase transitions and collective behavior that show exotic electronic and magnetic properties that are considered in the development of novel quantum technologies.

Quantum simulators provide a versatile testbed for studying many-body physics by mimicking the behavior of these quantum materials in a controlled environment \cite{Georgescu14}. By engineering the interactions between the constituent particles of the quantum simulator, one can create a system that displays properties similar to the target system, which would otherwise be difficult or impossible to realize in real materials. Examples of quantum simulators include ultracold atoms in optical lattices \cite{lewenstein2007}, nitrogen-vacancy centres \cite{casola2018}, trapped ions \cite{blatt2012}, or photonic systems \cite{Hartmann2016}.

Paradigmatic systems that are often realized in quantum simulator platforms and that allow us to study a range of interesting quantum many-body phenomena are quantum spin models like the Ising model considered in the main text \cite{parkinson2010}. The quantum Ising model describes interactions between spins on a lattice. The quantum state of the spin system is governed by the Ising Hamiltonian [see Eq.~\eqref{eq:ising}], which includes a coupling term between neighboring spins that tends to (anti-)align the spins depending on the sign of the coupling constant, and a magnetic field term that tries to align each spin along the direction of the field. The competition between the spin interaction and the magnetic field gives rise to different phases of matter and thus to a quantum phase transition.

Unlike classical phase transitions (such as the liquid-gas transition in water), a quantum phase transition occurs at zero temperature by varying a non-thermal parameter of the system like the magnetic field \cite{sachdev2011}. At the phase transition the properties of the system abruptly change which leads to non-analyticities in the ground state properties such as diverging susceptibilites and correlation lengths. The latter are also typical characteristics of continuous phase transitions which feature a power-law scaling of correlations and the emergence of universal behavior that is independent of the specific details of the system. For example, the transverse field Ising model exhibits a critical point at $J\!=\!g_x$ where the system transitions from a disordered paramagnetic phase where spins are aligned along the magnetic field in $x$-direction to an ordered ferromagnetic phase where spins are uniformly aligned in either the positive or negative $z$-direction. 

Quantum many-body ground states can also be studied and classified according to their entanglement properties \cite{Eisert10}. To that end, the von Neumann entanglement entropy can be used as a measure of the amount of entanglement between a subset of particles and the rest of the system. It is defined in terms of the reduced density matrix $\rho_A=\text{Tr}\ket{\psi}\bra{\psi}$ for any bipartition $A/B$ of the system
\begin{equation} 
    S^{A}_{\mathrm{ent}} = -\text{Tr}\left[ \rho_A\log\rho_A \right].
\end{equation}
The entanglement entropy can be regarded as a measure of how correlated a system is and thus, it is usually large for systems that are highly correlated. 
For ground states of local, gapped Hamiltonians it has been shown that the entanglement entropy scales proportionally to the boundary area of the partition and hence, these states are commonly referred to as area-law entangled states. On the other hand, gapless critical states result in a logarithmic scaling of the entanglement entropy. And finally, for systems with long-range interactions, the entanglement is usually spread over a large region of space. Hence, the entanglement entropy typically scales with the volume of the partition and the corresponding states are said to be volume-law entangled. Note that most quantum states of a many-body Hilbert space display a volume-law of entangelement. Nontheless, area-law entangled states have received widespread attention since they can be efficiently simulated using the matrix product state formalism \cite{hastings2007,schuch2008}.

\section{Computational Methods}

\subsection{\label{app:mps}Matrix product states (MPS)}

\subsubsection{MPS in quantum physics}

\begin{figure}[t!]
	\includegraphics[width=0.9\columnwidth]{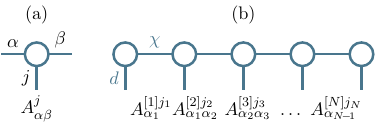}
	\caption{\label{fig:mps} (a) Diagrammatic notation of a rank-3 tensor. (b) Diagrammatic representation of MPS where $d$ denotes the local Hilbert space dimension and $\chi$ the bond dimension of the virtual legs that are contracted over.
	}
\end{figure}

Matrix product states (MPS) define a class of tensor networks (TN)  -- a computational tool developed to simulate quantum many-body systems~\cite{schollwock2011,orus2014}. 
Consider a lattice of $N$ sites with local on-site Hilbert space dimension $d_j$ on site $j$ [for spin-$1/2$ systems, $d_j=2$]. 
In general, the amplitudes, $\psi$, of a many-body wave function can be represented as a rank-$N$ tensor $\psi_{j_1,\dots ,j_N}$, where $j_i\in\{1,\dots,d_i\}$. A TN constitutes a decomposition of the rank-$N$ tensor into a product of lower-rank tensors. In particular, a MPS represents the wave function as a contraction over $N$ rank-3 tensors, cf.~Fig.~\ref{fig:mps}:
\begin{eqnarray}
	|\psi\rangle&\!=\!&\!\sum_{j_{1}, \ldots, j_{N}}\!\! \psi_{j_1,\dots ,j_N} \left|j_{1}, j_{2}, \ldots, j_{N}\right\rangle\\
	&\!=\!&\!\sum_{j_{1}, \ldots, j_{N}}\sum_{\alpha_{2}, \ldots, \alpha_{N}}\\
	& &\hspace*{1cm} A_{\alpha_{1} \alpha_{2}}^{[1] j_{1}} A_{\alpha_{2} \alpha_{3}}^{[2] j_{2}} \ldots A_{\alpha_{N} \alpha_{N+1}}^{[N] j_{N}}\left|j_{1}, j_{2}, \ldots, j_{N}\right\rangle,\nonumber
\end{eqnarray}
where $\left|j_{1}, j_{2}, \ldots, j_{N}\right\rangle$ denotes the basis states. The physical indices $j_i=\{1,\dots,d_i\}$ run over the local Hilbert space dimension.
The non-physical indices, denoted by $\alpha$, are referred to as bond indices; the corresponding bond dimension $\chi_i$ determines the number of parameters in the ansatz which scales as $N\chi^2d$ for a uniformly fixed bond dimension.

It has been shown that MPS define an efficient representation of states obeying the area-law of entanglement in one dimension, i.e., states whose entanglement entropy at any bond of the lattice is independent of the system size $N$ \cite{hastings2007,schuch2008}. For such states, the bond dimension $\chi$ becomes system-size independent and all MPS operations, such as computing overlaps, scale only linearly in $N$ [as compared to the exponential scaling for generic states]. A class of states that automatically fulfill the area-law are ground states of local, gapped Hamiltonians. In contrast, states for which the entanglement entropy depends on the system size, e.g.~linearly (volume-law states), or logarithmically (critical states), will inevitably require a bond dimension that increases with number of spins $N$ for an exact representation; this results in computational resources scaling, at worst, exponentially with $N$. In the latter class fall quantum states of systems taken far-from-equilibrium which feature a ballistic growth of the entanglement entropy~\cite{calabrese2005,kim2013}. Hence, the applicability of MPS is usually restricted to short-time dynamics or to time evolution generated by local Hamiltonians with times scaling at worst polynomially in the number of spins \cite{poulin2011}. These conditions are satisfied by our proposed control setup and thus, justify a description using MPS. In Sec.~\ref{sec:case2} and Sec.~\ref{app:case2} we also analyze the entanglement entropies of the quantum states during time evolution which show that they can indeed by faithfully approximated by MPS.

MPS come with a well-developed toolbox of algorithms for calculating ground states (e.g., DMRG) \cite{white1992}, computing time evolution (e.g., TEBD), and with efficient algorithms for computing overlaps and expectation values of local observables~\cite{schollwock2011,orus2014}. In our simulations, we use the two-site density matrix renormalization group (DMRG) algorithm provided in the TensorNetwork library \cite{tensornetwork_google} to compute ground states, and a custom simplified version of time-evolving block decimation (TEBD) to time-evolve quantum states. Since we are only employing unitaries generated by a sum of commuting operators, the time evolution can be simplified, leading to a considerable computational speed-up. Each time evolution step is carried out by applying a sequence of single/two-site operators along the spin chain. After every application of a two-site operator, a singular value decomposition is performed to maintain the MPS form and to reduce the bond dimension by truncating the matrix containing the singular values. The error introduced in this process, can be quantified by the truncation weight $\epsilon$, which is calculated from the norm of the discarded singular values $\lambda_{\alpha}^{[i]}$ at each bond $i$
\begin{equation} 
	\label{eq:trunc_error}
	\epsilon = 1-\prod_{i=1}^{N-1} \left(1-2\sum_{\alpha=\chi_{\mathrm{trunc}}}^{\chi_{\max}}|\lambda_{\alpha}^{[i]}|^{2}\right),
\end{equation}
and represents an upper bound for the total truncation error after one full TEBD sweep.

As a measure of entanglement we use the von Neumann entanglement entropy; it can be expressed in terms of the singular values $\lambda_{\alpha}^{[i]}$ for any bipartition of the system at bond $i$:
\begin{equation} 
	S^{i}_{\mathrm{ent}} = -\sum_{\alpha=1}^{\chi}|\lambda_{\alpha}^{[i]}|^{2} \log \left(|\lambda_{\alpha}^{[i]}|^{2}\right) .
\end{equation}

\vspace{-0.5cm}
\subsubsection{MPS in machine learning}

While MPS were originally developed within the condensed matter community to study the low-energy properties of quantum many-body systems, the variational expressivity of MPS and, more generally, tensor networks, has recently been harnessed to solve machine learning problems~\cite{miles2016,han2018,glasser2019b,martyn2020,selvan2020,liu2018,sun2020b,liu2021,stokes2019,glasser2019,guo2018,wang2020,liu2019,reyes2020,miles2018,cheng2019,wall2021,gao2020}. In one of the first examples of MPS-based machine learning, an MPS ansatz is used for the compression of the weight matrix in a linear classification task \cite{miles2016}. In this approach the classical data is first mapped to a ``spin" state defined on a higher dimensional space, which is then contracted with the weight MPS. The latter features one tensor (usually in the center) containing a dangling leg representing the resulting prediction of the model, e.g., the class probabilities. So far most applications of MPS-based machine learning involved supervised learning tasks such as classification \cite{miles2016,martyn2020,selvan2020,liu2018,sun2020b}, or unsupervised learning tasks such as generative modeling \cite{han2018,liu2021,stokes2019}, sequence modeling \cite{glasser2019,guo2018}, and anomaly detection \cite{wang2020}. 
Recently, MPS have also been utilized as a feature map for classical data in a quantum reinforcement learning (RL) framework \cite{chen2021}.

The MPS architecture can be optimized via conventional gradient descent, and the gradients can be obtained through backpropagation analogous to the optimization of neural network parameters. Alternatively, one can use a DMRG-style routine for MPS that only locally optimizes one or two tensors at a time, while sweeping back and forth through the MPS \cite{miles2016}. This algorithm has the advantage that the bond dimension can be adapted dynamically during the optimization and has also shown better stability for large system sizes ($N>200$) where backpropagation fails due to exponentially vanishing gradients \cite{sun2020,liu2021b}. Since we are dealing with intermediate system sizes $N<200$ in this study, we used backpropagation for calculating the gradients of the MPS tensors.

\vspace{-0.8cm}
\subsection{\label{app:dqn}Q-learning and DQN}
\vspace{-0.2cm}

Within the RL framework the agent chooses actions according to a strategy, called policy $\pi(a|s)$ -- a function that assigns a probability to every action $a$ depending on the current state $s$ of the environment \cite{sutton2018}. The goal is to find the optimal policy $\pi^\ast(a|s)$, i.e., the optimal action to take in any state $s$ that maximizes the expected return $R=\mathbb{E}_\pi \left[\sum_{t=0}^{T} r_{t} | s_{0}=s\right]$ starting from state $s$ and following the policy $\pi$. In this work, we consider episodic tasks involving a termination condition (e.g., a fidelity threshold) which the agent has to reach within a fixed number of steps $T$. Once the termination condition is satisfied, the episode is over and the environment is reset. This is in contrast to non-episodic tasks which continue indefinitely.

Q-learning is a model-free RL algorithm in which the agent learns an optimal policy $\pi^\ast(a|s)$ solely via observing environment transitions, i.e., without knowing or building a representation of the environment dynamics, and without access to any prior information about the system~\cite{watkins1992}. To every fixed policy $\pi$ (optimal or sub-optimal), we can assign a Q-function, defined as the expected return starting from state $s$, taking action $a$, and following the policy $\pi$ afterwards:
\begin{equation} 
	Q^\pi(s,a) = \mathbb{E}_\pi \left[\sum_{t=0}^{T} \gamma^t r_{t} | s_{0}=s, a_{0}=a\right].
\end{equation}
The discount factor $\gamma\in(0,1]$ gives a higher importance to immediate rewards and therefore ensures stability for continuing, non-episodic RL tasks.

In Q-learning, the optimal policy $\pi^\ast$ is found indirectly through learning the optimal Q-value function $Q^*(s,a)$ that gives the maximum expected cumulative discounted reward: 
\begin{equation}
	Q^*(s,a)=\max_\pi Q^\pi(s,a).
\end{equation}
Once the optimal Q-values are known, the optimal policy is deterministic: $\pi^{*}(s)=\arg \max _{a} Q^{*}(s, a)$, i.e., it is given by greedily taking actions according to the maximum optimal Q-value in each state.

\vspace{-0.5cm}
\subsubsection{\label{subsec:tabular_Q}Tabular Q-Learning}
\vspace{-0.3cm}

When the state space is discrete, the optimal Q-function can be learned using tabular Q-learning through an iterative update rule derived from the Bellman optimality equation \cite{watkins1992}
\begin{eqnarray}
	Q_{k+1}\left(s, a\right) \!&\leftarrow&\! Q_k\left(s, a\right)+\alpha\delta_k, \\
	\delta_k &=& r(s, a)+\gamma  \max _{a^{\prime}} Q_{k}\left(s^{\prime}, a^{\prime}\right) - Q_k\left(s,a \right), \nonumber
\end{eqnarray}
where $k$ denotes the iteration step of the algorithm, $\alpha \in(0,1]$ is the learning rate, and $\delta_k$ is the temporal difference error. Note that Q-learning requires isolated tuples $(s,a,r,s')$, known as transitions, and not complete trajectories. Moreover, due to the presence of the $\max$ function in the update-rule above, the algorithm is \textit{off-policy}: this means that the transitions can come from any policy (also old ones) -- and yet the new updated Q-function approaches the optimal $Q^\ast$. 

Convergence is guaranteed if each possible state-action pair $(s,a)$ can, in principle, be visited infinitely often. To fulfill this condition the agent has to \textit{explore} sufficiently different state-action pairs. At the same time, the agent should also \textit{exploit} the high-reward transitions, especially towards the end of training when the Q-value estimates have mostly converged to their true values. A common choice of behavior policy to follow during training that satisfies this Exploration-Exploitation dilemma, is an $\epsilon$-greedy policy:
\begin{equation}
	a=\begin{cases}
		\text{random action} & \text{with probability $\epsilon$}\\
		\arg \max _{a'} Q^\pi(s, a') & \text{otherwise}
	\end{cases},
\end{equation}
i.e.~the agent chooses a random action with some small probability $\epsilon$ and the greedy action maximizing the Q-value otherwise. The hyperparameter $\epsilon$ can be decreased, e.g., exponentially, starting from a value close to $1$ (exploration-dominated regime) at the beginning of training, to a small value, e.g., $\epsilon=0.01$, leading to less exploration and more exploitation as training progresses.

\vspace{0.5cm}
\subsubsection{\label{subsec:DQN}Deep Q-Learning}

For large or continuous state spaces, such as Hilbert spaces, the tabular Q-learning algorithm described above is inapplicable. In such cases, it is only possible to learn an approximation to the optimal Q-values, $Q_{\theta}(s, a) \approx Q^{*}(s, a)$, given by a parameterized function, e.g., a neural network \cite{mnih2015}. The parameters $\theta$ of the variational ansatz are then optimized by minimizing the expected mean-square temporal difference error
\begin{eqnarray}
	\label{eq:Qloss}
	L_{k}\!\left(\theta_{k}\right)&=&\mathbb{E}_{\left(s, a, r, s^{\prime}\right) \sim \mathcal{R}}\bigg[\Big(y_k  - Q_{\theta_{k}}\!\left(s, a\right)\Big)^{2}\bigg], \nonumber\\
	y_k &=& r+\gamma \max _{a^{\prime}} Q_{\bar\theta_{k}}\!\left(s^{\prime}, a^{\prime} \right).
\end{eqnarray}
The minibatch of transitions $(s, a, r, s^{\prime})$, used in each optimization step $k$, is uniformly sampled from a fixed-size replay buffer $\mathcal{R}$ that contains previously collected transitions from agent-environment interactions. Since Q-learning is an off-policy algorithm, transitions used for updating the Q-value do not have to coincide with the target policy allowing the use of experience replay. Thus, the subroutine of collecting environment transitions can be run independently and, if necessary, in parallel to the optimization subroutine, thus speeding up training. Therefore, the use of a replay buffer makes Q-learning more data-efficient than policy gradient methods.

Note that the RL loss function $L_k$ in Eq.~\eqref{eq:Qloss} is different from the loss in supervised learning, in that the regression target $y_k=r+\gamma \max _{a^{\prime}} Q_{\bar\theta_{k}}\!\left(s^{\prime}, a^{\prime} \right)$ itself depends on the parameterized Q-values that have to be learned; therefore, the target (i.e., the label) changes in the course of training. 
This running target makes DQN different from ordinary gradient descent, and is the reason for the lack of convergence guarantees in DQN. 
To stabilize deep Q-learning, a second \textit{target} Q-value network $Q_{\bar\theta_{k}}\!\left(s, a\right)$ is introduced whose parameters $\bar\theta$ are held fixed during the optimization step. The optimized parameters $\theta$ are periodically copied to the target network $\bar{\theta}\leftarrow \theta$.

Finally, we also employ Double Q-learning to reduce overestimation errors in the Q-values \cite{hasselt2015}. Here, the regression target is replaced by
\begin{equation}
	y_{k}^{\text {Double}} = r+\gamma\ Q_{\bar\theta_{k}}(s^{\prime}, \underset{a^{\prime}}{\operatorname{argmax}}\ Q_{\theta_{k}}\!\left(s^{\prime}, a^{\prime} \right)).
\end{equation}
The full training algorithm is called DQN and we show the corresponding pseudocode in Algorithm \ref{alg:qmps}.

\onecolumngrid
\vspace{+0.02cm}

\begin{algorithm}[H]
	
	{\caption{QMPS training}\label{alg:qmps}
		\begin{algorithmic}[1]
			\Require Target state $|\psi_\ast\rangle$, fidelity threshold $F^\ast$, maximum episode length $T$, number of training episodes $N_{\mathrm{eps}}$, learning rate $\alpha$, batch size $N_{\mathrm{batch}}$, discount factor $\gamma$, replay buffer size $N_{\mathrm{buff}}$, target network update frequency $n_{\mathrm{target}}$, exploration parameters ($\epsilon_{\text{init}}, \epsilon_{\text{final}}$)
			\State Initialize QMPS network $Q_{\theta}$ and copy parameters to target network $\bar{\theta}\leftarrow\theta$
			\State Reset RL environment (sample initial state $|\psi_0\rangle$)\vspace*{1mm}
			\State \# \textit{Fill replay buffer with random transitions}
			\For {$i=1,..,N_{\mathrm{buff}}$ } 
			\State Select random action $a\rightarrow\pm\hat{A}$
			\State Time evolve state $|\psi^{\prime}\rangle= \exp(\pm i \delta t_\pm \hat{A})|\psi\rangle$
			\State Compute reward $r={N}^{-1}\log(|\langle\psi^{\prime}|\psi_\ast\rangle|^2)$
			\State Append transition $(|\psi\rangle, a, r, |\psi^{\prime}\rangle)$ to replay buffer
			\State Set $|\psi\rangle = |\psi^{\prime}\rangle$
			\If{$r > F^\ast$ \textbf{or} $T$ is reached}
			\State Reset RL environment (sample new initial state $|\psi_0\rangle$)
			\EndIf
			\EndFor
			\State \# \textit{Start training}
			\For {$l=1,..,N_{\mathrm{eps}}$} 
			\State Reset RL environment (sample initial state $|\psi_0\rangle$)
			\State Compute decay exploration parameter: $\epsilon_l=\epsilon_{\text{final}}+ (\epsilon_{\text{init}} - \epsilon_{\text{final}})  \exp(-8l/N_{\text{eps}})$
			\For {$t=0,..,T$}
			\State \# \textit{Update network}
			\State Sample $N_{\mathrm{batch}}$ transitions $(|\psi\rangle, a, r, |\psi^{\prime}\rangle)$ from replay buffer
			\State Compute regression target $y = r+\gamma\ Q_{\bar\theta}(\psi^{\prime}, \operatorname{argmax}_{a^{\prime}}\ Q_{\theta}\!\left(\psi^{\prime}, a^{\prime} \right))$
			\State Compute gradients of $L(\theta)=\sum_{\text{batch}}(y  - Q_{\theta}\!\left(\psi, a\right))^{2}$ w.r.t.~parameters $\theta$
			\State Perform gradient descent step using ADAM
			\State Every $n_{\mathrm{target}}$ steps: Copy QMPS parameters $\theta$ to target QMPS network $\bar{\theta}\leftarrow\theta$\vspace*{1mm}
			\State \# \textit{RL environment step}
			\State Select action $\pm\hat{A_t}\leftarrow a_t=
			\begin{cases}
				\text{random action} & \text{with probability $\epsilon_l$}\\
				\text{argmax}_a Q_{\theta}(\psi_t, a) & \text{otherwise}
			\end{cases}$
			\State Time evolve state $|\psi_{t+1}\rangle= \exp(\pm i \delta t_\pm \hat{A}_{t})|\psi_t\rangle$
			\State Compute reward $r_t={N}^{-1}\log(|\langle\psi_{t+1}|\psi_\ast\rangle|^2)$
			\State Append transition $(|\psi_t\rangle, a_t, r_t, |\psi_{t+1}\rangle)$ to replay buffer
			\If {$r_t > F^\ast$ \textbf{or} $T$ is reached}{ \textbf{\textit{BREAK}}}
			\EndIf
			\EndFor
			\EndFor
	\end{algorithmic}}
\end{algorithm}
\twocolumngrid

\subsection{\label{app:qmps}Details of the QMPS architecture and training}

In all examples discussed in this manuscript the QMPS tensors are initialized as identity matrices with Gaussian noise ($\sigma\!=\!0.2$) added to all components both for the real and complex parts. The tensors are additionally scaled by a factor of 0.25. The neural network weights and biases are initialized with real Gaussian random numbers ($\sigma\!=\!0.1$). 
All parameter values of the QMPS framework are summarized in Tab.~\ref{table:parameter}. The values of the hyper-parameters including the time evolution step sizes $\delta t_\pm$ are obtained by performing a coarse grid-search, i.e.~we trained on a few different parameter values and select the ones which yield best performance results. Note that we adopt slightly different values for $\delta t_+$ and $\delta t_-$. This choice prevents the agent to simply undo an action by evolving with the inverse operator which helped stabilise training.

As mentioned in the main text a single QMPS agent is not able to reach arbitrarily high fidelities due to the discreetness of the action space, the constant step size, and the fixed maximum episode length. An additional challenge is posed by the large deviation in the expected return values for states at the beginning and the end of the episode:~The QMPS network is not able to resolve small differences in the reward which is however required close to the target state where the log fidelities approach zero. Therefore, we introduce a multi-stage learning scheme in Section \ref{sec:case1} where successive agents with tighter fidelity thresholds are trained starting from states which are pre-prepared from agents optimized on smaller thresholds. This training strategy also allows the step size to be chosen separately for each agent.

\subsubsection{Optimization}

The gradients of the neural network and the QMPS parameters can be computed via conventional backpropagation and, in principle, any automatic differentiation library can be employed for this task. However, we obtained a considerable speed-up (factor of $\sim\!10$) by implementing the gradient computation from scratch. 
The neural network takes as input the real-valued QMPS feature vector and therefore the parameters are chosen to be real-valued as well. On the other hand, restricting the QMPS tensors to real numbers greatly limited the expressivity of the ansatz. Hence, each tensor component is comprised of both a real and imaginary parameter. Due to the overall QMPS ansatz being not holomorphic (the absolute value is not complex differentiable), the real and imaginary parameters have to be updated independently by computing the gradient with respect to each of them separately.

\begin{table}[t!]
	\centering
	\begin{tabular}{| l | c |} 
		\hline
		\textbf{Parameter }& \textbf{Value} \\ 
		\hline \hline
		number of training episodes $N_{\text{eps}}$ & 40000 -- 80000  \\
		optimizer & ADAM \\ 
		learning rate $\alpha$ & $5\times 10^{-5} - 1\times 10^{-4}$ \\ 
		batch size & 32 -- 64\\ \hline
		RL discount factor $\gamma$ & $0.98$  \\ 
		RL buffer size & 8000 \\ 
		target network update frequency $n_\mathrm{target}$ & 10 \\ \hline
		initial exploration $\epsilon_{\text{init}}$ & 1.0 \\ 
		final exploration $\epsilon_{\text{final}}$ & 0.01 \\ 
		exploration decay $\epsilon_l$ & 
		$ \exp(-8\!\times\! l/N_{\text{eps}})$ \\ \hline
		
		QMPS bond dimension $\chi_{Q}$ & 4 -- 32 \\
		QMPS feature vector dimension $d_f$& 32 -- 72 \\ \hline
		
		NN number of hidden layers& 2 \\
		NN number of hidden neurons& 100 -- 200 \\ 
		NN nonlinearity& $\tanh$ \\ \hline
		
	\end{tabular}
	\caption{QMPS training hyperparameters.
	}
	\label{table:parameter}
\end{table}

\begin{table}[t!]
	\centering
	\begin{tabular}{| p{2.5cm} |p{2.9cm} |p{1.4cm} | p{1.4cm} |} 
		\hline
		\textbf{Parameter}&\hfil \textbf{Study~A}\newline (QMPS-1,~QMPS-2) & \textbf{Study B} & \textbf{Study C} \\ 
		\hline \hline
		system size $N$ &\hfil 4 &\hfil 32 &\hfil 16 \\ \hline
		single-particle fid.~threshold $F^\ast$ &\hfil (0.96, 0.992) &\hfil 0.99 &\hfil 0.97 \\ \hline
		max.~episode\newline length $T$ &\hfil 50 &\hfil 50 &\hfil 50 \\ \hline
		number of actions &\hfil 12 &\hfil 7 &\hfil 12 \\ \hline
		step size $\delta t_{+}$ &\hfil $\left(\frac{\pi}{8},\frac{\pi}{16}\right)$ &\hfil $\frac{\pi}{12}$ &\hfil $\frac{\pi}{12}$ \\ \hline
		step size $\delta t_{-}$ &\hfil $ \left(\frac{\pi}{13},\frac{\pi}{21}\right)$ &\hfil $\frac{\pi}{17}$ &\hfil $\frac{\pi}{17}$\\ \hline
		quantum~state \newline bond~dim. $\chi_\psi$ &\hfil 4 &\hfil 16 &\hfil 16 \\ \hline
	\end{tabular}
	\caption{RL environment parameters.}
	\label{table:env_parameter}
\end{table}

\subsubsection{\label{subsec:compute_resources}Compute resources}
For a system of size $N$, local Hilbert space dimension $d$, and uniformly fixed bond dimension $\chi$, the number of MPS parameters scale as $N\chi^2d$. The quantum state MPS time evolution (based on SVD and matrix multiplication) as well as the QMPS optimization (based only on matrix multiplication) scale linear in $N$ and at worst polynomial in $\chi$ and $d$. We have not fixed the bond dimensions of the MPS and QMPS to be uniform on all sites, but rather let both of them grow exponentially from the boundary up to a maximum uniform bond dimension in the middle of the MPS.

Note that the linear complexity scaling of the QMPS framework stands in contrast to the exponential scaling expected when using a conventional neural network architecture and training on the full wave function. Taking the $N=32$ case study as an example and assuming a batch size of 64, one would need to load $64\times 2^{32}=10^{11}$ parameters onto the CPU/GPU which would result in at least 1 TB of data. Additionally, we would have to store the neural network parameters which takes as an input the full wave function, and take into account the replay buffer which stores extra $\sim 8000$ states. Hence, from a viewpoint of memory resources alone, a tensor-network approach is required for the simulation of quantum many-body systems and the training thereof.  Merging the TN architectures with RL is what enables the ML-based control of 32 and more spins discussed in the main text.

For the hyperparameters chosen in this study, most time was spent in the optimization step which requires two forward passes and one backward pass on a batch of input states. 
Overall, one full episode of training (including 50 environment and optimization steps) for $N\!=\!32, \chi_Q\!=\!32, \chi_\psi\!=\!16, d_f\!=\!32$, and a batch size of 64 took 8.7 sec on a Intel Xeon Gold 6230 CPU and 1.5 sec on a NVIDIA Tesla P100 SXM2 GPU. Reducing the QMPS bond dimension to $\chi_Q\!=\!16$ leads to runtimes of 4.3 sec (CPU) and 1.5 sec (GPU) respectively. Let us note that the code has not been optimized for a GPU and with some modifications an even larger speedup can be expected. Therefore, larger system sizes should also be within reach in the near future. 

\subsection{QMPO architecture\label{qmpo}}

The QMPS architecture makes use of an MPS ansatz to extract relevant features from the input quantum state. Hence, we can interpret the QMPS as a set of quantum states (up to normalization); calculating the feature vector then amounts to computing the fidelity between the input state and each QMPS state. However, rather than learning parameterized ``quantum states'' $\ket{\theta_Q}$ and evaluating inner products, we could also learn parameterized operators (or observables) $\hat{\theta}_Q$ that act on the evolved physical quantum state instead. If the operators are further restricted to be hermitian, we are able to express the feature vectors as an expectation value of the operator $\hat{\theta}_Q$ and hence the Q-values are given by
\begin{equation}
    Q_{\theta}(\psi, a)=f_{\theta}(\langle \psi |\hat{\theta}_Q|\psi\rangle) ,
\end{equation}
where $f_{\theta}$ denotes the subsequent NN that the feature vectors are fed through.

\begin{figure}[t!]
    \centering
    \includegraphics[width=1.0\columnwidth]{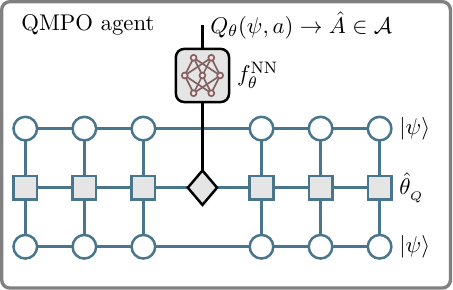}
    \caption{\label{fig:app:qmpo} {\bf QMPO framework}~based on trainable matrix product operator [gray squares]. The Q-values are computed by first calculating the expectation values of the QMPO $\hat{\theta}_Q$ with respect to the input quantum states $|\psi\rangle$. The extracted features are then fed through the subsequent neural network.
    }
\end{figure}

Similarly to the MPS representation of quantum states, we can decompose a hermitian operator $\hat{\theta}$ into a product of local tensors of rank 4, called matrix product operator (MPO) \cite{Pirvu2010}
\begin{eqnarray}\label{eq:mpo}
    \hat{\theta}_{i_1\ldots i_N}^{j_1 \ldots j_N}\!=\!\!\!\!\sum_{\alpha_{2}, \ldots, \alpha_{N}}
    O_{\alpha_{1} \alpha_{2}}^{[1] i_1 j_{1}} O_{\alpha_{2} \alpha_{3}}^{[2] i_2 j_{2}} \ldots O_{\alpha_{N} \alpha_{N+1}}^{[N] i_N j_{N}}.
\end{eqnarray}
By inserting an additional tensor with a dangling leg at the center of the MPO, we introduce an ansatz analogous to the QMPS that maps an input quantum state to a feature vector via computing expectation values [see Fig.~\ref{fig:app:qmpo}].

The QMPO tensors are initialized as identity matrices with Gaussian
noise ($\sigma = 0.5$), added both to the real and imaginary parts of each parameter $X^{i j}_{\alpha \alpha'}\!=\!\text{Re}[X^{i j}_{\alpha \alpha'}] + i \text{Im}[X^{i j}_{\alpha \alpha'}]$. The resulting tensors $X^{i j}_{\alpha \alpha'}$ are not yet hermitian and hence we choose $O_{\alpha \alpha'}^{i j} = \sum_k X_{\alpha \alpha'}^{i k} X_{\alpha \alpha'}^{* jk}$ as the hermitian operators $O$ in the MPO ansatz of Eq.~\eqref{eq:mpo}. 

We also find that normalizing the tensors $O$ before training substantially helps the optimization. To that end, we perform a singular value decomposition on each tensor $O_{\alpha \alpha'}^{i j}= \sum_k U_{\alpha \alpha'}^{i k} S_{\alpha \alpha'}^{k} V_{\alpha \alpha'}^{k j}$ and overwrite the singular values $S_{\alpha \alpha'}^{k} $ with uniformly sampled random numbers form the interval $[0, 2/\chi]$, where $\chi$ is the (local) bond dimension of the corresponding QMPO tensor $O_{\alpha \alpha'}^{i j}$. Hence, the new, normalized QMPO tensors are given by $\tilde{O}_{\alpha \alpha'}^{i j}= \sum_k U_{\alpha \alpha'}^{i k} \tilde{S}_{\alpha \alpha'}^{k} V_{\alpha \alpha'}^{k j}$, where the singular values have been replaced by $\tilde{S}$.

When training, we work with real parameters only. Therefore, we perform a Cholesky decomposition to retrieve the $X$ operators, i.e.,
\begin{equation}\label{eq:mpodecomp}
    \tilde{O}_{\alpha \alpha'}^{i j} = \sum_k \tilde{X}_{\alpha \alpha'}^{i k} \tilde{X}_{\alpha \alpha'}^{* jk}.
\end{equation}
We then define the optimizable parameters to be the real and imaginary parts of the $\tilde{X}$ operators. Training works analogously to the QMPS optimization, that is, via gradient descent and backpropagation. The only difference is that due to the hermiticity requirement of the operators $O$, we always need to perform the additional matrix multiplication step of Eq.~\eqref{eq:mpodecomp}.

The QMPO framework has some advantages over QMPS. First, the Q-value computation can now be interpreted as the measurement of an observable which can be performed efficiently on NISQ devices, if we restrict the observable to be local. Second, the expectation value does not vanish (or explode) exponentially in the system size as is the case for fidelities and hence, we expect that the training of this ansatz is more stable for larger system sizes and requires less hyper-parameter tuning (especially with respect to the parameter initialization). However, the QMPO ansatz is also computationally more demanding, e.g., computing the expectation value in Eq.~\eqref{eq:mpo} scales as $\chi^4$ whereas the calculation of the fidelity in the QMPS only scales as $\chi^3$ (assuming that the bond dimensions of the MPS and MPO are equal and uniform).

In Sec.~\ref{app:case1b} we report results obtained with the QMPO ansatz and compare it with other architecture choices.

\begin{figure*}[t!]
	\centering
	\includegraphics[width=1.0\textwidth]{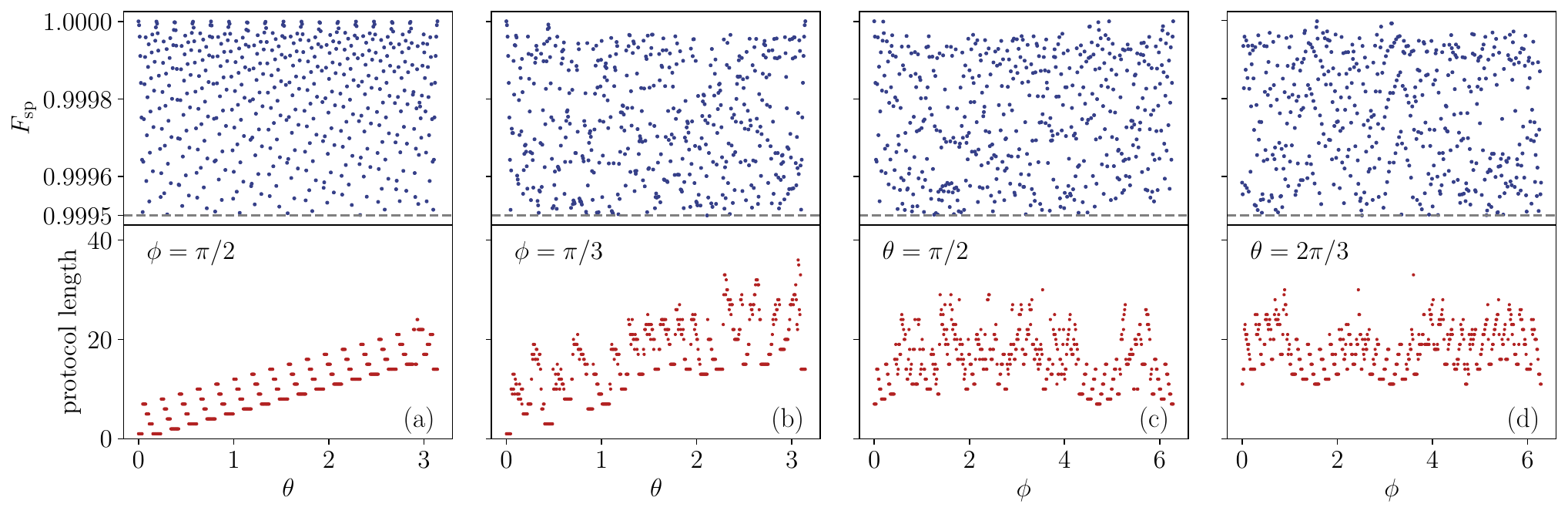}
	\caption{\label{fig:app:sp} {\bf Universal single-particle control ---}~Upper panel:~Single-particle fidelity between the final and $z$-polarized target state ($\theta\!=\!0$) as a function of the initial state parameters $\theta$ and $\phi$. In {\bf (a),(b)} $\phi$ is held fixed; in {\bf (c),(d)} $\theta$ is fixed instead. The agent is able to surpass the fidelity threshold $F^\ast_{\mathrm{sp}}\!=\!0.9995$, (vertical dashed line) for any state on the Bloch sphere. Lower panel: The corresponding number of protocol steps used by the QMPS agent to reach the target state. $N\!=\!64$ spins.
	}
\end{figure*}

\section{Details of the control studies\label{details}}

Parameters related to the RL environment and the spin systems of each control study can be found in Tab.~\ref{table:env_parameter}.

\subsection{\label{app:case1}Universal state preparation from arbitrary initial quantum states}

\subsubsection{\label{app:case0}Single-particle control}

To provide another benchmark of the QMPS framework, we test it in a single-particle control setting. Our goal is to prepare a specific state (here chosen to be the spin-up state) from \textit{any} other single-particle state. We translate this setup to the many-body regime by considering $N\!=\!64$ spins uniformly polarized in one direction on the Bloch sphere, which can be exactly approximated with an MPS of bond dimension $\chi_\psi\!=\!1$. Note that an arbitrary single-particle spin state can always be expressed as $\ket{\psi}\!=\!\cos (\theta / 2)|0\rangle\!+\!\mathrm e^{i \phi} \sin (\theta / 2)|1\rangle$, where $0\!\leq\!\theta\!\leq\!\pi$ and $0\!\leq\!\phi\!<\!2\pi$.

We train a QMPS agent starting each episode from a uniformly sampled state on the Bloch sphere, with a fixed single-particle fidelity threshold of $F^\ast_{\mathrm{sp}}\!=\!0.9995$ (many-body fidelity $F^\ast\!\sim\!0.97$), and an action set that is composed only of single-particle rotations $\mathcal{A}\!=\!\{\hat{X},-\hat{X},\hat{Y},-\hat{Y},\hat{Z},-\hat{Z}\}$. Figure~\ref{fig:app:sp} shows the achieved fidelities between the final and target state for different initial states represented by the angles $\theta$ and $\phi$. We find in all cases that the QMPS agent is able to successfully reach the fidelity threshold and hence is capable of performing universal single-particle state preparation. We also plot the protocol length starting from each initial state, which, as expected, increases as the distance between the initial and the $z$-polarized target state ($\theta\!=\!0$) becomes larger [Fig.~\ref{fig:app:sp}(a)].

\subsubsection{\label{app:case1b}Universal ground state preparation for $N=4$ spins}
\begin{figure}[t!]
	\centering
	\includegraphics[width=1.0\columnwidth]{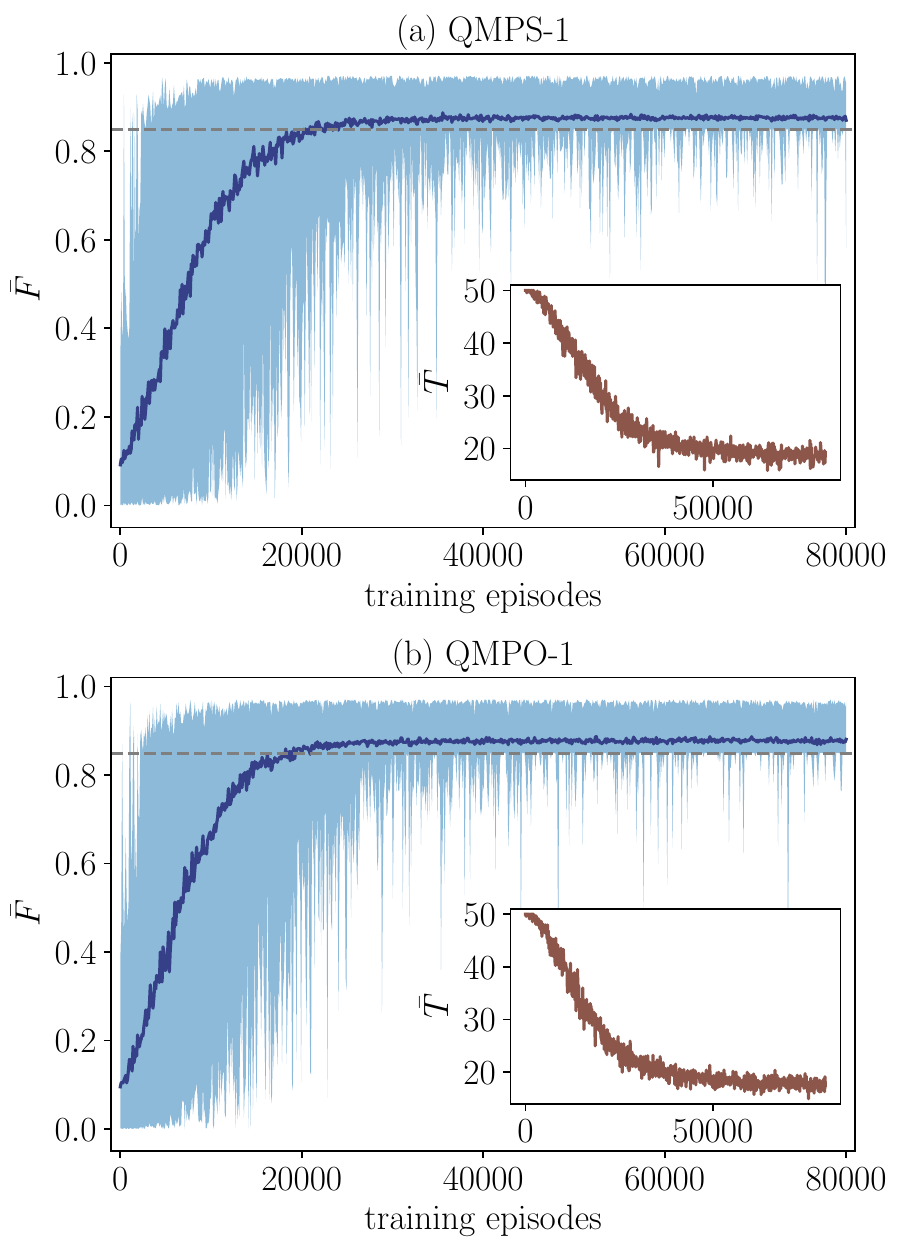}
	\caption{\label{fig:app:cs1}{\bf Universal four-qubit control ---} {\bf (a)}  Learning curves of the QMPS-1 agent trained on a many-body fidelity threshold of $F^\ast\!\sim\!0.85$ (gray-dashed line). The achieved final fidelity $\bar{F}$ is shown averaged over a window of 100 training episodes. The light-blue data indicates the range between the best and the worst fidelity values. Inset:~The number of episode steps $\bar{T}$ averaged over 100 episodes. The maximum number of steps per episode was set to 50. {\bf (b)} Learning curves for training a QMPO agent on the same problem as in (a). The final performance suggests that the QMPO architecture can be used as an alternative ansatz for state preparation tasks.} $N\!=\!4$ spins.
\end{figure}

In Fig.~\ref{fig:app:cs1} (a) we provide the learning curves of the first QMPS-1 agent which was trained to prepare the Ising ground state with a many-body fidelity threshold of $F^\ast\!\sim\!0.85$ [Sec.~\ref{sec:case1}]. When testing the trained \mbox{QMPS-1} agent on a set of $10^3$ random initial states, in $99.8\%$ of instances the fidelity threshold is attained within the 50 allowed number of steps. In Fig.~\ref{fig:app:cs1} (b) we show the corresponding learning curves of a QMPO agent [see Sec.~\ref{app:qmps} for the architecture and optimization procedure details] trained on the same problem. For $10^3$ randomly sampled initial states, we reach a success rate of $100\%$ which suggests that the QMPO ansatz provides an alternative, expressive ansatz for the state preparation scenario at hand.

\begin{figure*}[t!]
	\centering
	\includegraphics[width=1.0\textwidth]{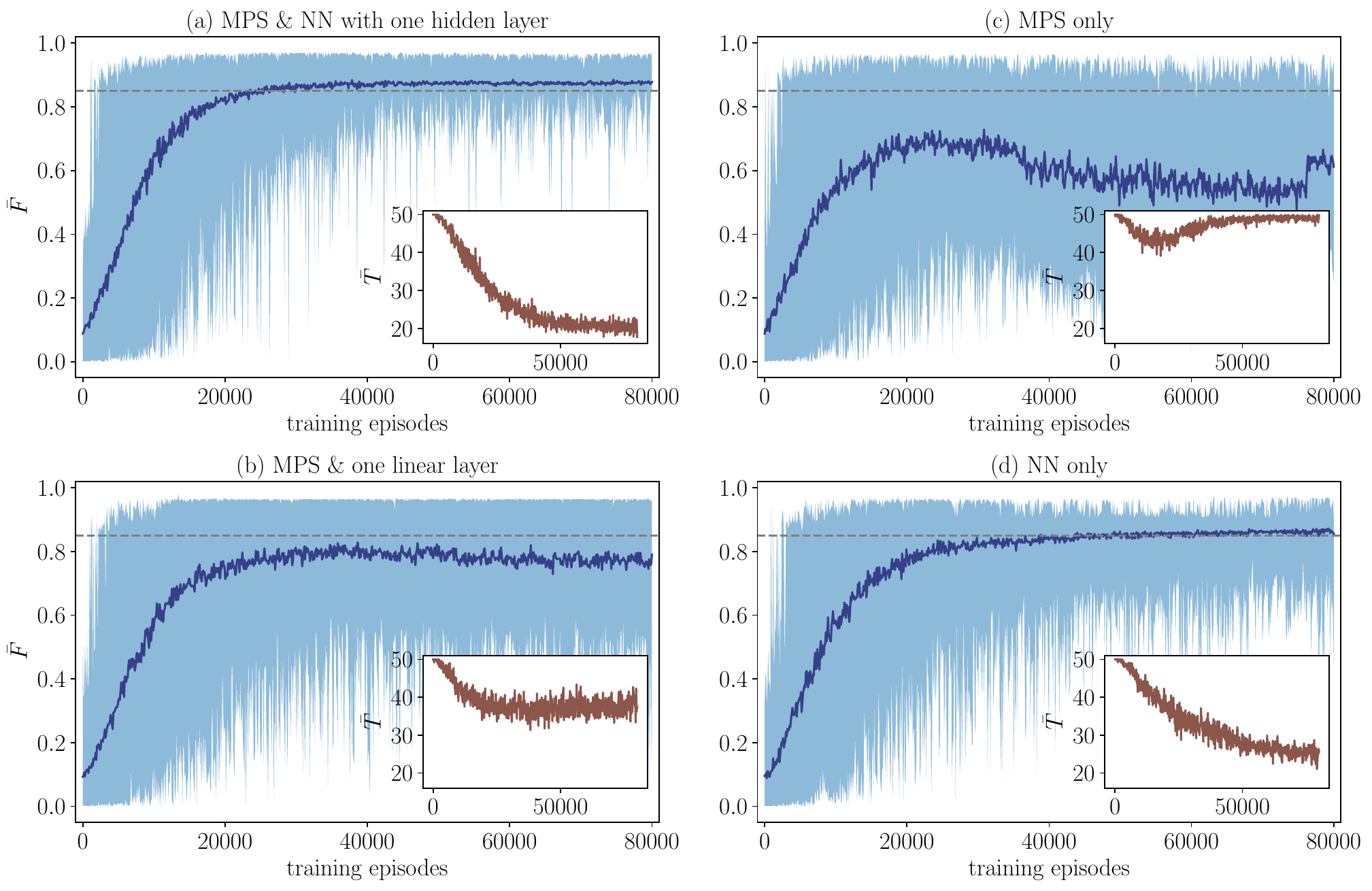}
	\caption{\label{fig:app:cs1b}{\bf Universal four-qubit control ---}~Comparison of learning curves for different QMPS-1 agent architecture choices. The agents are trained on a many-body fidelity threshold of $F^\ast\!\sim\!0.85$ (gray-dashed line). The achieved final fidelity $\bar{F}$ is shown averaged over a window of 100 training episodes. The light-blue data indicates the range between the best and the worst fidelity values. Inset:~The number of episode steps $\bar{T}$ averaged over 100 episodes. The maximum number of steps per episode was set to 50. {\bf (a)} Learning curves of a hybrid architecture of an MPS and a NN with a single hidden layer of dimension 100. The final state preparation success rate is $97.9\%$ computed over $10^3$ randomly sampled initial states. This stands in contrast to the success rate of $99.8\%$ achieved by the QMPS-1 agent from the main text which involved a NN with two hidden layers. Therefore, a deeper NN can increase the expressiveness of the overall ansatz. {\bf (b)} Training curves of a hybrid architecture of an MPS and a single final linear layer. The corresponding achieved success probability is $35.9\%$ and indicates that a non-linear NN is required for successful training of the QMPS agent for the given task. {\bf (c)} Learning curves for a QMPS architecture composed only of an MPS. Training becomes unstable and the final success rate is $1.6\%$. {\bf (d)} Learning curves for a NN (two hidden layers; each of dimension 100) which is trained on the full quantum state wave function. The final success probabilities compute to $94.3\%$. Note that by doubling the hidden dimension to 200, we can increase the success rate to $96.8\%$. $N\!=\!4$ spins.
	}
\end{figure*}

Next, we investigate how the trainability and the final success rates of the \mbox{QMPS-1} agent depend on the architecture choice. For comparison, the success probability of $99.8\%$ and the learning curves of Fig.~\ref{fig:app:cs1} were obtained on a hybrid MPS+NN architecture where the NN contained two hidden layers, each of dimension 100. Figure~\ref{fig:app:cs1b}(a) instead shows the learning curves for a hybrid MPS+NN architecture with a single hidden layer of dimension 100. In this case, the agent is only able to successfully prepare the target state for $97.9\%$ of the random initial states which suggests that a deeper NN can indeed increase the expressiveness of the overall ansatz. Figure~\ref{fig:app:cs1b}(b) displays the learning curves of an agent composed of an MPS followed by a linear layer. The average fidelity converges to a lower value than the fidelity threshold and the final success rate computes to $35.9\%$ only. This indicates that for the task of universal four-qubit control the non-linearities in the NN greatly increase the expressiveness and are required for successful training. Finally, in Fig.~\ref{fig:app:cs1b}(c) we plot the learning curves of a QMPS architecture consisting only of an MPS. We found that training becomes unstable across all considered hyperparameters and random seeds of the optimization. The final success probabilites are $1.6\%$. Note that under certain simplifications of the control problem (e.g., smaller fidelity threshold, restriction to real initial wave functions), we were able to successfully train an MPS-only architecture. However, the average episode lengths were in general larger and hence less optimal than the ones we obtain from the hybrid MPS+NN ansatz.

The small Hilbert space dimension of the $N=4$ qubit control problem has the advantage that we can compare the QMPS framework to an RL training scheme in which we optimize a conventional feed-forward NN on the full wave function data. To that end, we simulate the quantum state exactly, i.e., without making use of the MPS formalism, and feed the full $2^4$ component wave function vector into a NN with two hidden layers and hidden dimensions of 100 each. Since the wave function components are complex-valued numbers, we double the size of the NN input vector ($2\times 2^4 = 32$) to account for the real and imaginary parts of the wave function. Figure~\ref{fig:app:cs1b}(d) shows the resulting learning curves of the best performing Q-network which we choose after a coarse grid-like hyperparameter search and out of three different random seeds of the optimization. The final success rate is evaluated to $94.3\%$ and thus substantially lower than the success probability of $99.8\%$ we obtained from the hybrid QMPS architecture that involves a NN of the same size. Doubling the NN hidden dimensions of the Q-network to 200 gave rise to a success rate of $96.8\%$ instead. Hence, even for small system sizes, the QMPS ansatz can lead to a performance enhancement compared to the more traditional NN-only architecture. Note that the QMPS ansatz for the $N=4$ qubit problem represents only a small computational overhead over the NN-only ansatz; in fact the number of optimizable MPS parameters for this problem computes to 592 while the number of parameters contained in the NN are given by 14612 (hidden dimension of 100) and 49212 (hidden dimension of 200). This suggests that the MPS ansatz indeed represents an efficient and natural architecture when learning from quantum states and can be beneficial even in the limit of small system sizes where an MPS description is usually not necessary.

\begin{figure}[t!]
	\centering
	\includegraphics[width=1.0\columnwidth]{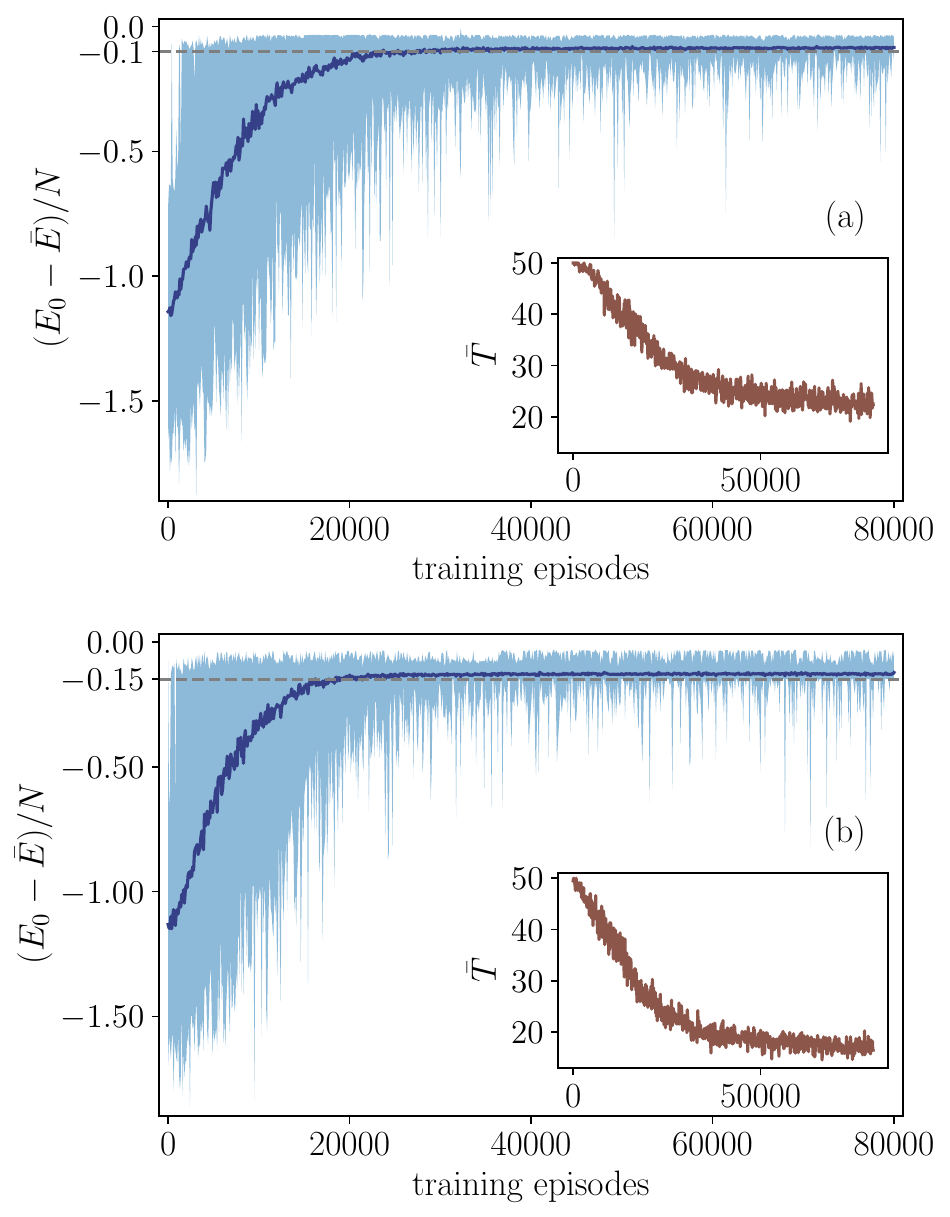}
	\caption{\label{fig:app:cs1c}{\bf Universal four-qubit control ---}~Learning curves of QMPO agents where the energy density is chosen as the reward function, i.e., $r_t\!=\!(E_0\!-\!E_t)/N$ with $E_0$ being the true ground state energy. The energy thresholds (gray-dashed lines) are set to {\bf (a)} $(E_0\!-\!E^*)/N\!=\!-0.1$ and {\bf (b)} $(E_0\!-\!E^*)/N\!=\!-0.15$ respectively. The achieved final energy density $(E_0\!-\!\bar{E})/N$ is shown averaged over a window of 100 training episodes. The light-blue data indicates the range between the best and the worst reward values. Inset:~The number of episode steps $\bar{T}$ averaged over 100 episodes. The maximum number of steps per episode was set to 50.}
\end{figure}

Finally, we provide additional results obtained when training a QMPO agent with a modified reward function:~Instead of using the log-fidelity between the target $\ket{\psi_*}$ and the current state $\ket{\psi_t}$, we compute the energy, i.e., the expectation value of the Ising Hamiltonian $H$ with respect to the current state $E_t = \langle \psi_t |H|\psi_t\rangle$. The reward is then defined as $r_t\!=\!(E_0\!-\!E_t)/N$ where $E_0$ is the true ground state energy of $H$. Note that this reward description might be advantageous in situations where the target state is unknown or the training is performed directly on a quantum device. Figure~\ref{fig:app:cs1c} shows the learning curves obtained for two different energy thresholds $(E_0\!-\!E^*)/N$ of $-0.1$ (a) and $-0.15$ (b) respectively. To benchmark the final performance of the QMPO agents, we again sample $M\!=\!10^3$ random initial states and find that for $98.9\%$ (a) and $99.7\%$ (b) of those instances the energy threshold is surpassed when acting with the optimized QMPO protocols. To compare these results with the ones attained for the log-fidelity reward, we also compute the average final fidelity $\bar{F}\!=\!M^{-1}\sum_i^M|\langle\psi_*|\psi^i_T\rangle|^2$ at the end of each episode which yields $\bar{F}=0.88$ (a) and $\bar{F}=0.79$ (b). For comparison the fidelity threshold used in Fig.~\ref{fig:app:cs1} was set to $F^\ast\!\sim\!0.85$.

\subsection{\label{app:case2}Preparation of a polarized product state from paramagnetic ground states for $N=32$ spins}

This section provides further details on the case study presented in Sec.~\ref{sec:case2}.

To speed training up, we restrict the action space to $7$ actions for this control setup with $\mathcal{A} = \{\hat{Y},\hat{Z},-\hat{Z},\hat{X}\hat{X},-\hat{X}\hat{X},\hat{Y}\hat{Y},-\hat{Y}\hat{Y}\}$. This action set was determined by first training on a smaller system size ($N=8$) using all actions and then selecting only those that appear in the final optimal protocols. 

\begin{figure}[t!]
	\centering
	\includegraphics[width=0.95\columnwidth]{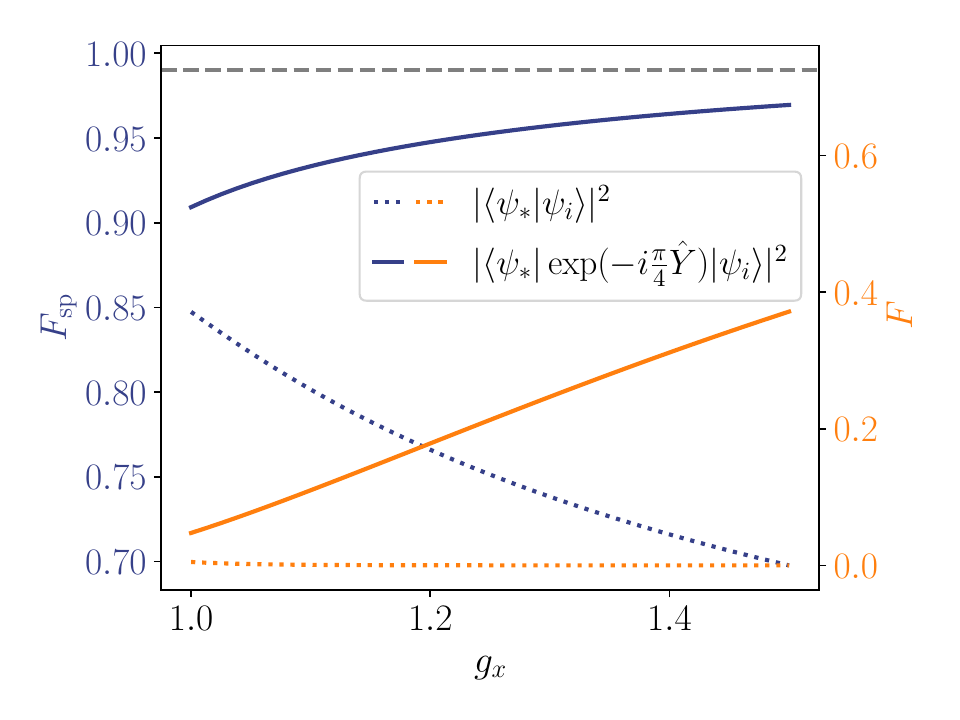}
	\caption{\label{fig:app:fid} {\bf Transverse-field Ising control ---}~Single-particle fidelity (blue, left $y$-axis) and many-body fidelity (orange, right $y$-axis) between the target $z$-polarized state and an initial Ising ground state for different values of the transverse field $g_x$ (dotted curves). 
		The solid lines show the respective fidelities of the target state and the initial ground state after applying a single-particle $\hat Y$-rotation with $\delta t_{-}\!=\!\pi/4$. 
		The QMPS agent is able to improve on the trivial rotation and prepare the target state with single-particle fidelities $F_{\mathrm{sp}}\!>\!0.99$ (many-body fidelities $F_{\mathrm{sp}}\!>\!0.72$) [gray dashed line]. $N\!=\! 32$ spins.
	}
\end{figure}

To get an intuition about the difficulty of this control setup, we proceed as follows: 
(i) we demonstrate that the initial states are sufficiently far (in the Hilbert space distance) from the target state, e.g., by computing the fidelity as a function of $g_x$, cf.~Fig.~\ref{fig:app:fid} [dashed lines]. This corresponds to a protocol where the agent does not take any action.
(ii) an alternative protocol can be produced by noticing that the initial states are paramagnetic, while the target is a $z$-polarized state, and thus a $\pi/2$-rotation about the $y$-axis presents a good candidate for the optimal protocol [it is indeed optimal in the limit $g_x\to\infty$]. The corresponding fidelities are shown in Fig.~\ref{fig:app:fid} [solid lines]. 
Compared to these fidelities, the threshold for the RL agent is given by the horizontal dashed line; it gives a lower bound on the performance of the QMPS protocols. Notice that, the QMPS agents are able to considerably improve on the initial fidelity and outperform the trivial $\hat Y$-rotations for the considered range of transverse field values. 

\begin{figure}[t!]
	\centering
	\includegraphics[width=0.95\columnwidth]{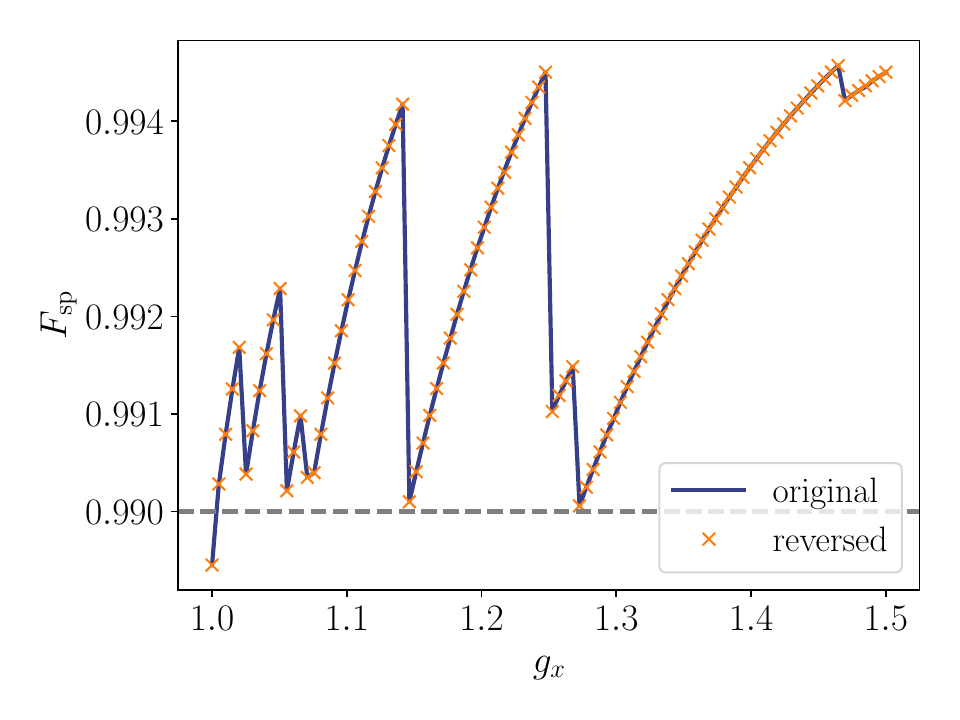}
	\caption{\label{fig:app:reversed} {\bf Transverse-field Ising control ---}~Final single-particle fidelities when starting from Ising ground states with transverse field values $g_x$ and preparing the $z$-polarized target state (blue curve). The orange points show the fidelity when reversing the state preparation scenario, i.e.~one starts from the polarized state and applies the inverse QMPS protocol to reach the corresponding Ising ground state. The fidelities achieved by the reversed protocol are comparably high. Therefore they justify that, in this case study, the trained QMPS agent can be employed for the inverse state preparation task as discussed in the main text. $N\!=\! 32$ spins.
	}
\end{figure}

The QMPS agent was trained to prepare a specific target state (the $z$-polarized state) starting from a class of Ising ground states. In principle, the obtained protocols can be inverted to achieve the opposite, i.e.~prepare \textit{any} paramagnetic Ising ground state from the $z$-polarized state. This is often the objective in quantum computing or simulation tasks where the system starts out in a simple product state and is then brought into the state that has to be investigated or that encodes the solution to a problem. With a single trained QMPS agent one can generate optimal controls that, when reversed, prepare a variety of different states that can then be used for computation. Note however, that the final state reached by using the original QMPS protocol does not exactly coincide with the target state since we do not achieve a perfect fidelity of unity. It is therefore not clear whether the inverse protocol, when starting from the exact target state, prepares the original initial states with an equally high fidelity or whether it does considerably worse. In Fig.~\ref{fig:app:reversed} we provide the achieved single-particle fidelities when preparing Ising ground states from the $z$-polarized state by reversing the optimal QMPS protocols and compare them to the fidelities of the original state preparation routine [cf.~Fig.~\ref{fig:case2a} of Sec.~\ref{sec:case2}]. We find that the fidelities do not differ significantly which justifies that inverse state preparation using the QMPS protocols is possible in this particular control scenario.

\begin{figure*}[t!]
	\centering
	\includegraphics[width=1.0\textwidth]{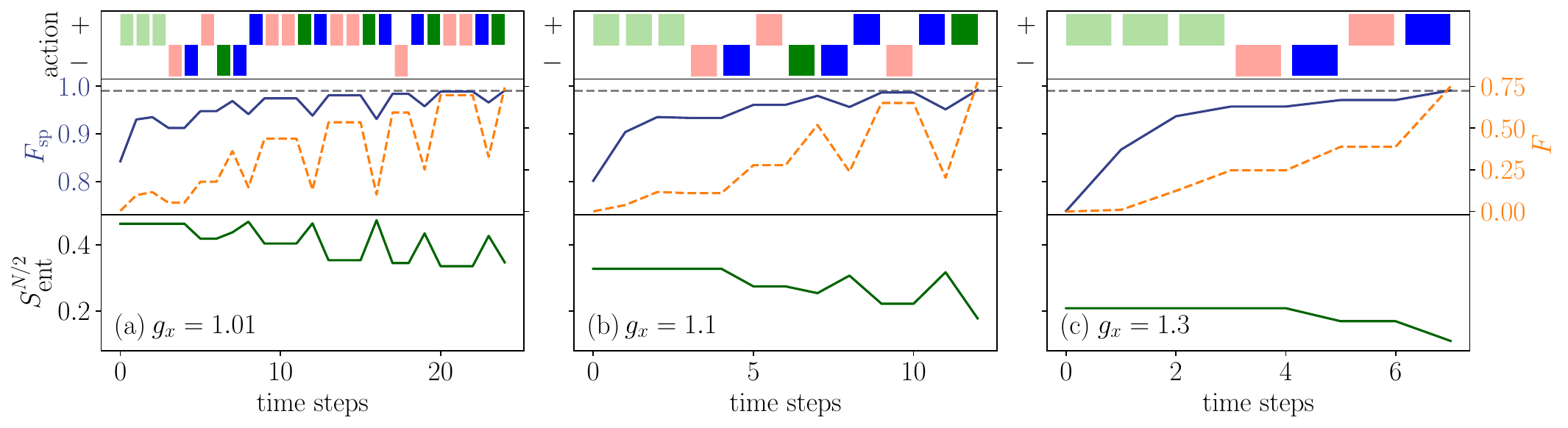}
	\caption{\label{fig:app:protocols}{\bf Transverse-field Ising control ---}~Upper panels:~Final protocols of the QMPS agent when starting from a ground state with {\bf (a)} $g_x=1.01$, {\bf (b)} $g_x=1.1$, and {\bf (c)} $g_x=1.3$. Middle panel:~Single-particle fidelities (blue, left $y$-axis) and many-body fidelities (blue, right $y$-axis) between the evolved states and the $z$-polarized target state at each protocol step. The fidelity threshold $F^\ast_{\mathrm{sp}}\!=\!0.99$ ($F^\ast\!\sim\!0.72$) is indicated by a gray dashed line. Lower panel:~The corresponding half-chain von Neumann entanglement entropy calculated at each step of the protocol. The applied unitaries do not create an excessive amount of entanglement allowing the time evolved states to be described with a relatively low bond dimension. $N\!=\! 32$ spins.}
\end{figure*}

\begin{figure}[t!]
	\centering
	\includegraphics[width=0.95\columnwidth]{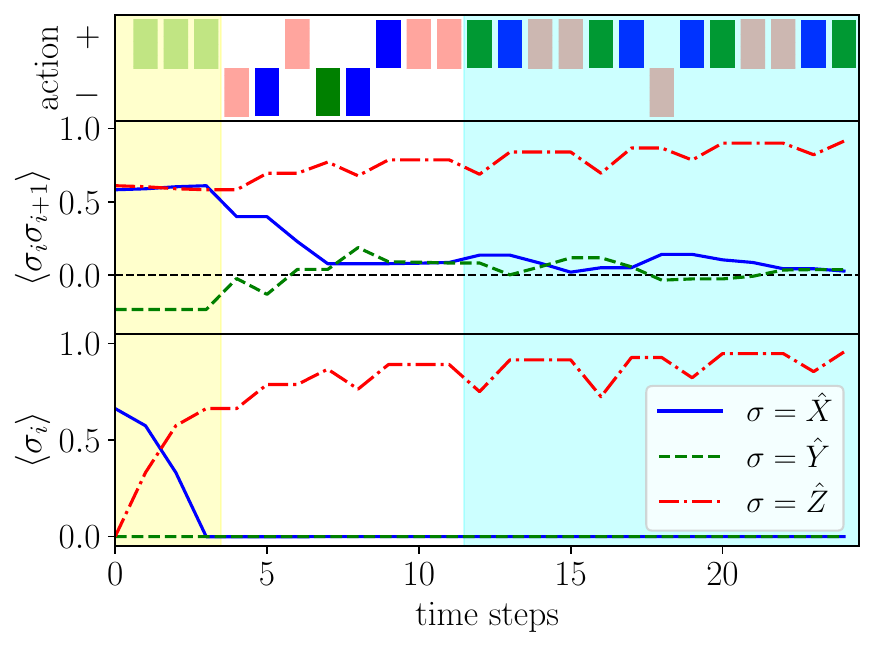}
	\caption{\label{fig:app:correlations} {\bf Transverse-field Ising control ---}~Analysis of the optimal QMPS protocol obtained when starting from the initial ground state at $g_x=1.01$. Shown are the local spin-spin correlations $\langle \sigma_i \sigma_{i+1}\rangle$ and the local magnetization $\langle \sigma_i\rangle$ along each direction at the center of the spin chain ($i=15$). The yellow shaded segment indicates the initial $\hat{Y}$-rotations which align the spin along the $z$-axis; the blue shaded area points to a generalized Euler-angle-like many-body rotation which reduces unwanted correlations and disentangles the state. $N\!=\! 32$ spins. See Video~\protect\hyperlink{video:1}{1}.
	}
\end{figure}

\begin{figure}[t!]
	\centering
	\includegraphics[width=0.95\columnwidth]{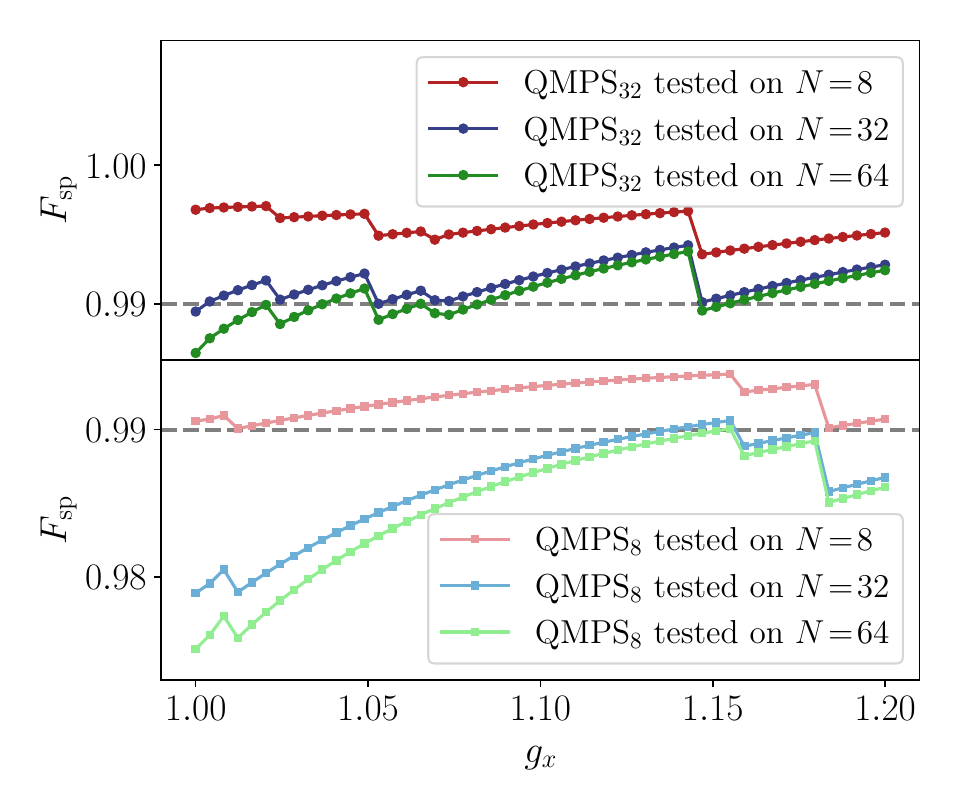}
	\caption{\label{fig:app:systemsize}{\bf Transverse-field Ising control ---}~Upper panel:~Final single-particle fidelities achieved when testing the protocols of a QMPS agent trained on $N\!=\!32$ spins (blue curve) on a smaller system size of $N\!=\!8$ (red curve) and on the larger $N\!=\!64$ spin system (green curve). The increase in the single-particle fidelity for $N\!=\!8$ suggests that optimized protocols can be transferred to smaller system sizes for this particular control setup. Lower panel: The opposite scenario, where the protocols of a QMPS agent trained on $N\!=\!8$ spins (light red) are tested on larger systems of $N\!=\!32$ spins (light blue) and $N\!=\!64$ spins (light green). The fidelity threshold (gray dashed line) cannot always be maintained for the larger system size especially close to the critical point at $g_x\!\sim\!1$.
	}
\end{figure}

In Fig.~\ref{fig:app:protocols} we display three optimal QMPS protocols starting from initial ground states at $g_x\!=\!1.01,1.1,1.3$, respectively. Interestingly, in all three cases, the agent learns to initially apply a $\pi/2$-rotation about the $y$-axis which is decomposed into three consecutive protocol steps since we fix the duration of each applied unitary to be $\delta t_+ \!=\!\pi/12$. As discussed in the main text, the agent is able to successfully prepare the target state also for initial ground states outside of the training interval, i.e., for $g_x\!>\!1.1$. Note however, that the predicted Q-values of the QMPS agent can be quite different from the true return when tested outside of the training interval, yet the policy learned by the agent can still produce meaningful optimal protocols. In the bottom panels of Fig.~\ref{fig:app:protocols} we show the half-chain von Neumann entanglement entropy $S_{\mathrm{ent}}$ of the encountered states when evolving according to the optimal protocols. The entanglement entropy stays small and hence, allows the time evolved system to be simulated with a relatively small bond dimension $\chi_\psi$ (for training we set $\chi_\psi\!=\!16$). Note however, that the entanglement entropy is not fully reduced to zero at the end of the protocol which is especially the case for the states close to the critical point. This can be attributed to the logarithmic scaling of the entanglement entropy in the initial critical state, with the subsystem size $N/2$ due to the presence of long-range correlations. While the agent is able to reduce local correlations effectively [Fig.~\ref{fig:app:correlations}, middle panel], the short protocol is not capable of destroying all long-range correlations which persist in the final state. The prepared state, therefore, has a finite many-body overlap with the separable target state [see orange curves/axis in Fig.~\ref{fig:app:protocols} showing the many-body fidelity], which explains the discrepancy in the final entanglement entropies.

\begin{figure}[t!]
	\centering
	\includegraphics[width=0.95\columnwidth]{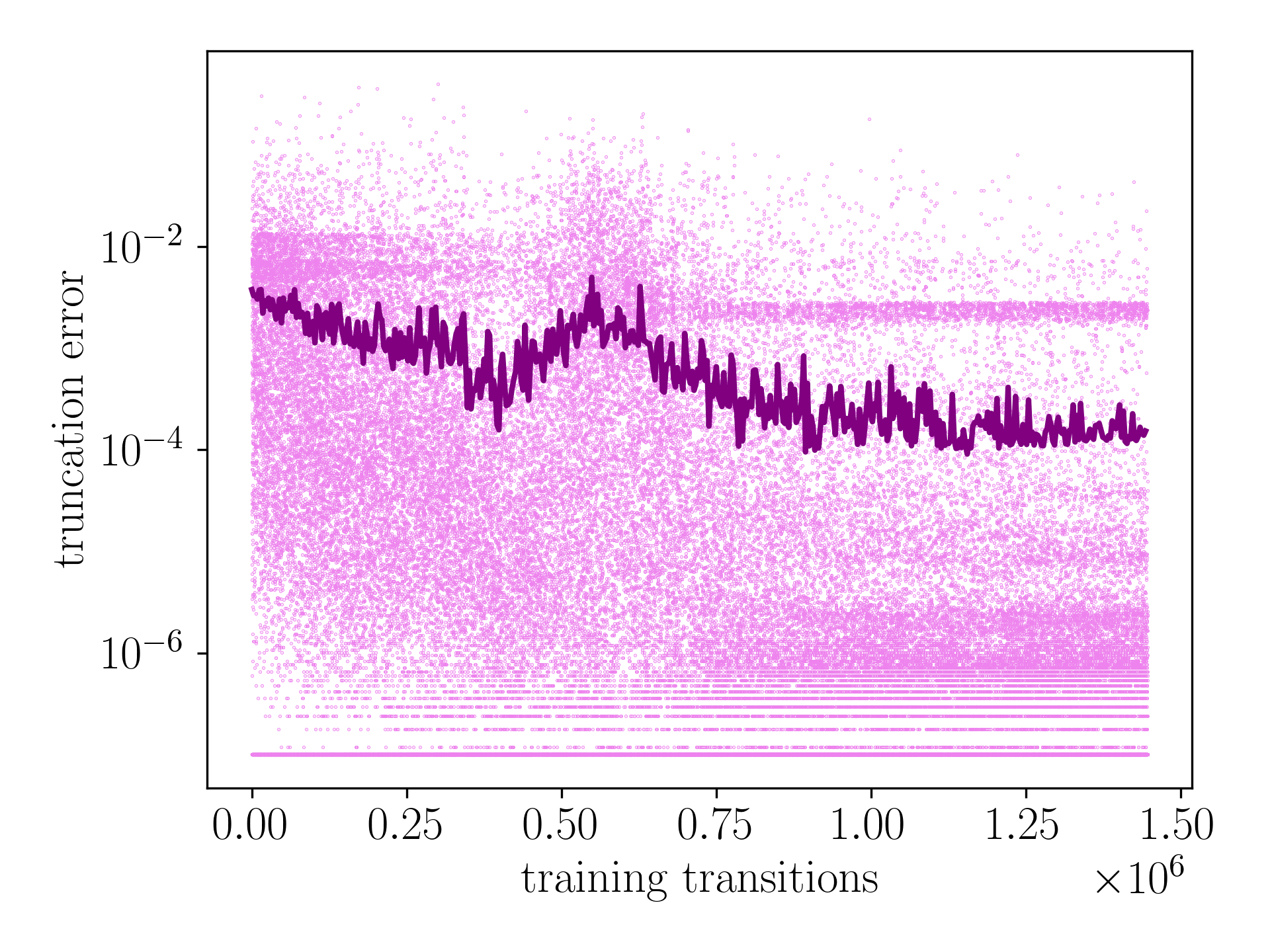}
	\caption{\label{fig:app:trunc}{\bf Transverse-field Ising control ---}~Truncation error, cf.~Eq.~\eqref{eq:trunc_error}, for the evolved states encountered during training of the QMPS agent. The dark purple curve shows the truncation error averaged over a window of $3000$ transitions. Training was performed with a quantum state bond dimension of $\chi_\psi\!=\!16$; $N\!=\!32$ spins.
	}
\end{figure}

In Fig.~\ref{fig:app:correlations} we plot the local expectation value of the magnetization along each direction and the local spin-spin correlations at the center of the chain which reveal the role of each unitary occurring in the protocol sequence corresponding to the state preparation task shown in Fig.~\ref{fig:app:protocols}(a) [$g_x=1.01$]. As already mentioned, the first three $\hat Y$-rotations align the state along the $z$-axis bringing the expectation values of the $\hat X$ and $\hat Y$ component to zero. The role of the remaining unitaries is to decrease unwanted correlations and consequently to disentangle the state. Note that the $\langle\hat X \hat{X}\rangle$ and $\langle\hat Y\hat Y\rangle$ correlations approach zero in an intertwined manner which is caused by an intricate combination of $\hat X \hat{X}$ and $\hat Y\hat Y$ disentangling gates and single-particle $\hat Z$-rotations. The latter are important for finely realigning the state after each disentangling operation and as such preventing the correlations from diverging [see Video~\protect\hyperlink{video:1}{1}].

Next, we analyze how well the optimal protocols perform on systems with a different number of spins. To this end, we test the protocols optimized for $N\!=\!32$ spins (QMPS$_{32}$) on $N\!=\!8,64$ spin systems, and vice-versa: protocols obtained after training on $N\!=\!8$ spins (QMPS$_{8}$) are tested on the larger $N\!=\!32,64$ systems [Fig.~\ref{fig:app:systemsize}]. We find that the fidelity threshold can still be reached when applying the QMPS$_{32}$ protocols on smaller system sizes. However, the opposite is not true: the QMPS$_{8}$ protocols, in general, give rise to fidelities below the threshold when tested on the $N\!=\!32,64$ spin systems. These system-size (in)dependence suggests that, for this particular control setup, one can devise suitable pretraining techniques for large system sizes, based on the behavior of agents successfully trained on smaller systems. Moreover, the QMPS agent tends to find optimal protocols which appear robust to changes in the system size, and the control task likely admits a solution also in the thermodynamic limit.

\begin{figure}[t!]
	\centering
	\includegraphics[width=1.0\columnwidth]{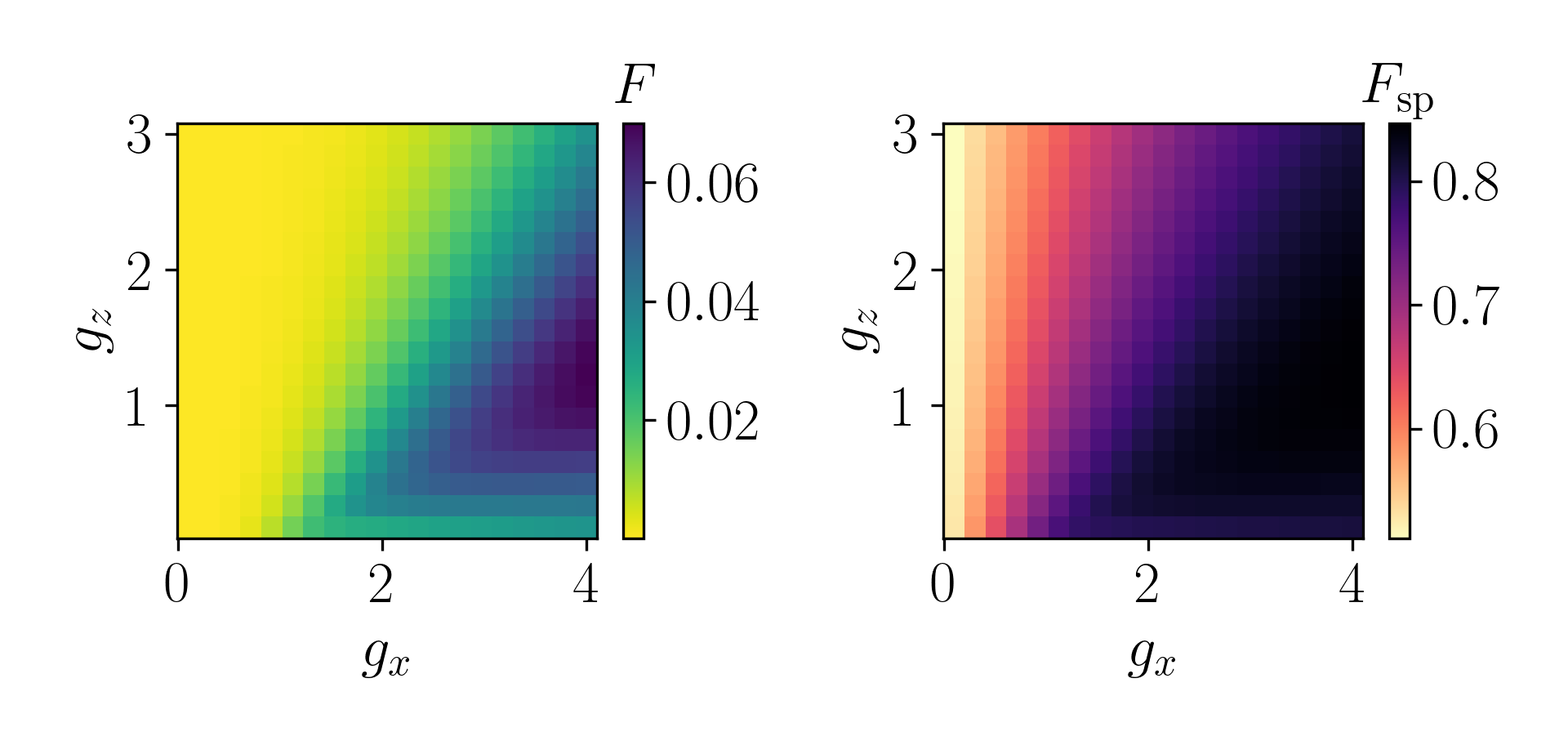}
	\caption{\label{fig:app:case3a}{\bf Self-correcting mixed-field Ising control ---}~Many-body fidelity $F$ (left) and single-particle fidelity $F_{\mathrm{sp}}$ (right) between the critical-region target state and the initial ground states at transverse and longitudinal field values $g_x,g_z$. The trained QMPS agent prepares the target state with many-body and single-particle fidelities $F\!>\!0.61, F_{\mathrm{sp}}\!>\!0.97$ respectively, and therefore considerably improves on the initial fidelity values. $N\!=\!16$ spins.
	}
\end{figure}

\begin{figure}[t!]
	\centering
	\includegraphics[width=1.0\columnwidth]{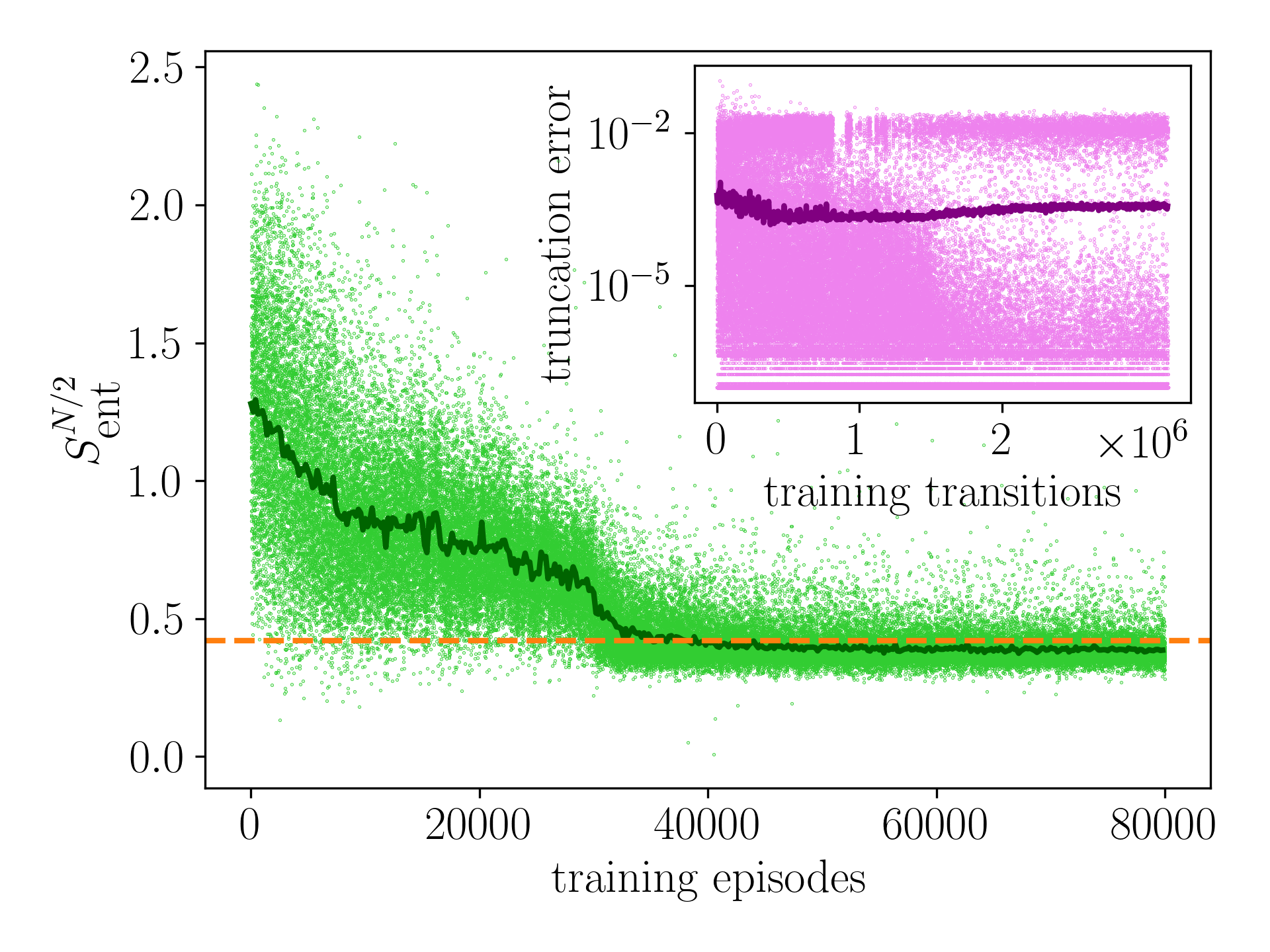}
	\caption{\label{fig:app:case3b}{\bf Self-correcting mixed-field Ising control ---}~Half-chain von Neumann entanglement entropy of final states during training. The dark green curve denotes the average over 200 episodes and the orange dashed line indicates the entanglement entropy of the critical-region target state. The entropies decrease as learning progresses and converge to a value close to that of the target state. Inset:~Truncation error averaged over a window of $3000$ transitions (dark purple). Training was performed with a quantum state bond dimension of $\chi_\psi\!=\!16$; for testing $\chi_\psi\!=\!32$ was used. $N\!=\!16$ spins.
	}
\end{figure}

\begin{figure*}[t!]
	\centering
	\includegraphics[width=1.0\textwidth]{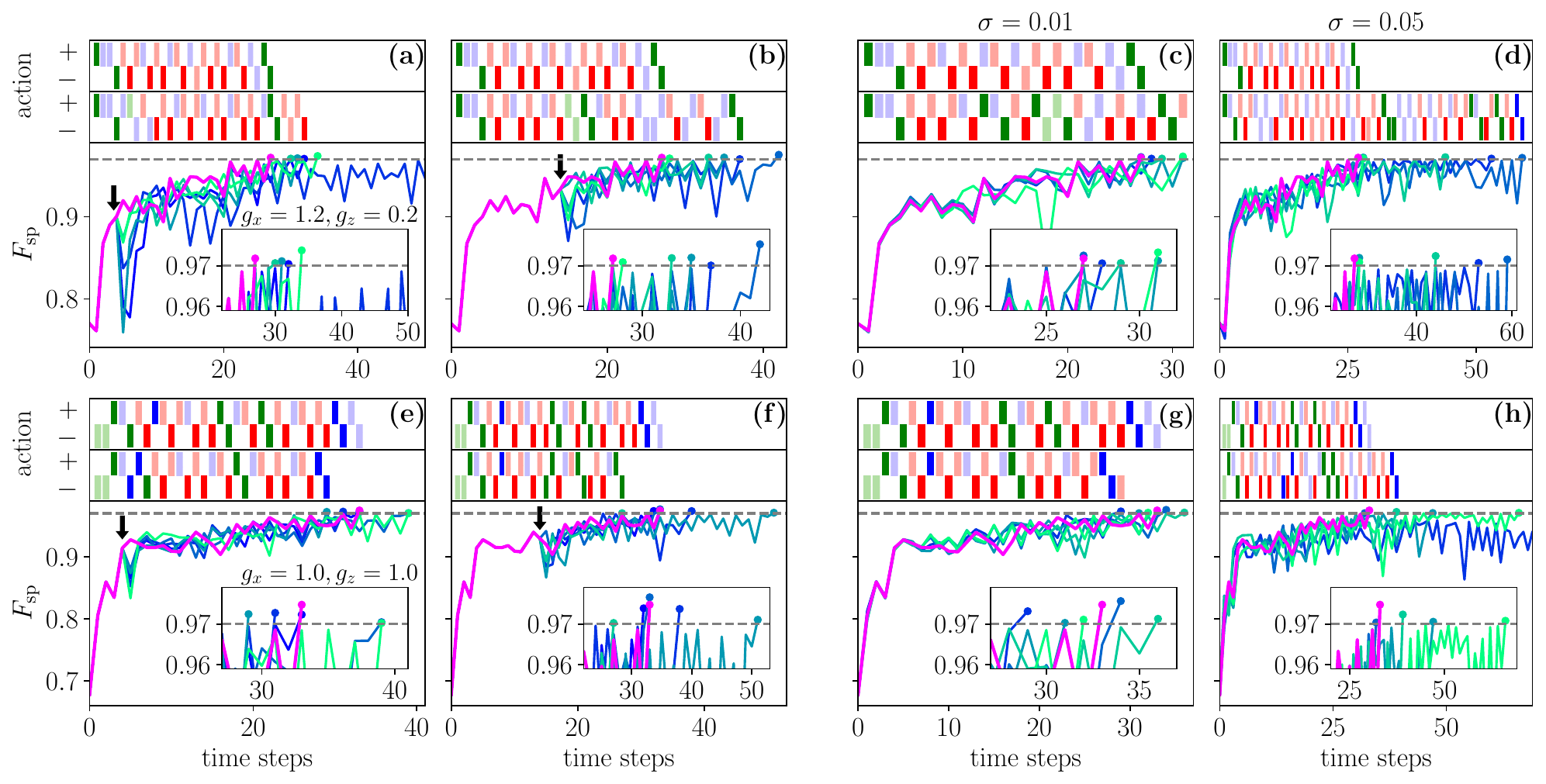}
	\caption{\label{fig:app:case3c}{\bf Self-correcting mixed-field Ising control ---}~Exemplary QMPS protocols and time-dependence of the single-particle fidelity when starting from an initial ground state within the training region at $J\!=\!-1,g_x\!=\!1.2,g_z\!=\!0.2$ {\bf (a)-(d)}, and outside the training region at $J\!=\!-1,g_x\!=\!1.0,g_z\!=\!1.0$ {\bf (e)-(h)}. For each subplot the upper panel displays the optimal QMPS protocol without perturbations, the middle panel presents an exemplary protocol subject to noise or perturbations, and the bottom panel shows the single-particle fidelities at each time step for different protocols [the original, unperturbed QMPS protocol is always indicated by the magenta line]. The single-particle fidelity threshold of $F^\ast_{\mathrm{sp}}\!=\!0.97$ ($F^\ast\!\sim\!0.61$) is denoted by a gray dashed line. In {\bf (a)-(b), (e)-(f)} the QMPS protocol is modified at time step $t=5$ ($t=15$) [indicated by a black arrow] by taking 5 different random actions. Afterwards, the system is again evolved according to the QMPS agent leading to 5 distinct trajectories [blue lines]. In all but one case the QMPS agent is able to correct for the mistake and successfully reaches the fidelity threshold. In {\bf (c)-(d), (g)-(h)} white Gaussian noise with standard deviation $\sigma\!=\!0.01$ {\bf (c),(g)}, or $\sigma\!=\!0.05$ {\bf (d),(h)} is added to the time step duration $\delta t_\pm$. The system is evolved with 5 different random seeds [blue lines]. The QMPS agent is again able to adapt its protocol and successfully reaches the fidelity threshold in most cases.
		$N\!=\!16$ spins. 
		See Videos~\protect\hyperlink{video:2}{2} and~\protect\hyperlink{video:3}{3}.
	}
\end{figure*}

Finally let us comment on the quantum state entanglement and the related MPS bond dimension. During the initial exploratory stage of training, random unitaries are applied to the system leading in general to a ballistic growth of the entanglement entropy. In this case the fixed bond dimension of $\chi_\psi\!=\!16$ is not always sufficient to capture the evolved quantum states giving rise to the large truncation errors, shown in Fig.~\ref{fig:app:trunc}. These truncation errors, however, naturally decrease as training progresses and the action selection becomes more deterministic while the Q-function converges close to the optimal one. While for training a bond dimension of $\chi_\psi\!=\!16$ was used, testing was performed with $\chi_\psi\!=\!32$ for which the truncation errors vanished to machine precision. This check ensures the stability of the QMPS protocols to changes in the accuracy of the MPS approximation.

\subsection{\label{app:case3}Learning robust critical-region state preparation for $N=16$ spins}

This section provides further details on the case study presented in Sec.~\ref{sec:case3}.

To compare the achieved fidelities of the QMPS agent reported in Fig.~\ref{fig:case3a}(a), we show the fidelities between the initial mixed-field Ising ground states and the critical-region target state before any controls are applied in Fig.~\ref{fig:app:case3a}. The QMPS agent is able to reach the target with single-particles fidelities $F_{\mathrm{sp}}\!>\!0.97$ (corresponding to a many-body fidelity of $F\!\sim\!0.61$).

The half-chain von Neumann entanglement entropy of the quantum states during training is displayed in Fig.~\ref{fig:app:case3b}. Similar to the previous case study, during the initial stages of training the encountered states are highly entangled due to the randomness of the action selection. Once the agent learns how to reliably prepare the target state, the entropies decrease and are scattered closely around the target state value [orange dashed line]. We emphasize that critical states possess a logarithmic correction to the area-law of entanglement, which makes their preparation a non-trivial task. For training we used a relatively small bond dimension of $\chi_\psi\!=\!16$ which led to the finite truncation errors shown in the inset of Fig.~\ref{fig:app:case3b}. However, training is still successful and all subsequent tests were performed by setting $\chi_\psi\!=\!32$ which did not affect the QMPS protocols or the final achieved fidelities.

\begin{figure*}[t!]
	\centering
	\includegraphics[width=1.0\textwidth]{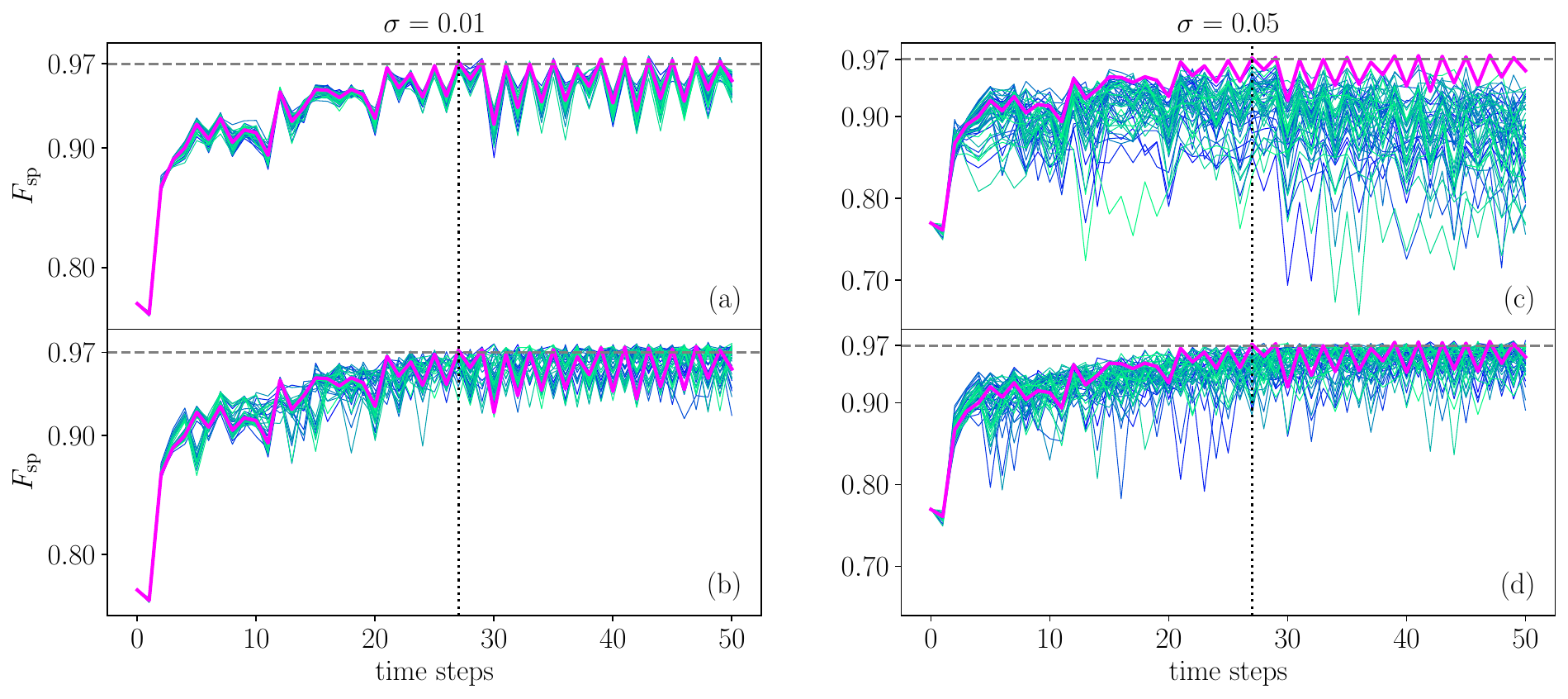}
	\caption{\label{fig:app:case3d}{\bf Self-correcting mixed-field Ising control ---}~Time-dependence of the single-particle fidelity when adding Gaussian random noise with standard deviations $\sigma=0.01$ {\bf (a)-(b)}, and $\sigma=0.05$ {\bf (c)-(d)} to the time step duration $\delta t_{\pm}$ and starting from an initial ground state at $J\!=\!-1,g_x\!=\!1.2,g_z\!=\!0.2$. For each figure, we sampled 100 different trajectories (each corresponding to a different random seed). The original, unperturbed QMPS protocol is always indicated by the magenta line. The single-particle fidelity threshold of $F^\ast_{\mathrm{sp}}\!=\!0.97$ ($F^\ast\!\sim\!0.61$) is denoted by a horizontal gray dashed line and the number of steps required to reach the threshold in the noise-free case is indicated by a vertical black dotted line. In {\bf (a),(c)} we always evolve according to the fixed, unperturbed QMPS protocol we obtained from the noise-free simulation. The percentage of successfully prepared target states within 50 steps is $90\%$ in the case of weak noise ($\sigma=0.01$) and $0\%$ in the case of strong noise ($\sigma=0.05$). In {\bf (b),(d)} we use the adaptive QMPS agent to generate different protocols for each distinct run (compare to Fig.~\ref{fig:app:case3c}(a)-(d)). The respective success percentages are $100\%$ ($\sigma=0.01$) and $74\%$ ($\sigma=0.05$). Hence, in both instances, the self-correcting agent is able to improve over the fixed, noise-free protocol.
	}
\end{figure*}

Next, we provide further examples showcasing the ability of the QMPS agent to self-correct its protocols on-the-fly when the time evolution is noisy or perturbed. In Fig.~\ref{fig:app:case3c}, we consider two different initial ground states, one within the training region ($J\!=\!-1,g_x\!=\!1.2,g_z\!=\!0.2$) [(a)-(d)] and one outside ($J\!=\!-1,g_x\!=\!1.0,g_z\!=\!1.0$) [(e)-(f)], and analyze the success of state preparation subject to different protocols. In the upper panels of each subplot in Fig.~\ref{fig:app:case3c} we display the actions of the optimal protocol [top] and one exemplary protocol that has been perturbed [bottom]. The lower panel always shows the single-particle fidelities at each step of the protocols.

First, we take the optimal QMPS protocols and modify it at time step $t\!=\!5$ ($t\!=\!15$) by taking 5 random suboptimal actions instead. Afterwards the system is again evolved according to the greedily acting QMPS agent giving rise to 6 distinct trajectories (the magenta line corresponds to the unperturbed one). In most cases the agent is able to correct for the mistake by adapting the subsequent protocol and reaches the fidelity threshold nonetheless. However, in one instance [Fig.~\ref{fig:app:case3c}(a)], the resulting QMPS protocol does not converge and the agent fails to prepare the target state. Hence, the agent is not able to generalize to states generated by this particular protocol sequence, and likely predicts wrong Q-values that steer the agent eventually away from the target state. This is, however, not surprising since the agent has only been trained on states within a small part of the many-body Hilbert space and therefore, it cannot be expected to devise successful protocols from arbitrary states.

Note that for the initial state outside of the training interval [(e),(f)], the perturbation of the original QMPS protocol gave rise to a shorter protocol, i.e.~the fidelity threshold is reached in a fewer number of steps. Hence, in this case the original protocol is not a local minimum of the control landscape. However, this is not surprising, since the QMPS agent has not been trained on this initial state and therefore, the predicted Q-values are not guaranteed to have converged to the true optimal values.

Finally, we study the robustness of the QMPS agent to a randomized time step duration $\delta_\pm$ by adding white Gaussian noise with standard deviation $\sigma\!=\!0.01,0.05$ to it [Fig.~\ref{fig:app:case3c}(c),(d),(g),(h)]. We evolve the system with 5 different random seeds giving again rise to 6 distinct trajectories (the magenta line corresponds to the unperturbed one). For each of the 5 randomized time evolutions, the QMPS agent has to eventually adapt its protocol by performing a different sequence of actions. It successfully prepares the target state in all but one case which falls outside the training region [Fig.~\ref{fig:app:case3c}(h)]. Note here as well that in some instances the agent is able to devise shorter control protocols compared to the original unperturbed ones. Hence, the randomized step duration can have a positive effect on the control problem allowing for faster state preparation protocols.

In Fig.~\ref{fig:app:case3d} we compare the achieved fidelities in the presence of noise when we evolve with the adapted protocols (bottom) and with the original, noise-free protocol (top) starting from an initial state $J\!=\!-1,g_x\!=\!1.2,g_z\!=\!0.2$ within the training region. We again consider Gaussian random noise with standard deviations $\sigma\!=\!0.01,0.05$ and repeat the time evolution with 100 different random seeds for each of the two cases. When the noise is weak ($\sigma=0.01$), the fixed, unperturbed protocol gives rise to qualitatively similar fidelity curves regardless of the seed (see Fig.~\ref{fig:app:case3d}(a)). We found that for 90 out of the 100 runs, the protocol successfully prepares the target state, that is, the fidelity threshold is reached within 50 steps. However, in the cases where the fidelity threshold is not surpassed, the final fidelities are close to the target value of $F^\ast_{\mathrm{sp}}\!=\!0.97$. If we instead allow the QMPS agent to adapt to the perturbed states, we obtain the fidelity curves in Fig.~\ref{fig:app:case3d}(b). The resulting protocols give rise to qualitatively different trajectories that diverge more towards later time steps (compare to Fig.~\ref{fig:app:case3c}(a)-(d)). However, in this case the agent successfully prepares the target state for each of the 100 runs. Hence, for sufficiently weak noise strengths, the original unperturbed protocol is expected to give qualitatively similar results to the noise-free dynamics. However, even in this example the self-correcting agent has a measurable advantage over the fixed protocol.

This situation changes when we consider the case of strong noise ($\sigma=0.05$) as shown in Fig.~\ref{fig:app:case3d}(c)-(d). The fixed, unperturbed protocol leads to diverging fidelity curves already after a few steps (top). In fact, the success probability for the simulated 100 runs is 0. In contrast, the adaptive agent prepares the target state successfully for 74 out of the 100 instances within the 50 allowed time steps. Moreover, the agent clearly tries to steer the quantum states towards high fidelity regions. This example therefore demonstrates that the self-correcting agent is able to improve over the original, noise-free protocol when the dynamics is being perturbed.

\section{\label{app:case4}NISQ Implementation of the QMPS Framework}

\subsection{QMPS to quantum circuit mapping\label{subsec:circuit_mapping}}

In the following we illustrate the QMPS to circuit mapping on the example of a $N\!=\!4$ spin/qubit system. The QMPS state $\ket{\theta_Q^\ell}$ is represented as
\begin{equation}\label{eq:qmps2}
    \ket{\theta_Q^\ell}=\sum_{j_1, \ldots, j_4} \sum_{\alpha_1, \alpha_2,  \alpha_3} A_{\alpha_1 }^{[1] j_1} A_{\alpha_1 \alpha_2}^{[2] j_2} A_{\alpha_2 \alpha_3}^{[3] j_3;\ell} A_{\alpha_3 }^{[4] j_4}\left|j_1, \ldots, j_4\right\rangle,
\end{equation}
where we have already contracted the feature tensor with its neighboring tensor $A^{[3]}$ and $\ell$ denotes the feature vector index. Our goal is to rewrite the QMPS state as a quantum circuit $\ket{\theta_Q^\ell} = U^{\ell}_{\theta} \ket{0}$, where the state preparation unitary $U^{\ell}_{\theta}$ is composed of several gates $U^{\ell}_{\theta}=G^{\ell}_4\cdots G^{\ell}_1$.

First, we transform the QMPS into the left canonical form via successive QR or singular value decompositions such that
\begin{align}
    &\sum_{j_1 \alpha_1} A_{\alpha_1 }^{[1] j_1} A_{\alpha_1 }^{[1] j_1 *}=1, \\
    &\sum_{j_i \alpha_i} A_{\alpha_{i-1} \alpha_i}^{[i] j_i} A_{ \alpha_{i-1}^{\prime} \alpha_i}^{[i] j_i *}=I_{\alpha_{i-1} \alpha_{i-1}^{\prime}}, \qquad (i=2,3) \\ 
    &\sum_{j_4} A_{\alpha_3}^{[4] j_4} A_{\alpha_{3}^{\prime}}^{[4] j_4 *}=I_{\alpha_3 \alpha_3^{\prime}} \label{eq:unitary}.
\end{align}
In principle, the resulting tensors $A^{[i]}$ also depend on the feature index $\ell$ after performing the canonicalization. However, in what follows we omit the index $\ell$ and assume that all subsequent steps are performed for each of the values of $\ell$ separately.

\begin{figure}[t!]
    \centering
    \includegraphics[width=0.75\columnwidth]{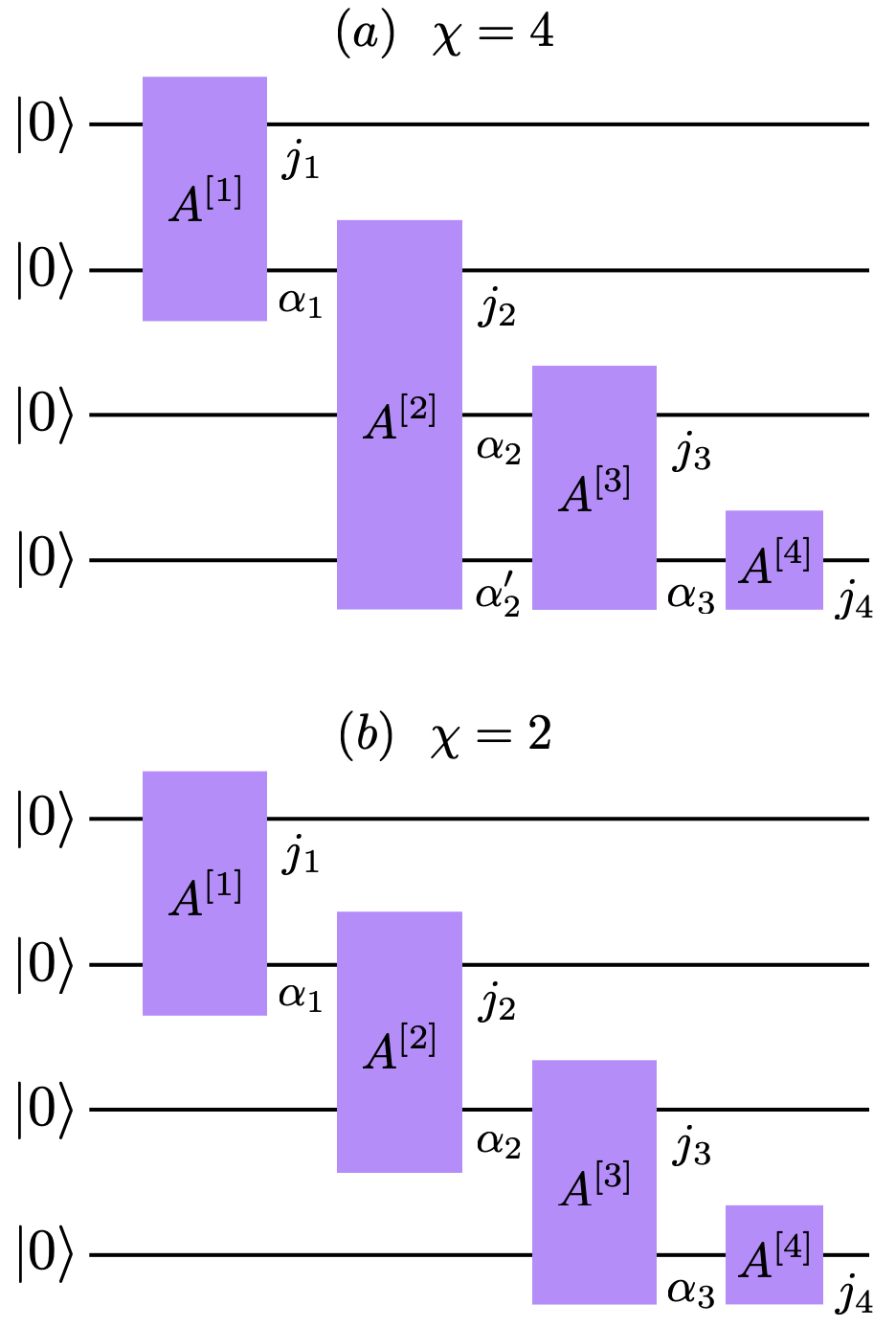}
    \caption[MPS to circuit mapping]{\label{fig:qmps_circ}MPS to circuit mapping for the $N=4$ MPS of Eq.~\eqref{eq:qmps2} in left orthogonal form. (a) An MPS with bond dimensions $2-4-2$. (b) The truncated MPS with bond dimensions $2-2-2$. 
    }
\end{figure}

\begin{figure*}[t!]
    \centering
    \includegraphics[width=1.0\textwidth]{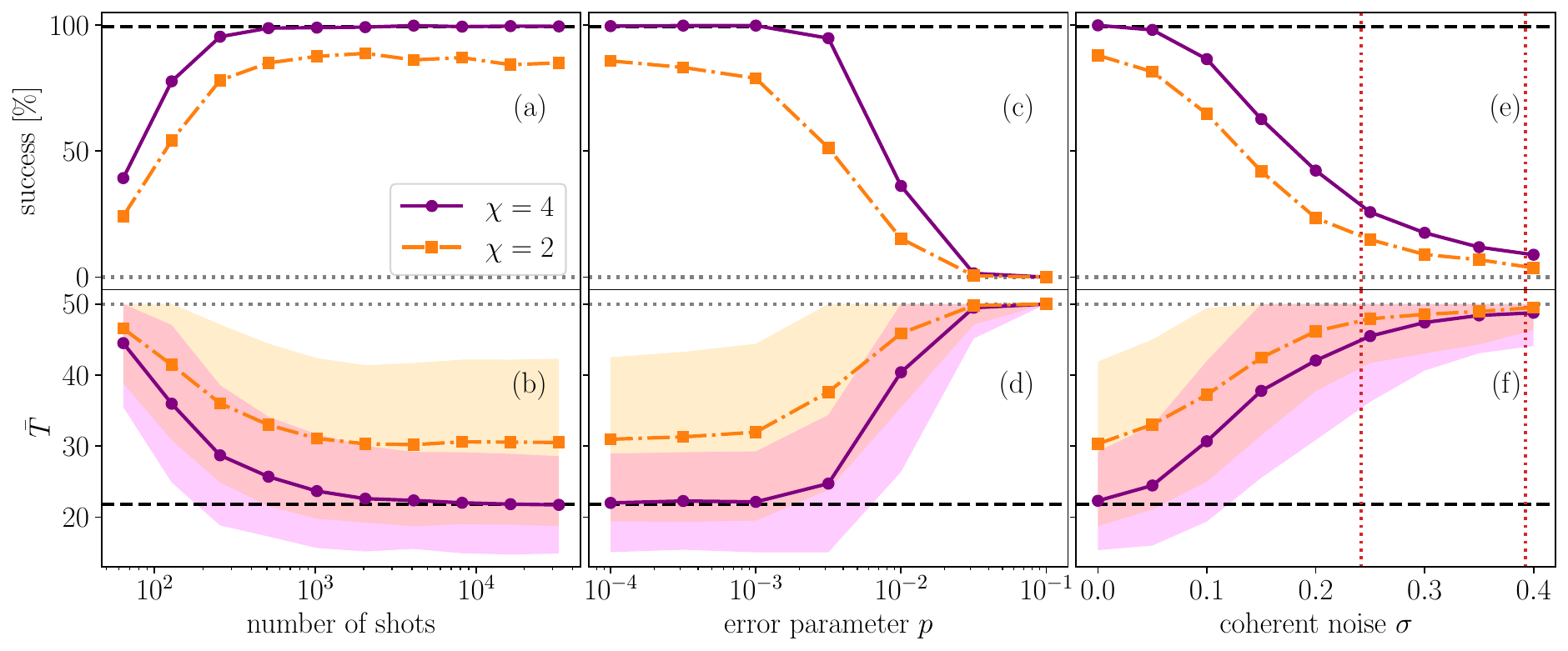}
    \caption{\label{fig:app:case4}{\bf Universal four-qubit control --- (a),(b)}~We sample 1000 random initial states and apply the QMPS circuit framework with a varying number of measurement shots. In (a) we display the percentage of runs in which the target state is successfully reached (i.e., the fidelity threshold of $F^\ast\!\sim\!0.85$ is surpassed after at most 50 protocol steps). The success rate under exact computation (without sampling) is shown as a black dashed line. The success probability when acting random is zero (gray dotted line). We provide both, the results for the full $\chi=4$ QMPS circuit (purple solid line) and the truncated $\chi=2$ QMPS (orange dash-dotted line). We find that as low as $\sim 500$ shots are sufficient for reaching success rates close to $100\%$. (b) The corresponding average number of required protocol steps $\bar{T}$ for reaching the fidelity threshold. The standard deviation is indicated by the shaded areas. The black dashed line corresponds again to the average value computed via exact techniques, the gray dashed-dotted line indicates the maximum number of allowed episode steps (50). {\bf (c),(d)}~The success rate and average protocol length when adding an amplitude and phase damping noise channel with error parameter $p$ after each gate. The noise parameter for all single-qubit gates is always fixed to $p_1 = 10^{-4}$. The number of measurements shots is set to 4096. For error parameters $p<10^{-3}$ we are able to retain high success probabilities. {\bf (e),(f)}~The success percentage and average protocol length when adding (coherent) Gaussian random noise with standard deviation $\sigma$ to the time step duration $\delta t_\pm$ of each action. For comparison, the original, unperturbed time step sizes $\delta t_+=\pi/8$ and $\delta t_-=\pi/13$ which the agent was trained on are indicated by the vertical dotted lines. This type of noise tests the ability of the agent to self-correct protocols in an online fashion.}
\end{figure*}

The quantum circuit mapping of $\ket{\theta_Q^\ell}$ is depicted in Fig.~\ref{fig:qmps_circ}(a). We can interpret the rightmost tensor $A^{[4]}$ as a single-qubit unitary, i.e., $G_4 = A_{\alpha_3}^{[4] j_4}$ as it satisfies Eq.~\eqref{eq:unitary}. Similarly, we can rewrite the adjacent tensor $A_{\alpha_2 \alpha_3}^{[3] j_3}$ with dimensions $4\times 2\times 2$ as a two-qubit unitary after reshaping the index $\alpha_2$: $G_3 = A_{\alpha_2,\alpha_2'}^{[3] j_3,\alpha_3}$. The next tensor $A_{\alpha_1 \alpha_2}^{[2] j_2}$ represents an isometry with input dimension 2 and output dimensions $4\times 2 $. Hence, we need to extend the columns of $A_{\alpha_1,0,0}^{[2] j_2,\alpha_2,\alpha_2'}$ by padding it with the $(2^3\!-2)$ orthonormal vectors $X$ in the kernel of $A^{[2]^\dagger}$. The resulting square matrix $G_2 = [X\ A^{[2]}]$ is then chosen as the three-qubit unitary. Finally, we can apply the same steps to the remaining isometry $A_{\alpha_1}^{[1] j_1}$, i.e., we pad the columns of $A_{0,0}^{[1] j_1,\alpha_1}$ with the $2^2$ dimensional kernel of $A^{[2]^\dagger}$ and interpret the resulting matrix as the two-qubit gate $G_1$.

Using the above mapping, an MPS circuit with bond dimension $\chi=2^n$ always contains at least one $(n+1)$-qubit gate. Thus, the $N\!=\!4$ QMPS with bond dimension $\chi\!=\!4$ results in a circuit including a three-qubit gate. However, the native gates realized in most present-day quantum computers contain at most two-qubit unitaries. Therefore, gates acting on more than two qubits first have to be decomposed into two-and single-qubit gates. Performing the decomposition in an exact manner is usually expensive, requires the use of optimization techniques, and often leads to very deep circuits nonetheless. With the short coherence times and large error rates of current quantum devices, it therefore quickly becomes infeasible to
execute MPS circuits of bond dimension $\chi>2$. Hence, we need alternative circuit mappings that give rise to at most two-qubit gates in the final circuit. The simplest approach is to truncate the given QMPS to a bond dimension $\chi=2$ MPS (see Fig.~\ref{fig:qmps_circ}(b) for the corresponding circuit). However, if the truncation errors are too large, the resulting circuit will not be an accurate description of the true quantum state anymore. Several approximative methods have been proposed to bridge this gap and prepare high fidelity states while restricting to the use of only two-qubit gates~\cite{barratt2021,lin2021,ran2020,Rudolph2022,Dov2022,Foss-Feig2022,Wall2022}. Note that all of these approaches can also be applied to the QMPS ansatz. For the remainder of this work we will consider the previously described exact mappings of the full $\chi=4$ and truncated $\chi=2$ QMPS as shown in Fig.~\ref{fig:qmps_circ}.

\subsection{Additional results on the noise-robustness}

This section reports further results of the QMPS circuit framework introduced in Sec.~\ref{NISQ} which is tested on the universal ground state preparation task of Sec.~\ref{sec:case1}.

First, we investigate how the number of measurement shots for sampling the fidelity in Eq.~\eqref{eq:fidelity} affects the performance of the QMPS protocols under an ideal (noise-free) simulator. To that end, we sample 1000 random initial states and compute the percentage of successfully prepared states (i.e., those runs for which the fidelity threshold is reached within 50 protocol steps). Moreover, we also store the corresponding average protocol length $\bar{T}$ and show the results in Fig.~\ref{fig:app:case4}(a)-(b). Surprisingly, as few as $500$ shots are already sufficient to reach success rates of close to unity. There are several reasons for this robust performance:~First, in the cases where the agent predicts a wrong action due to sampling noise, it can easily correct for the mistake in subsequent time steps since it learned to prepare the target state from \textit{any} quantum state. Second, although the quantum circuit output is noisy, we find that the subsequent neural network does not enhance the noise and still outputs reasonable values. Finally, since the optimal action is always determined by the argmax of the Q-values, noise in the output does not affect the chosen actions as long as its magnitude is sufficiently small. Hence, we can achieve high success probabilities even in the presence of sampling noise. This stands in contrast to policy gradient techniques where the network outputs the action values itself and a noisy output can therefore lead to faulty protocols.

The corresponding average protocol length $\bar{T}$ shown in Fig.~\ref{fig:app:case4}(b) converges to its value under exact computations (black dashed line) only after about $10^4$ shots. Notice that this is still within the feasible regime for many modern quantum devices. Note moreover that the performance of the truncated QMPS circuit (orange lines in Fig.~\ref{fig:app:case4}) is considerably worse and indicates that we indeed require a bond dimension of $\chi=4$ to faithfully represent the QMPS agent.

Next, we study the effects of a combined amplitude and phase damping noise channel on the performance of the QMPS circuit. Similar to the experiments involving depolarizing noise discussed in the main text, we add the amplitude and phase damping channel after each action gate and QMPS gate, and set the error parameters equal, i.e.,~$p=p_{\text{amp}}=p_{\text{phase}}$. Due to the lower error rates of single qubit gates, we fix the single-qubit gate errors to $p_1 = 10^{-4}$ and plot the success rate as a function of the two and three qubit error parameters $p$ after sampling 1000 different random initial states [Fig.~\ref{fig:app:case4}(c)]. Figure~\ref{fig:app:case4}(d) displays the corresponding average number of protocol steps. For error rates $p<10^{-3}$ we are able to reach success probabilities close to unity. However, they quickly deteriorate for larger error parameter values.

Finally, we analyze the robustness of the QMPS agent to coherent gate errors, similar to the discussion in Section~\ref{sec:case3}. Coherent errors arise when the actual, executed gate differs from the gate that has to be applied. These errors can often be mitigated by calibrating the devices carefully. However, frequent calibration is expensive and therefore coherent errors can usually not be eliminated fully. We simulate coherent gate errors for each of the 12 different actions by adding mean-zero Gaussian random noise of standard deviation $\sigma$ to the time step duration $\delta t_\pm$. In contrast to the discussion in Section~\ref{sec:case3}, each action gate is fixed, although the angles of rotation $\delta t$ are shifted compared to the original step size the agent was trained on. We again sample 1000 random initial states and show the state preparation success rates for varying standard deviations $\sigma$ in Fig.~\ref{fig:app:case4}(e). For each of the 1000 runs we also use a different random seed when sampling the gate noise. We observe that for standard deviations $\sigma\!<\!0.5$, the QMPS agent is still able to self-correct the protocols and reach the target state nonetheless. However, for larger amounts of noise the agent is not capable of reliably preparing the target state for all of the initial states. 

We expect that the performance of the QMPS agent can likely be improved by performing some additional optimization on the quantum device taking into account the exact noise model and error rates.

\end{document}